\documentclass[%
 dvipdfmx,
 reprint,
unsortedaddress,
showpacs,preprintnumbers,
amsmath,amssymb,
aps, 
prd,
10pt,
floatfix
]{revtex4-1}

\usepackage[dvips]{graphicx}
\usepackage{dcolumn}
\usepackage{bm}
\usepackage[hidelinks,colorlinks=true,urlcolor={blue!50!black},linkcolor={blue!50!black},citecolor={blue!50!black}]{hyperref}
\usepackage[caption=false,subrefformat=parens,labelformat=parens]{subfig}
\usepackage[usenames,dvipsnames]{xcolor}

\def\be{\begin{equation}}
\def\ee{\end{equation}}

\def\({\left(}
\def\){\right)}

\def\[{\left[}
\def\]{\right]}

\begin{document}


\title{Late-time vacuum phase transitions: Connecting sub-eV scale physics with cosmological structure formation}

\author{Amol V. Patwardhan}
\email{apatward@ucsd.edu}
\affiliation{Department of Physics, University of California, San Diego, La Jolla, California 92093-0319, USA}
\author{George M. Fuller}
\email{gfuller@ucsd.edu}
\affiliation{Department of Physics, University of California, San Diego, La Jolla, California 92093-0319, USA}
\affiliation{Kavli Institute for Cosmological Physics, The University of Chicago, Chicago, Illinois 60637, USA}





\date{\today}

\begin{abstract}
We show that a particular class of postrecombination phase transitions in the vacuum can lead to localized overdense regions on relatively small scales, roughly $10^6$ to $10^{10}\, M_\odot$, potentially interesting for the origin of large black hole seeds and for dwarf galaxy evolution. Our study suggests that this mechanism could operate over a range of conditions which are consistent with current cosmological and laboratory bounds. One byproduct of phase transition bubble-wall decay may be extra radiation energy density. This could provide an avenue for constraint, but it could also help reconcile the discordant values of the present Hubble parameter ($H_0$) and $\sigma_8$ obtained by cosmic microwave background (CMB) fits and direct observational estimates. We also suggest ways in which future probes, including CMB considerations (e.g., early dark energy limits), 21-cm observations, and gravitational radiation limits, could provide more stringent constraints on this mechanism and the sub-eV scale beyond-standard-model physics, perhaps in the neutrino sector, on which it could be based. Late phase transitions associated with sterile neutrino mass and mixing may provide a way to reconcile cosmological limits and laboratory data, should a future disagreement arise.
\end{abstract}


\maketitle


\section{Introduction} \label{sec:intro}
 
In this paper we investigate the potential consequences of new sub-eV scale physics, specifically the cosmological implications of a vacuum phase transition occurring after photon decoupling. The experimental revelation of neutrino mass and flavor mixing physics, and the puzzle of the origin of neutrino masses, provide speculative license for this investigation \cite{Raffelt:1987kx,Fuller:1992lr,Kolb:1992dq,Frieman:1992ly,Barbieri:2005qf,Bamba:2008pd,Bhatt:2010ve}, and lower energy-scale phase transitions in the early Universe have been considered before \cite{Primack:1980uq,Wasserman:1986lr,Hill:1989fj,Press:1990vn,Fuller:1992lr,Kolb:1992dq,Frieman:1992ly,Luo:1994rt,Dutta:2009lr}. While our considerations are generic and need not pertain exclusively to the neutrino sector, the work presented here attempts to connect this speculative vacuum physics both with emerging observational probes and with unresolved problems in cosmology, in particular the origin of the seeds for supermassive black holes.

Advances in observational astronomy in the last few decades have allowed us to probe objects and structure at high redshifts, opening up opportunities to examine the state of the Universe at remote epochs. For example, observations have provided evidence for the existence of ultra-massive black holes ($\sim$$10^9 \text{--} 10^{10}\, M_\odot$) at high redshifts ($z \approx 5\text{--}7$) \cite{Fan:2001qy,Fan:2003lr,Willott:2003fk,Mortlock:2011lr}, corresponding to epochs where the Universe was only of order a billion years old. Explaining the existence of these objects, in the context of a standard $\Lambda$CDM cosmological model (i.e., a cosmological constant + cold dark matter) with Gaussian initial fluctiations (presumably from inflation), remains a challenge, and there have been several attempts to address this question \cite{Efstathiou:1988lr,Haiman:2001fk,Yoo:2004fk,Shapiro:2005qy, Khlopov:2005uq,Volonteri:2005uq,Volonteri:2006fj,King:2006kx,Li:2007yq,Kawakatu:2009vn,Sijacki:2009rt}.

Wasserman, in Ref.~\cite{Wasserman:1986lr}, suggested a novel way of dealing with this problem, although the primary motivation behind that study was an attempt to explain the organization of large-scale structure. That work outlined a mechanism through which a first-order vacuum phase transition could gravitationally bind comoving regions with scales small compared to the horizon size. In this paper we revisit this Wasserman mechanism. We modify this mechanism, highlight the role played by the current observed vacuum (dark) energy $\Lambda$, and show how it renders the binding process significantly more difficult to accomplish. Nevertheless, we demonstrate that this mechanism can produce nonlinear regime fluctuations on scales roughly $10^6\text{--}10^{10}\, M_\odot$, at high redshifts ($z\,\sim\, 3\text{--}10$, for a phase transition occurring at $z\,\sim\, 50\text{--}500$), and is subject to constraint by current and future observations. We limit our analysis to phase transitions in the postrecombination era so as to bypass the complications associated with damping of density perturbations via radiation diffusion.

In Sec.~\ref{sec:bg} we provide an overview of the local dynamics and conserved quantities in the expansion of the Universe, and in Secs.~\ref{sec:trans} and \ref{sec:nusc} we build on this and describe the Wasserman mechanism and the physics of cosmic vacuum phase transition nucleation in this context. Issues surrounding fluctuation binding and growth are discussed in Sec.~\ref{sec:binding}. Observational and experimental constraints and probes are outlined in Sec.~\ref{sec:cons}, and conclusions are given in Sec.~\ref{sec:concl}, along with speculations about possible connections to neutrino physics. 


\section{Background} \label{sec:bg}

\begin{figure}[b]
	\centering
	\includegraphics[width=0.3\textwidth]{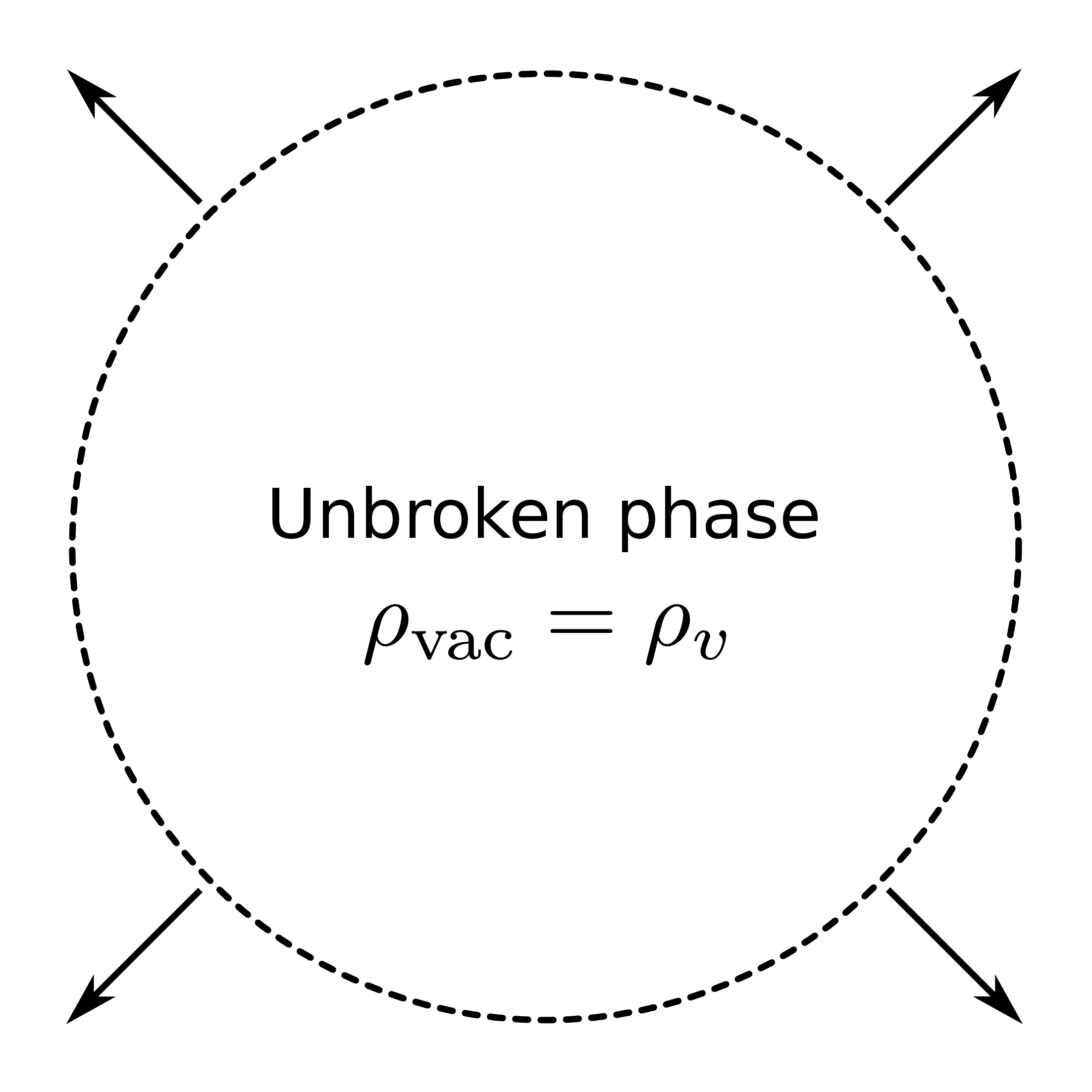}
	\caption{A comoving coordinate sphere expanding with the Hubble flow prior to the onset of the phase transition}
	\label{fig:sphere}
\end{figure}

We take the Universe prior to the phase transition to be homogeneous and isotropic and described by the Friedmann-Lema\^itre-Robertson-Walker (FLRW) metric. At any time $t$, the proper distance $d(t)$ of a point on an imaginary spherical shell of comoving radius $r$ (e.g., Fig.~\ref{fig:sphere}), from its center, is given by $d(t) = r\, a(t)$, where $a(t)$ is the scale factor. The Hubble parameter is $H(t) \equiv \dot{a}(t)/a(t)$, where $\dot{a} \equiv da/dt$ is the derivative of the scale factor with respect to FLRW coordinate time $t$. The location of the center of the shell can be chosen arbitrarily because of the spacetime symmetry, and is conveniently taken to be at the origin of our coordinate system.

The evolution of the scale factor with time is given by the Friedmann equation, which can be derived from the homogeneity and isotropy symmetry of this spacetime via Birkhoff's theorem. In the context of an FLRW spacetime, for regions small compared to the causal horizon length, i.e., $d\(t\) \ll H^{-1}(t)$, Birkhoff's theorem implies that the \lq\lq total mechanical energy\rq\rq\ of a comoving spherical shell is conserved. For a spherical shell of coordinate radius $r$, this condition can be written as
\be
	\frac{1}{2} m\, r^2\dot{a}^2(t) + \(-\frac{G (4/3)\, \pi\, r^3 a^3(t)\, m\, \rho} {r\, a(t)}\) = E,
\ee
where $G \equiv 1/m_P^2$ is the gravitational coupling constant, and $m_P \approx 1.22 \times 10^{22}$ MeV is the Planck mass. The two terms on the left-hand side of the equation can be interpreted as the kinetic and the \lq\lq gravitational potential\rq\rq\ energies of a test particle of negligible mass $m$ on the spherical shell. Here $\rho$ is the total mass-energy density. Dividing by $m r^2/2$, we obtain the familiar Friedmann equation
\be \label{eq:Friedmann}
	\dot{a}^2(t) - \frac{8\pi G}{3} \rho\, a^2(t) = -k,
\ee
where we take $k = -2E/mr^2$. The constant $k$ is related to the spatial Ricci curvature scalar and can take the values $\pm 1$ or $0$. The energy density $\rho$ includes contributions from nonrelativistic ($\rho_\text{NR}$), relativistic ($\rho_R$), and vacuum ($\rho_\text{vac}$) energy densities. Observational data suggest that our Universe is \lq\lq critically dense\rq\rq\ \cite{Hou:2012zr,Planck-Collaboration:2013zr,Hinshaw:2013ys,Bennett:2013fr,Sievers:2013lr}, corresponding to $k=0$ in Eq. (\ref{eq:Friedmann}), i.e., zero total energy on any comoving spherical shell. We can thus write
\be
	\dot{a}^2(t) - \frac{8\pi G}{3} \(\rho_\text{NR} + \rho_{R} + \rho_\text{vac}\) a^2(t) = 0.
\ee

The various components of energy density differ in the manner in which they depend on the scale factor: $\rho_\text{NR} \propto a^{-3}$ (follows from mass conservation), whereas $\rho_R \propto a^{-4}$ (a consequence of Stefan-Boltzmann law), and $\rho_\text{vac}$ does not depend on the scale factor at all. Consequently, if we define the scale factor to be $a(t_i) \equiv 1$ at some initial time $t_i$, then at any subsequent time $t$ we have
\be \label{eq:initial}
	\dot{a}^2(t) - \frac{8\pi G}{3} \[\frac{\rho_\text{NR}(t_i)}{a^3(t)} + \frac{\rho_{R}(t_i)}{a^4(t)} + \rho_\text{vac}\] a^2(t) = 0.
\ee

The relative mix of the various energy densities contributing to the gravitational potential will then dictate how a comoving volume evolves with time.

\section{Transition dynamics} \label{sec:trans}

Following Wasserman~\cite{Wasserman:1986lr}, we assume that a first-order phase transition in the vacuum takes place at some epoch after photon decoupling. The transition causes separation of phases via a bubble nucleation process \cite{Coleman:1977lr,Coleman:1977qy,Callan:1977fk}, leading to relatively small, initially spherical density fluctuations, distributed more or less evenly in space. The vacuum energy density $\rho_\text{vac}$ is assumed to drop across the bubble wall, from an initial value $\rho_v$ in the unbroken phase to its currently observed value $\rho_\Lambda \approx 3.5$ keV/cm$^3$ in the broken phase. 

\begin{figure}[htb]
	\centering
	\includegraphics[width=0.4\textwidth]{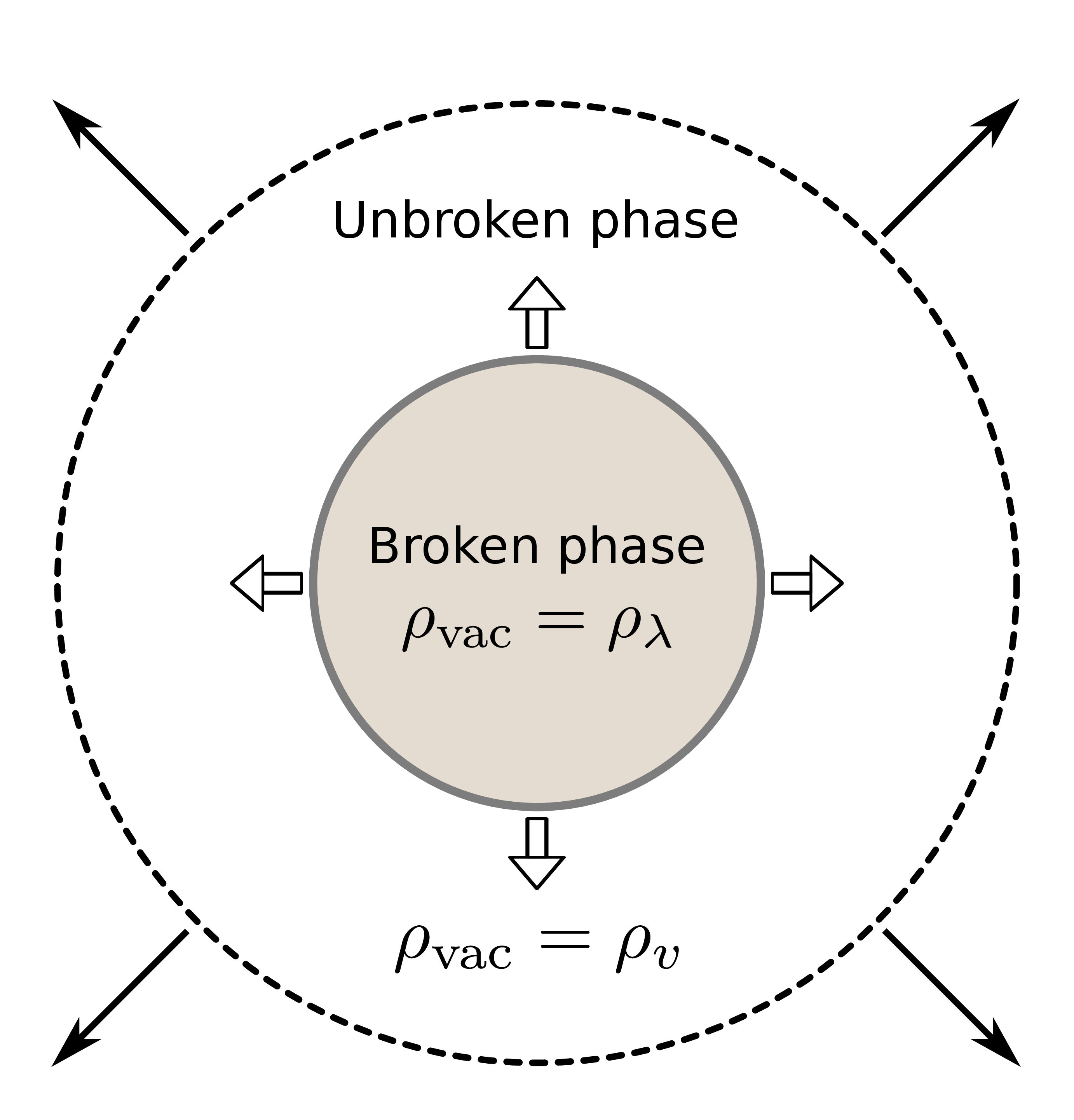}
	\caption{A bubble of the broken phase (shaded) nucleates inside a comoving spherical shell (dashed line) and expands relativistically. As it expands, it sweeps up most of the vacuum energy from the unbroken phase onto its wall (thick gray line).}
	\label{fig:bubble}
\end{figure}

Consider the evolution of a bubble that nucleates at time $t = t_\text{nuc}$, at the center of a comoving spherical shell, i.e., at $r = 0$. As shown in Refs.~\cite{Coleman:1977lr,Coleman:1977qy,Callan:1977fk,Ignatius:1994uq}, the spherical bubble containing the broken phase expands outwards, quickly ramping up to relativistic speed, and \lq\lq sweeping up\rq\rq\ the vacuum energy ($\rho_v - \rho_\Lambda$) from the surrounding unbroken phase onto its boundary (e.g., see Fig.~\ref{fig:bubble}). Another, perhaps more correct, interpretation of this phenomenon would be to identify the difference between the vacuum energies of the unbroken and the broken phase as a kinetic energy, which resides at the expanding phase boundary/bubble wall (not to be confused with the kinetic energy of the comoving shells).

\subsection{Shell crossing}

Suppose the expanding bubble wall crosses a comoving spherical shell of coordinate radius $r$ at some time $t_c(r) > t_\text{nuc}$. For $t < t_c(r)$, the equation of motion for the comoving spherical shell has exactly the same form as Eq. (\ref{eq:initial}) with $t_i = t_\text{nuc}$ [let us define $a(t_\text{nuc}) \equiv 1$], and $\rho_\text{vac} = \rho_v$. However, for $t > t_c(r)$ the expanding bubble wall carrying the swept-up vacuum energy has escaped the interior of the comoving sphere. The equation of motion of the comoving sphere therefore changes, as its kinetic energy is the same, but its gravitational potential energy is now smaller in magnitude. We now define the proper distance from the origin to the comoving shell as $d(t) = r\, a(t;r)$, where $a(t;r)$ is the modified scale factor for that comoving shell, inside the bubble volume. Using Birkhoff's theorem, the equation for energy conservation can now be written as (suppressing the $m r^2$ factors)
\be
\begin{split} \label{eq:crossing}
	\frac{1}{2}&\dot{a}^2(t;r) - \frac{4\pi Ga^3(t;r)}{3a(t;r)} \[\frac{\rho_{\text{NR},n}}{a^3(t;r)} + \frac{\rho_{R,n}}{a^4(t)} + \rho_\Lambda\] \\
	&= \frac{4\pi Ga^3(t_c(r);r)(\rho_v-\rho_\Lambda)}{3a(t_c(r);r)},
\end{split}	
\ee
where the symbols labeled with the subscript \lq\lq$,n$\rq\rq\ are being evaluated at $t = t_\text{nuc}$, i.e., at the onset of the phase transition. As the energy swept up on the bubble wall escapes the interior of the comoving sphere, the vacuum energy term in the gravitational potential drops from $\rho_v$ to $\rho_\Lambda$, making the potential less negative. The comoving shell thus becomes temporarily unbound, acquiring a positive total energy which appears on the right-hand side in Eq. (\ref{eq:crossing}). Upon simplification, we obtain the evolution equation for the modified scale factor $a(t;r)$,
\be \label{eq:crossed}
\begin{split}
	\dot{a}^2(t;r) = &\frac{8\pi G}{3} \[\frac{\rho_{\text{NR},n}}{a^3(t;r)} + \frac{\rho_{R,n}}{a^4(t)} + \rho_\Lambda\]a^2(t;r) \\
					&+ \frac{8\pi G}{3}(\rho_v-\rho_\Lambda)a_c^2(r),
\end{split}
\ee
where $a_c(r) \equiv a(t_c(r)) = a(t_c(r);r)$.

\begin{figure}[b]
		\centering
		\includegraphics[width=0.4\textwidth]{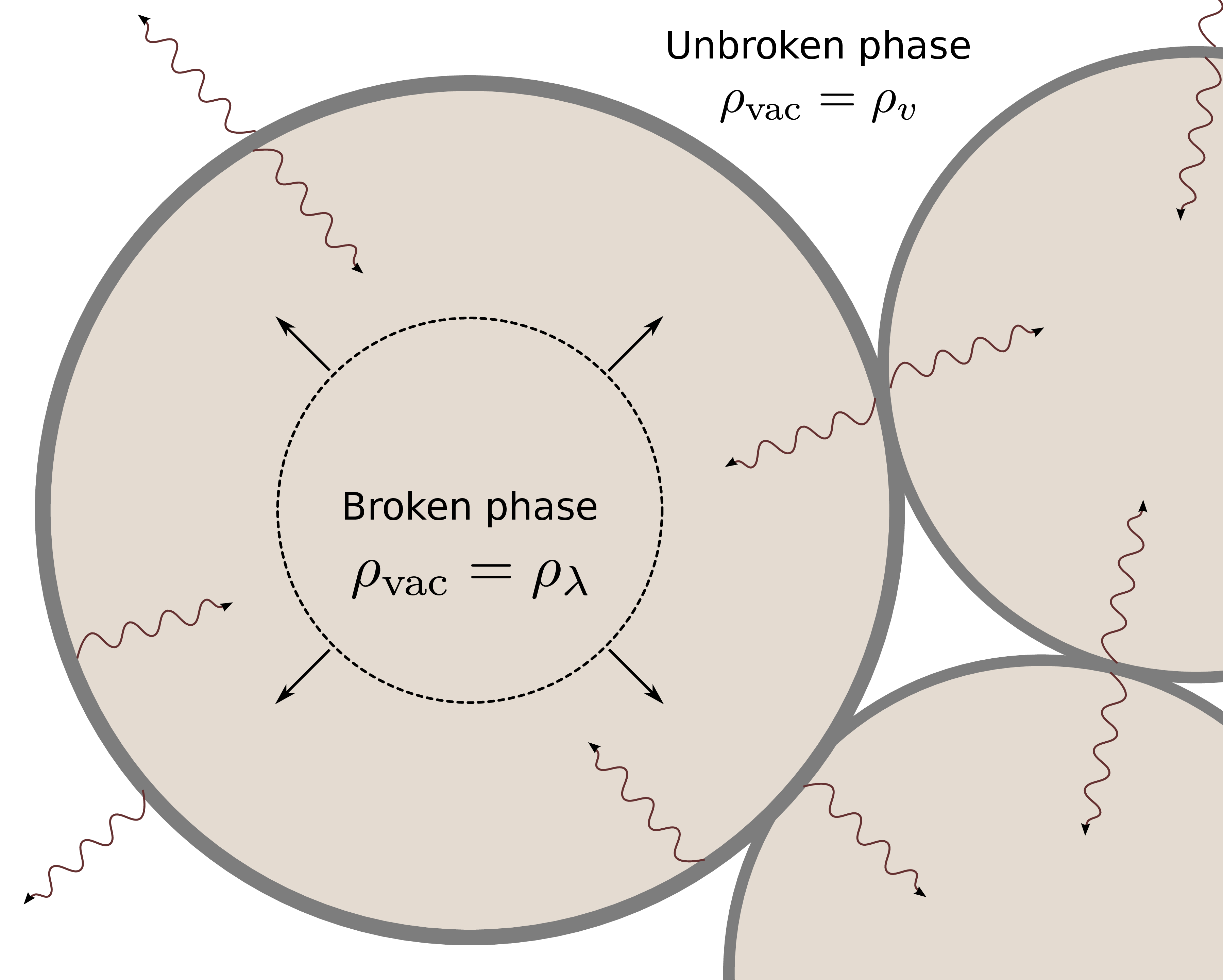}
		\caption{The expanding bubble crosses the comoving sphere and collides with an adjacent bubble. We assume that the bubble walls disintegrate as they collide, radiating away the swept-up vacuum energy (i.e., the kinetic energy of the bubble walls).}
		\label{fig:crossing}
\end{figure}

\subsection{Energy redistribution}

Suppose the expansion of the bubble wall stops at $t = t_f > t_c(r)$ when it collides with an adjacent expanding bubble (e.g., Fig.~\ref{fig:crossing}). Let us assume (again, in tune with Ref.~\cite{Wasserman:1986lr}) that the bubble walls disintegrate, and the swept-up energy is redistributed via emission of some kind of radiation that quickly fills up the fluctuation region uniformly with a relativistic energy density $\rho_v-\rho_\Lambda$ (which then begins to redshift away with the scale factor). We shall (for the most part) remain agnostic about the identity of this radiation, only assuming that all its couplings are weak enough to allow us to neglect any nongravitational effects. Admittedly, this is just one among several possible outcomes of such a vacuum phase transition, the dynamics of which are governed by the underlying physics. References \cite{Giblin:2010lr,Johnson:2010fk,Giblin:2013uq} have explored some of the other possibilities through numerical simulations of bubble collisions. Once the fluctuation region has been filled up, we can use Birkhoff's theorem again and write the condition for energy conservation as
\be \label{eq:redist}
\begin{split}
	\frac{1}{2}&\dot{a}^2(t;r) - \frac{4\pi G}{3} \bigg[\frac{\rho_{\text{NR},n}}{a^3(t;r)} + \frac{\rho_{R,n}}{a^4(t)} \\ &+ (\rho_v-\rho_\Lambda)\frac{a^4(t_f)}{a^4(t)} + \rho_\Lambda \bigg]a^2(t;r) \\
	= &\frac{4\pi Ga_c^2(r)(\rho_v-\rho_\Lambda)}{3}-\frac{4\pi Ga^3(t_f)(\rho_v-\rho_\Lambda)}{3a(t_f)}.
\end{split}	
\ee

The energy density $(\rho_v-\rho_\Lambda)a^4(t_f)/a^4(t)$ representing this leftover radiation now starts contributing to the gravitational potential, making it more negative. Therefore, to balance the books, the total energy, i.e., the right-hand side of Eq. (\ref{eq:redist}), also picks up an additional negative contribution $-4\pi G(\rho_v-\rho_\Lambda)a^2(t_f)/3$. Since $a(t_f) > a_c(r)$, it is immediately obvious that the total energy is now negative, which opens up the possibility of the spherical fluctuations becoming gravitationally bound and evolving away from the background FLRW metric, growing significantly overdense with time. Simplifying the above equation yields
\be \label{eq:binding}
\begin{split}
	\dot{a}^2(t;r) = &\frac{8\pi G}{3} \bigg[\frac{\rho_{\text{NR},n}}{a^3(t;r)} + \frac{\rho_{R,n}}{a^4(t)} + (\rho_v-\rho_\Lambda)\frac{a^4(t_f)}{a^4(t)} + \rho_\Lambda \bigg] \\&\times a^2(t;r) + \frac{8\pi G}{3}(\rho_v-\rho_\Lambda)(a_c^2(r)-a^2(t_f)).
\end{split}	
\ee

It may be noted that, in each of our equations of motion, all relativistic energy densities are seen to redshift with the universal scale factor $a(t)$, rather than the local scale factor $a(t;r)$. This is because as long as the radiation is sufficiently weakly coupled, any bubble-sized inhomogeneities in radiation density quickly get smeared out on time scales much shorter than the Hubble time. The relationship between the typical bubble size and the Hubble scale is elucidated in the following section.

\section{Nucleation scale} \label{sec:nusc}

Previous studies have identified a characteristic size for the bubbles of the broken phase at the end of a first-order relativistic cosmological phase transition~\cite{Coleman:1977lr,Coleman:1977qy,Callan:1977fk,Hogan:1983fk}, provided certain conditions are met (such as the nucleating action depending on cosmological background quantities and having a certain functional form). It has been shown that the ratio of the typical bubble size at the end of the phase transition to the Hubble length $H_c^{-1}$ at that epoch, is given by
\be \label{eq:delta}
	\delta \approx \[4B_1 \ln{\frac{m_P}{T_c}}\]^{-1},
\ee
where $T_c$ is the critical temperature (i.e., the temperature of the photon background at the onset of the phase transition). $B_1$ is the logarithmic derivative of the nucleating action in units of cosmological time $t$, and can be shown to be of $\mathcal{O}(1)$ or bigger (we have used $B_1 = 1$ in all our calculations). Equation (\ref{eq:delta}), in its stated form, is strictly valid only for radiation-dominated epochs, but the differences between that and the more correct expression occur in the argument of the logarithm, and are therefore rendered insignificant because $m_P \gg T_c$. The ratio of the typical bubble size to the Hubble length can thus be seen to scale only logarithmically with the critical temperature, and for $T_c \sim 0.01\text{--}0.1$ eV (which is the range we are interested in), the suppression factor is typically $\delta \sim 1/300B_1$. The typical spatial extent of the vacuum bubbles at the end of the phase transition can then be identified as
\be \label{eq:nusc}
	R_f \equiv \delta \, H_c^{-1} \sim \[4B_1 \ln{\frac{m_P}{T_c}}\]^{-1} \[\frac{8\pi G}{3}\rho_c\]^{-1/2},
\ee 
where the total energy density $\rho_c$ at the time of phase transition sets the Hubble scale at that epoch. The total mass enclosed within the fluctuation region is then given by $M_f \approx (4/3) \pi R_f^3 \rho_{\text{NR},n}$, setting a conservative upper limit on the mass that one might expect to collapse via such a mechanism. It turns out that the typical mass within a bubble volume ranges from $M_f \sim 5\times10^8\, M_\odot$ to $M_f \sim 3\times10^{11}\, M_\odot$ across our parameter space. The reason for this being a conservative limit is that, in practice, only a fraction of this mass may become gravitationally bound, as demonstrated in the following section.

\section{Fluctuation binding and growth} \label{sec:binding}

In Sec.~\ref{sec:trans}, we described how the process of energy redistribution via bubble nucleation can modify the equations of motion of a comoving spherical shell within the fluctuation region. At the end of the phase transition, the Friedmann equation was seen to pick up the additional term $(8\pi G/3)(\rho_v-\rho_\Lambda)(a_c^2(r) - a^2(t_f))$, which with Birkhoff's theorem could be interpreted as a binding energy on account of it being negative.

It can be shown that under specific circumstances, this binding energy can cause the expansion of the comoving regions to slow down and eventually stop. The effectiveness of this mechanism can be characterized by identifying a time scale over which locally overdense regions are formed. We can choose this to be the value of the FLRW time coordinate at which the expansion of a particular comoving shell stops, i.e., when the time derivative of the local scale factor [Eq. (\ref{eq:binding})] associated with that shell drops to zero. We call this the \lq\lq halting time\rq\rq\ for that comoving shell---this is analogous to the definition of \lq\lq turnaround time\rq\rq\ in conventional models of structure growth. Alternatively, we could look for time scales over which fluctuations of different sizes become nonlinear, by finding the value of the time coordinate at which $\delta\rho/\rho_\text{NR} = 1$. Here, $\delta\rho$ is the density perturbation, defined as the difference between the nonrelativistic matter density inside a bound comoving sphere and the average nonrelativistic matter density throughout the Universe, i.e.,
\be \label{eq:nonlin}
	\delta\rho(t;r) \equiv \rho_\text{NR}' - \rho_\text{NR} = \frac{\rho_{\text{NR},n}}{a^3(t;r)} - \frac{\rho_{\text{NR},n}}{a^3(t)}.
\ee

It is evident that the binding energy depends on the comoving radius $r$ [the dependence coming from the $a_c^2(r) - a^2(t_f)$ factor], and therefore the associated collapse time scales (halting time or time-to-nonlinearity) are also functions of $r$. And since $r$ is directly related to the mass scale of the comoving volume, it allows us to estimate the time scales over which comoving regions of different mass tend to become overdense.

\subsection{Toy model analysis}

\begin{figure}[htb]
		\centering
		\includegraphics[width=0.4\textwidth]{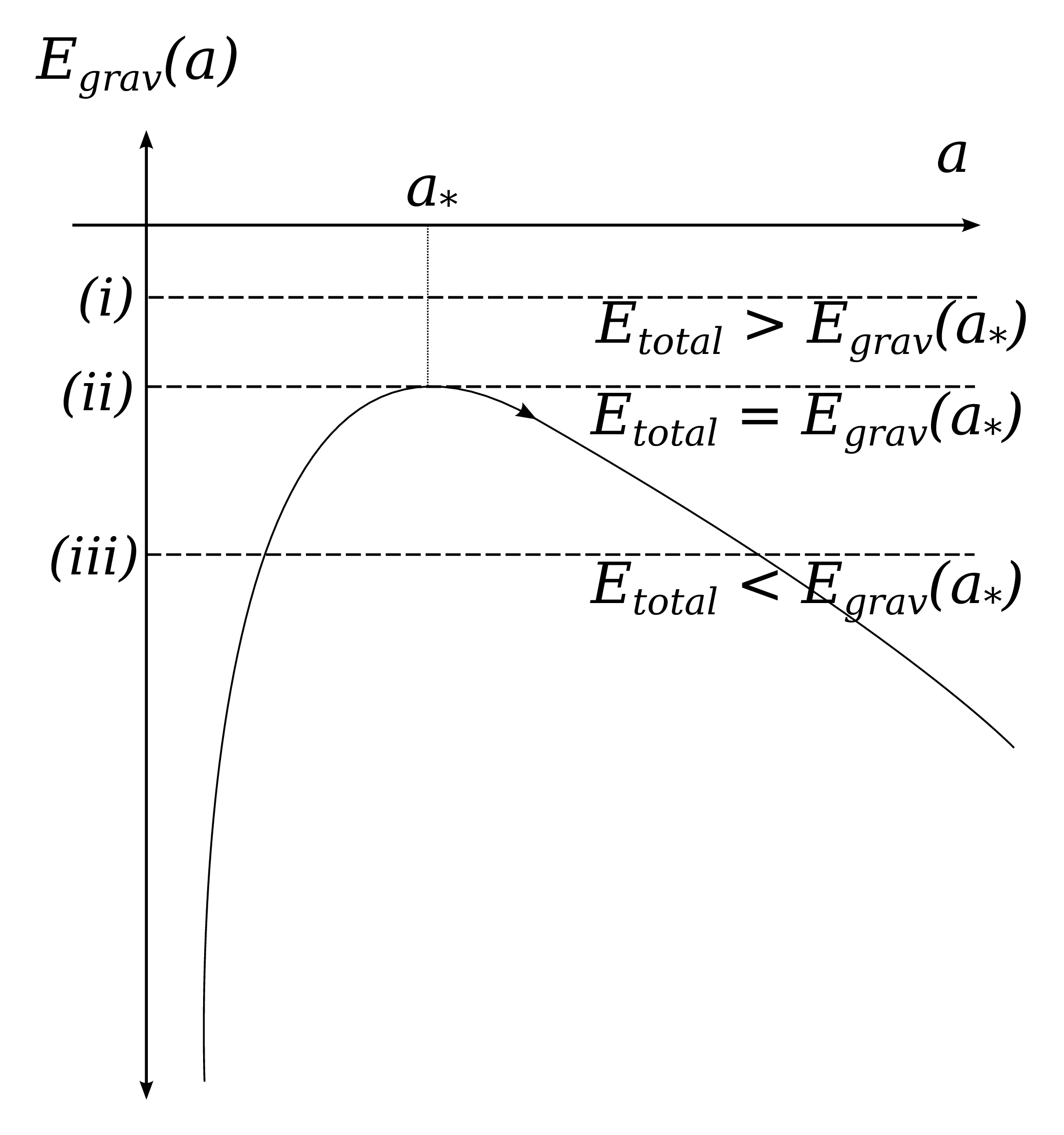}
		\caption{Cartoon illustrating the energetics associated with fluctuation binding. For binding to be accomplished, the kinetic energy has to go to zero, which can only happen if the total energy $E_\text{total}$ (horizontal dashed line) intersects the potential energy curve ($E_\text{grav}$, solid line). Cases (ii) and (iii) represent configurations which allow binding [with (ii) being right at the threshold], whereas case (i) depicts a situation where binding cannot be accomplished in spite of a negative total energy.}
		\label{fig:binding}
\end{figure}

Using a simple toy model we can demonstrate that, for the range of temperatures we are interested in, binding by this mechanism would not be possible unless the vacuum energy density prior to the phase transition were at least a few orders of magnitude higher than its present value. To simplify matters, let us ignore the contributions to the Friedmann equation coming from relativistic energy densities (not totally unreasonable, since these redshift away quickly and thus become insignificant at late times). With that approximation, Eq. (\ref{eq:binding}) reduces to
\be
\begin{split}
	\dot{a}^2(t;r) = &\frac{8\pi G}{3} \(\frac{\rho_{\text{NR},n}}{a(t;r)} + \rho_\Lambda a^2(t;r)\) \\
					&- \frac{8\pi G}{3}(\rho_v-\rho_\Lambda)[a^2(t_f)-a_c^2(r)] \\
					= &\frac{8\pi G}{3} \(\frac{\rho_{\text{NR},n}}{a(t;r)} + \rho_\Lambda a^2(t;r) - \rho_{BE}\),
\end{split}
\ee
where we have defined $\rho_{BE} \equiv (\rho_v-\rho_\Lambda)[a^2(t_f)-a_c^2(r)]$. Now the expansion of the comoving sphere halts when the right-hand side of the above equation goes to zero. For that to happen, $\rho_{BE}$ has to be at least as large as the minimum value attained by $\rho_{\text{NR},n}/a(t;r) + \rho_\Lambda a^2(t;r)$ (e.g., see Fig.~\ref{fig:binding}). Taking the derivative of this expression with respect to $a(t;r)$ and setting it to zero implies that this minimum value is attained when $a(t;r) = (\rho_{\text{NR},n}/2\rho_\Lambda )^{1/3}$, which gives the minimum required $\rho_{BE}$, 
\be
	(\rho_{BE})_\text{min} = \frac{3}{2^{2/3}}\rho_{\text{NR},n}^{2/3}\, \rho_\Lambda ^{1/3}.
\ee

It is more instructive to represent this in terms of the ratio $\rho_{BE}/\rho_\Lambda$. The corresponding minimum value of this ratio necessary for binding is then given by
\be
	\(\frac{\rho_{BE}}{\rho_\Lambda }\)_\text{min} = \frac{3}{2^{2/3}}\(\frac{\rho_{\text{NR},n}}{\rho_\Lambda }\)^{2/3}.
\ee

By using the observed closure fractions $\Omega_\text{NR}$ and $\Omega_\Lambda$ at the current epoch, the ratio $\rho_{\text{NR},n}/\rho_\Lambda$ at the time of phase transition can be expressed as a function of the critical temperature alone. $\rho_\Lambda$ is just a constant energy density, whereas $\rho_\text{NR}$ at any time is proportional to the cube of the temperature (inverse cube of the scale factor). Consequently, we can write
\be
\frac{\rho_{\text{NR},n}}{\rho_\Lambda} = \[\frac{\Omega_\text{NR}}{\Omega_\Lambda}\]_0\(\frac{T_c}{T_0}\)^3,
\ee
with the subscript \lq\lq 0\rq\rq\ being used to denote the values at the current epoch. The minimum required $\rho_{BE}/\rho_\Lambda$ in terms of the critical temperature is then given by
\be
\(\frac{\rho_{BE}}{\rho_\Lambda }\)_\text{min} = \frac{3}{2^{2/3}}\[\frac{\Omega_\text{NR}}{\Omega_\Lambda}\]_0^{2/3}\(\frac{T_c}{T_0}\)^2.
\ee

Since $T_0 \approx 0.23$ meV, the minimum $\rho_{BE}/\rho_\Lambda$ required for binding turns out to be $\sim$$10^4$ for $T_c \sim 0.01\text{--}0.1$ eV. Recalling that $\rho_{BE}$ is simply $(\rho_v-\rho_\Lambda)[a^2(t_f)-a_c^2(r)]$, it immediately follows that the ratio $\rho_v/\rho_\Lambda$ has to be quite large for this mechanism to bring about binding. It can be shown that the quantity $a^2(t_f) - a_c^2(r)$ can at most be of the same order as the ratio of transition width to Hubble time, i.e., $\delta$, as defined in Sec.~\ref{sec:nusc}. Quantitatively, it follows that $a^2(t_f) - a_c^2(r)$ is of order $10^{-2}$ or smaller, implying that $\rho_v/\rho_\Lambda$ has to be of order $10^6$ or bigger. The numbers change slightly [by some $\mathcal{O}(1)$ factor] when relativistic energy densities are included in the calculation, but the principle remains the same. This equips us with the foresight to immediately exclude certain regions from our parameter space where fluctuation binding via this mechanism is theoretically impossible to accomplish.

\subsection{Numerical calculations}

\begin{figure}[b]
	\centering
	\subfloat[$r = 0.1 \, (\delta \, H_c^{-1})$]{\label{fig:ctr1}\includegraphics[width=0.5\textwidth]{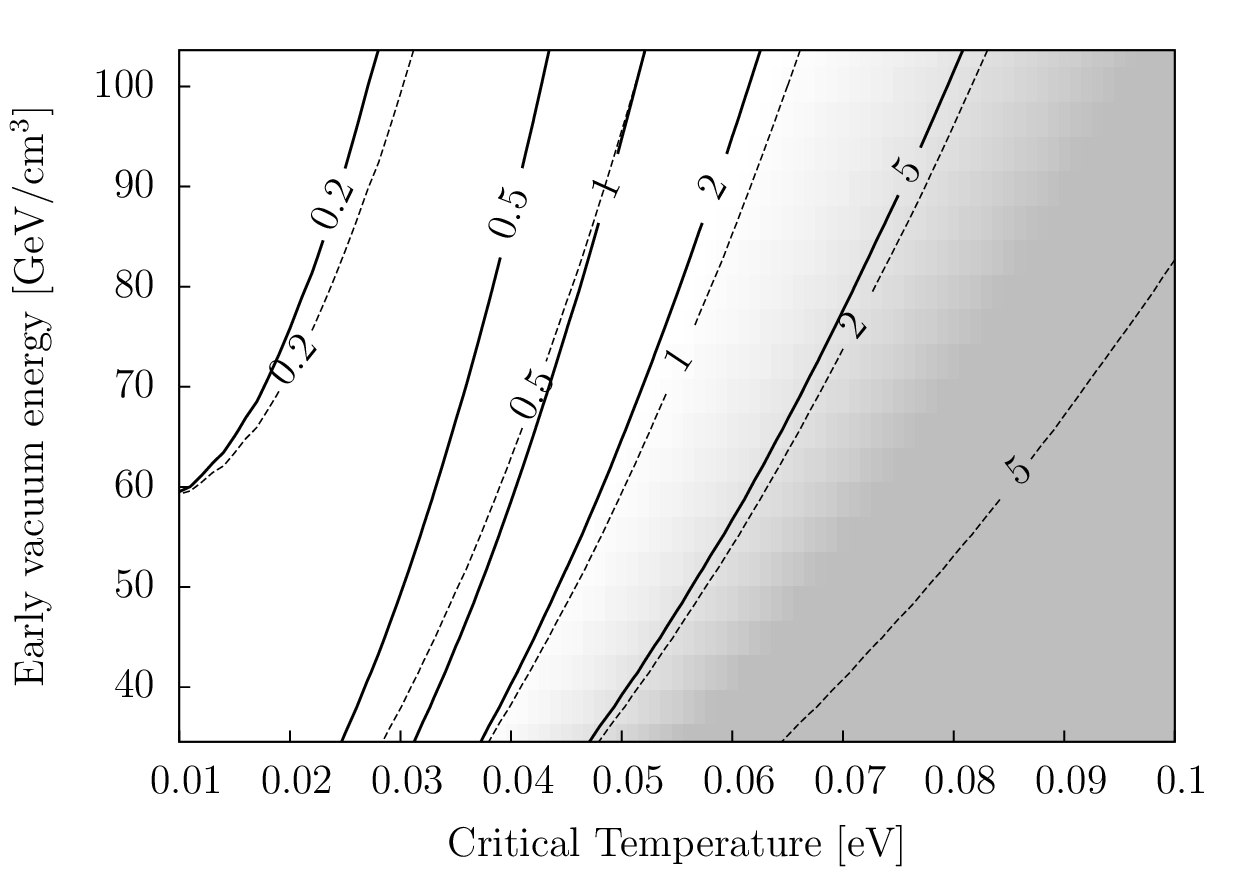}} \\
	\subfloat[$r = 0.5 \, (\delta \, H_c^{-1})$]{\label{fig:ctr5}\includegraphics[width=0.5\textwidth]{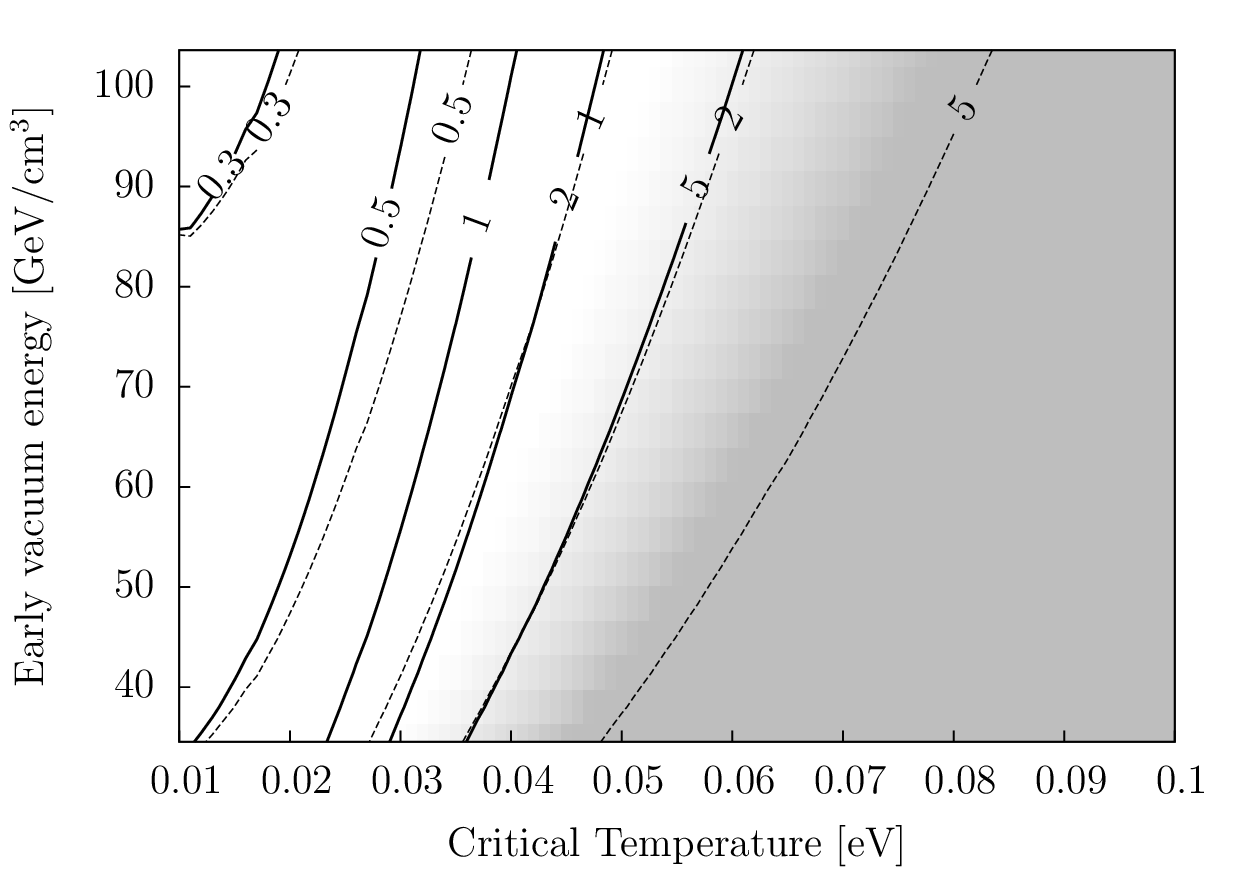}}
	\caption{Contours of halting time (solid) and time-to-nonlinearity (dashed), labeled in gigayears, across a parameter space spanned by critical temperature and early vacuum energy density. Plots (a) and (b) correspond to two different comoving shells, labeled with their respective coordinate radii in relation to the typical nucleation scale $\delta \, H_c^{-1}$. Shaded areas indicate regions of the parameter space where comoving fluctuations of the specified radii cannot become gravitationally bound, either within a $\mathcal{O}(1\text{ Gyr})$ time scale, or at all.}
	\label{fig:ctr}
\end{figure}

Numerical results can be obtained by first integrating Eq. (\ref{eq:crossed}) from $t_c(r)$ to $t_f$. This allows us to explicitly compute the quantity $a_c^2(r) - a^2(t_f)$ which can then be used in Eq. (\ref{eq:binding}). Integrating Eq. (\ref{eq:binding}) then allows us to compute the time scales defined earlier, i.e., the halting time, or the time-to-nonlinearity. The results of these computations depend on the two extra parameters of our model---the critical temperature $T_c$ and the early vacuum energy density $\rho_v$. We can then look for regions of the parameter space where the above time scales are of the order of $1$ Gyr or lower (a $1$ Gyr cosmological time corresponds to a redshift $z \approx 5.5$, whereas $2$ Gyr and $0.5$ Gyr correspond to $z \approx 3$ and $z \approx 10$, respectively).

Figure \ref{fig:ctr} shows contours of halting time and time-to-nonlinearity plotted against the critical temperature $T_c$ in eV; and the early vacuum energy density $\rho_v$ in GeV/cm$^3$ (for comparison, the current observed vacuum energy density is about $3.5 \times 10^{-6}$ GeV/cm$^3$). For a given set of parameter values, the aforementioned time scales may be calculated for different comoving shells, i.e., at different coordinate radii within the fluctuation volume [plots (a) and (b) in Fig.~\ref{fig:ctr}]. These plots reveal that binding is favored at lower critical temperatures and higher values of early vacuum energy density. Also it can be observed that, within a fluctuation volume, the inner regions (smaller comoving radii) become bound on shorter time scales compared to the outer regions (larger radii).

\begin{figure}[b]
	\centering
	\includegraphics[width=0.5\textwidth]{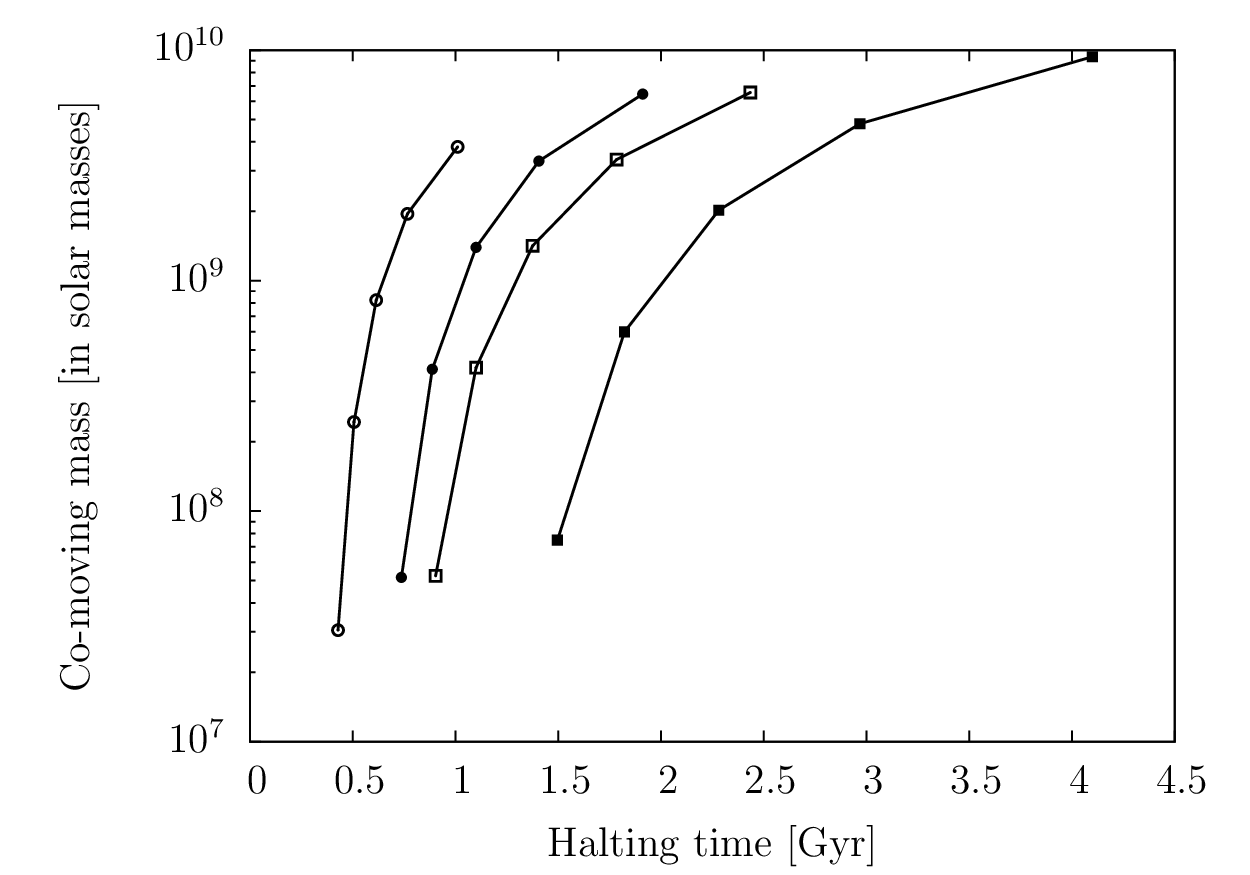}
	\caption{Halting times corresponding to different mass scales, for certain selected parameter values. The four curves (from left to right) correspond to the following parameter values: $T_c = 0.03$ eV, $\rho_v = 55.3$ GeV/cm$^3$; $T_c = 0.03$ eV, $\rho_v = 38$ GeV/cm$^3$; $T_c = 0.05$ eV, $\rho_v = 100.2$ GeV/cm$^3$; and $T_c = 0.05$ eV, $\rho_v = 76$ GeV/cm$^3$, respectively. Evidently, lower critical temperatures and higher early vacuum energy densities imply stronger binding.}
	\label{fig:mass}
\end{figure}

Since the coordinate radius is directly related to the mass enclosed within the comoving volume [via $M \approx (4/3) \pi r^3 \rho_{\text{NR},n}$], we can associate a collapse time scale with a comoving mass at each point in the parameter space. Figure \ref{fig:mass} shows how halting times vary with comoving mass for a few selected combinations of parameter values. It can be seen that comoving regions of mass up to $10^9\, M_\odot$ could become significantly overdense on a time scale of order $1$ Gyr via this mechanism.

It should be emphasized that such time scales calculated by employing the Friedmann equations are only supposed to serve as a guide to the eye. Once the fluctuations become significantly overdense, physics at smaller scales comes into play and can lead to fragmentation of the gravitationally bound comoving volume. But even in that case, the viability of this mechanism in helping create seed black holes of mass $10^3\text{--}10^6\, M_\odot$ at early redshifts cannot be ruled out. The objective behind this analysis is simply to demonstrate that formation of small, overdense regions is a likely outcome of such a vacuum phase transition process, resulting in smaller time scales associated with the formation of supermassive objects, as compared to conventional cosmological models.

\section{Observational constraints}\label{sec:cons}

\subsection{Contributions to closure fraction}

As demonstrated in the previous section, bringing about binding via such a mechanism requires that the density of early vacuum energy be several orders of magnitude bigger than its current value. This immediately raises the question of how this would affect the observed closure fractions of the different energy densities at various epochs. For instance, the impact of the early vacuum energy could be assessed based on its contribution to the closure density at photon decoupling.

\begin{figure}[t]
	\centering
	\includegraphics[width=0.5\textwidth]{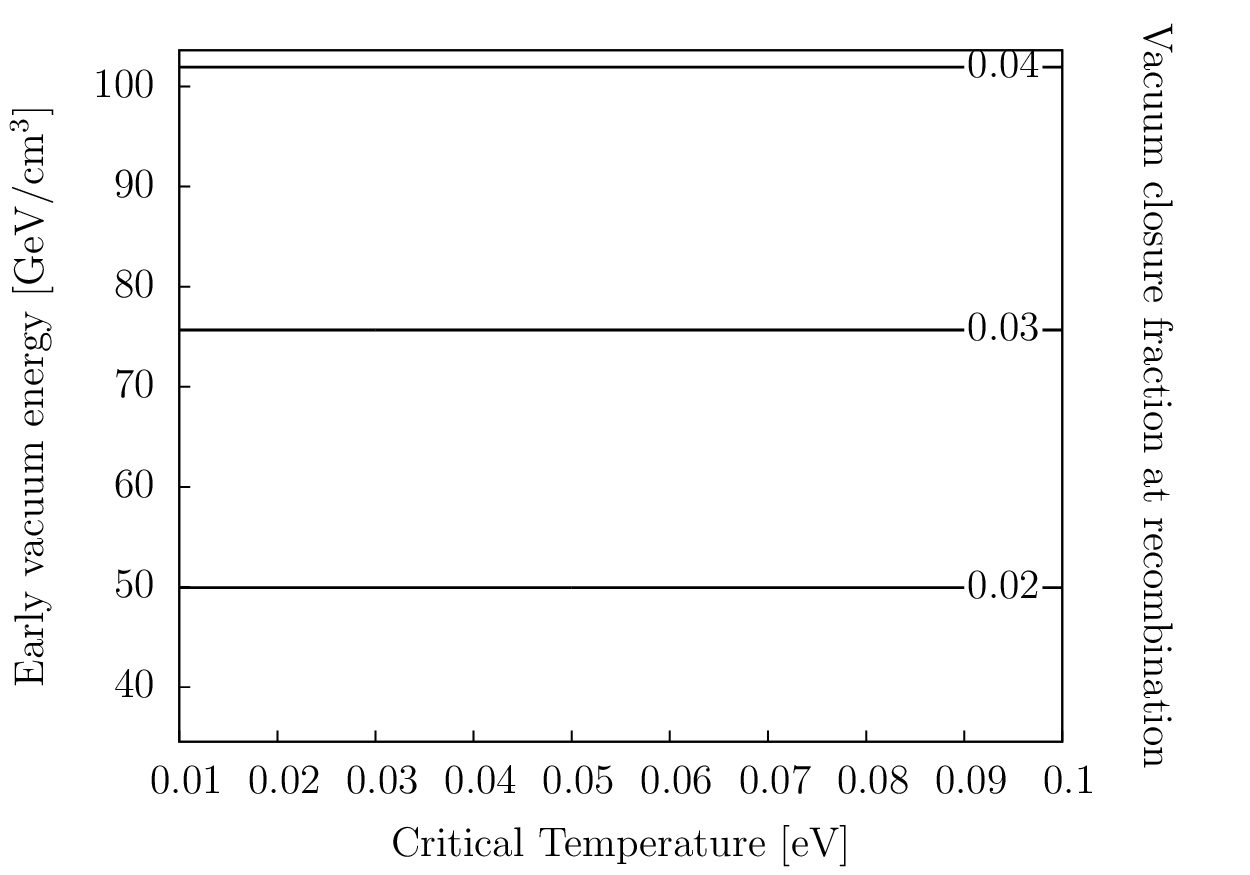}
	\caption{Contours labeled with the early vacuum energy closure fraction at photon decoupling, plotted across a parameter space spanned by critical temperature and early vacuum energy density.}
	\label{fig:closRC}
\end{figure}

As can be seen from Fig.~\ref{fig:closRC}, in our model, the closure fraction of early vacuum energy at photon decoupling is independent of the critical temperature, and typically varies between $1\text{--}4$ percent in the regions of the parameter space that are of interest to us. There have been a few attempts to constrain the closure fraction of early vacuum energy at recombination using CMB data \cite{Doran:2006mz,Calabrese:2011gf,Calabrese:2011ly,Reichardt:2012ys,Hou:2012zr,Pettorino:2013ve,Planck-Collaboration:2013zr,Sievers:2013lr,Bielefeld:2013yq}. At present, these limits, although model dependent, are in the same ballpark as our calculated numbers (i.e., at the level of a few percent of the critical density). Future observations will likely have the potential to impose stronger constraints on our parameter space. A large value of early vacuum energy density could also affect the best-fit values of other recombination parameters, such as $N_\text{eff}$.

\begin{figure}[b]
	\centering
	\includegraphics[width=0.5\textwidth]{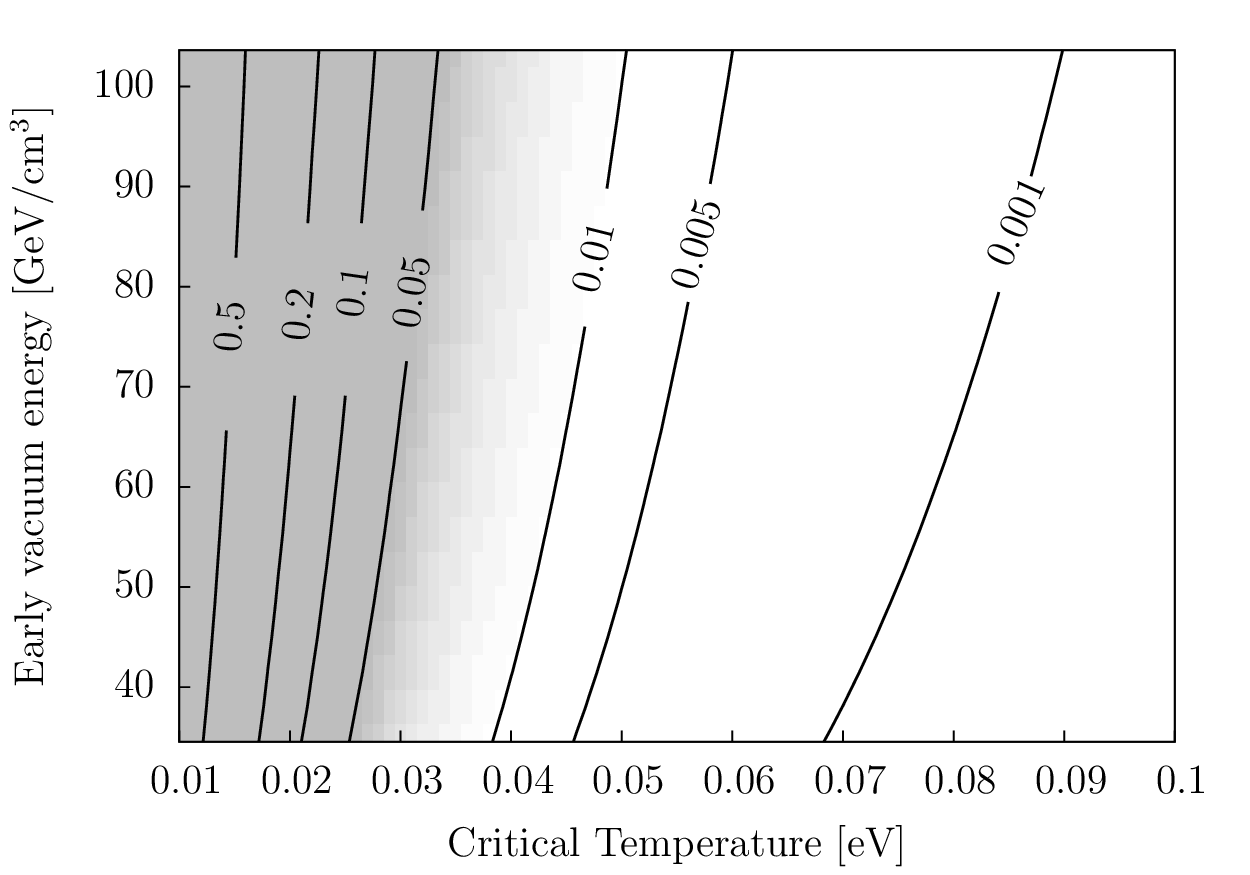}
	\caption{Contours labeled with the closure fraction, at the present epoch, of the leftover radiation from bubble collisions, plotted across a parameter space spanned by critical temperature and early vacuum energy density. Shaded areas represent regions of the parameter space that are disfavored by current observational data.}
	\label{fig:closure}
\end{figure}

As described in Sec.~\ref{sec:trans}, we assume that the early vacuum energy in our model gets swept up onto the bubble walls, before disintegrating via conversion to some unknown relativistic particles. It is then possible to calculate the closure fraction of this leftover radiation at the present epoch, as illustrated in Fig.~\ref{fig:closure}.

There exist very strong constraints on the temperature and spectral shape (and consequently, the energy density) of the photon background \cite{Fixsen:2009qy}, and our calculated closure fraction throughout the parameter space appears too high to be consistent with these. This allows us to rule out the possibility of photons constituting a significant fraction of the leftover radiation. The radiation may also not contain, in appreciable amounts, other standard-model particles that interact electromagnetically (e.g., charged leptons), since that could leave an imprint on the CMB spectrum in the form of Compton $y$-distortions. However, constraints on other forms of relativistic energy density (\lq\lq dark radiation\rq\rq---such as active/sterile neutrinos, or perhaps something more exotic) at the current epoch are not as strong. In fact, it has been argued \cite{Wyman:2013mz} that extra radiation energy density can reconcile discordant values of $H_0$ and $\sigma_8$ inferred from CMB and other, more direct inferences, respectively.

The Wasserman mechanism that lies at the heart of our analysis requires that the swept-up vacuum energy decay into relativistic particles in order to bring about binding. However, this does not preclude the possibility of these particles becoming nonrelativistic at late times as the Universe cools, in a manner akin to the cosmic background neutrinos. In such a situation, these particles could contribute to the cold dark matter density at late times. Uncertainties associated with the CMB-determined $\Omega_{CDM}$ best-fit values hover typically around $1\text{--}2$ percent of the critical density \cite{Hou:2012zr,Planck-Collaboration:2013zr,Calabrese:2013fr,Hinshaw:2013ys,Bennett:2013fr,Sievers:2013lr} at the $1\sigma$ level, allowing for some room to maneuver in this regard.

\subsection{Hubble parameter}

Another way to express the contribution of the leftover radiation at late times is by looking at the impact it has on the Hubble parameter at various epochs. It follows from the Friedmann equation that the Hubble parameter at a redshift of $z$ is related to the closure density as
\be \label{eq:Hubble}
	H^2(z) = \frac{8\pi G}{3}\rho(z),
\ee
where $\rho(z)$ is the closure density of the Universe at that epoch, consisting of contributions from nonrelativistic matter, radiation and vacuum energy densities. Consequently, the additional energy density coming from the leftover radiation can influence the Hubble parameter by contributing to the right-hand side of Eq. (\ref{eq:Hubble}).

\begin{figure}[b]
	\centering
	\includegraphics[width=0.5\textwidth]{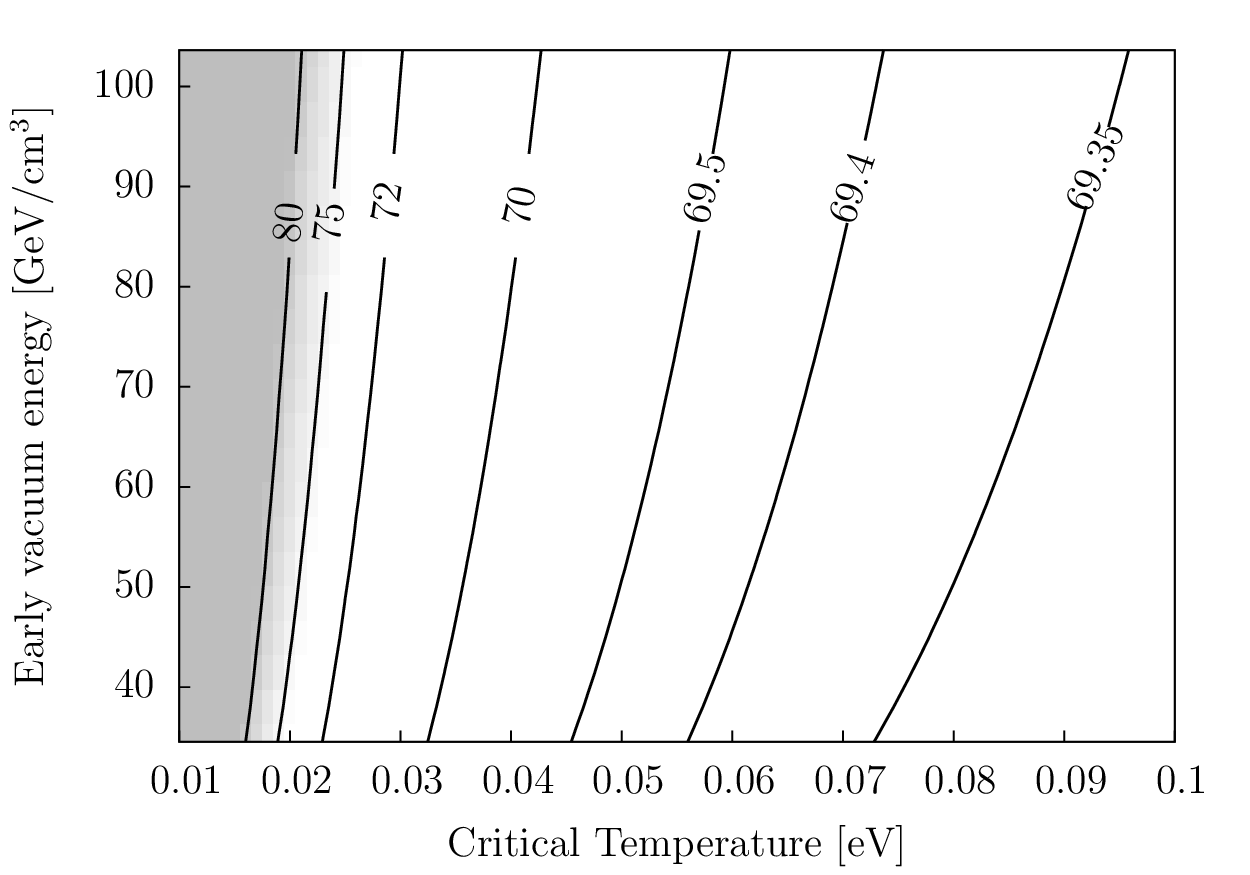}
	\caption{Contours labeled with the Hubble parameter, calculated at the present epoch in our model ($H_0$, in $\text{km s}^{-1} \text{Mpc}^{-1}$), plotted across a parameter space spanned by critical temperature and early vacuum energy density. Shaded areas represent regions of the parameter space where the calculated $H_0$ values are not in good agreement with current observational data.}
	\label{fig:H0}
\end{figure}

Figure \ref{fig:H0} shows what the contours of the current Hubble parameter would look like in our model across the parameter space. Nonrelativistic and vacuum energy densities used for this calculation were adopted from WMAP9+eCMB+BAO+H$_0$ results \cite{Hinshaw:2013ys,Bennett:2013fr}. Neutrinos were approximated as massless. The impact of the leftover radiation can be gauged by observing the shift in $H_0$ from its best-fit $\Lambda$CDM value of $69.32 \text{ km s}^{-1} \text{Mpc}^{-1}$, determined using data from the above surveys.

Quite clearly, the difference between the calculated $H_0$ in our model and the CMB-derived best-fit, is minimum in the lower right-hand corner of the parameter space, i.e., at higher critical temperatures and smaller values of early vacuum energy density. And as Fig.~\ref{fig:closure} demonstrates, this also corresponds to a smaller leftover radiation closure fraction. The calculated $H_0$ values may also be compared to direct low-redshift observational estimates using Type~Ia supernovae and Cepheid variable stars \cite{Riess:2011lr,*Riess:2011lrE,Freedman:2012fk}.

\begin{figure}[b]
	\centering
	\includegraphics[width=0.5\textwidth]{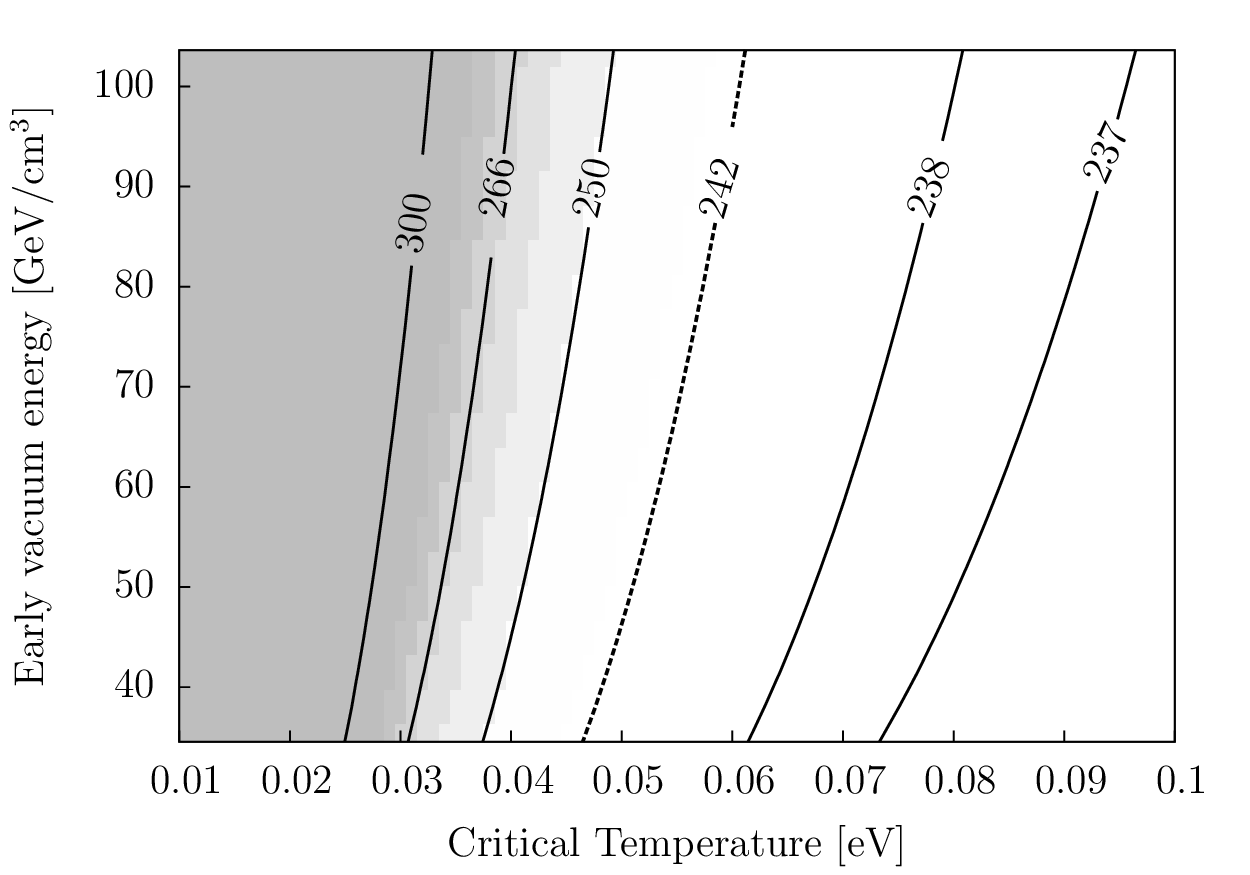}
	\caption{Contours labeled with the Hubble parameter, calculated at a redshift of $z = 2.36$ in our model ($H_{2.36}$, in $\text{km s}^{-1} \text{Mpc}^{-1}$), plotted across a parameter space spanned by critical temperature and early vacuum energy density. The dashed line at $242\text{ km s}^{-1} \text{Mpc}^{-1}$ marks the 95\% C.L. limit given by Ref.~\cite{Font-Ribera:2013uq}, while the shaded areas represent regions of the parameter space where the calculated $H_{2.36}$ values are significantly at odds (3$\sigma$ or more) with that result.}
	\label{fig:Hz}
\end{figure}

It is also possible to measure the Hubble parameter at relatively high redshifts, e.g., $z\approx 2.36$, using Quasar-Lyman $\alpha$ forest cross-correlations. Font-Ribera \textit{et al.}, in Ref.~\cite{Font-Ribera:2013uq}, calculate $H(z = 2.36) = 226 \pm 8\text{ km s}^{-1} \text{Mpc}^{-1}$. Extrapolating the $\Lambda$CDM best-fit value from CMB observations (WMAP9+eCMB+BAO+H$_0$) to that redshift gives $H(z = 2.36) \approx 236\text{ km s}^{-1} \text{Mpc}^{-1}$, consistent with the above result to within error. The same procedure can be applied to our cosmological model to calculate the corresponding $H(z = 2.36)$ values across the parameter space, as shown in Fig. \ref{fig:Hz}.

It can be seen that a large chunk of our parameter space is consistent with direct observations at that redshift to within $2\text{--}3 \sigma$. However, comparing Figs.~\ref{fig:H0} and \ref{fig:Hz} tells us that the higher we go in redshift, the tighter these constraints get. This is understandable, since the contribution from the leftover radiation is much more significant at higher redshifts (e.g., see Fig. \ref{fig:rhomod}). Other techniques with future very large telescopes may be able to obtain the Hubble parameter at even higher redshift, $z\sim 5$ \cite{Yuan:2013qy}. Such observations in the future would represent a promising avenue for further constraining our model.

\subsection{Scale factor evolution and age of the Universe}

\begin{figure}[b]
	\centering
	\includegraphics[width=0.5\textwidth]{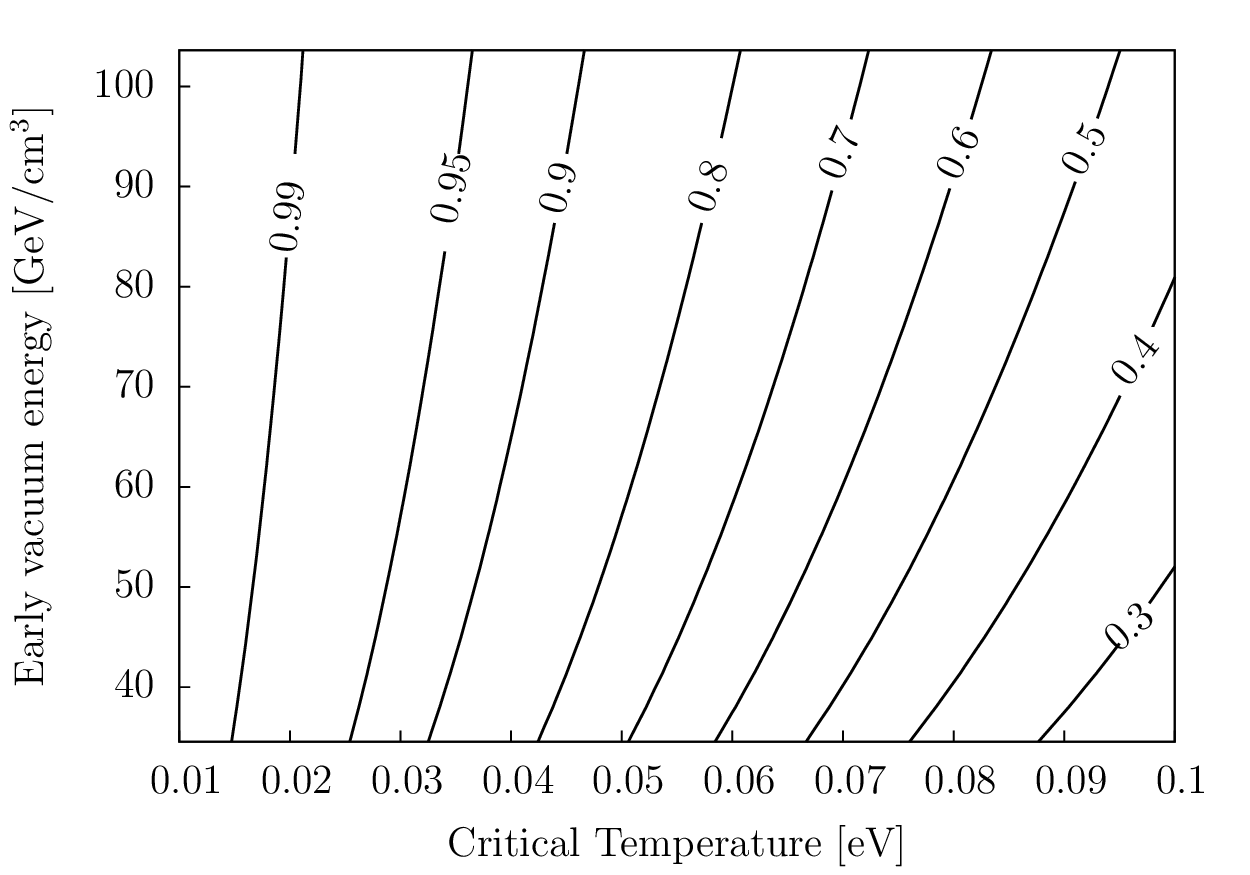}
	\caption{Contours labeled with the early vacuum energy closure fraction at the onset of the phase transition (i.e., at $T = T_c$), plotted across a parameter space spanned by critical temperature and early vacuum energy density.}
	\label{fig:ClosureTc}
\end{figure}

We have shown that the closure fractions of early dark energy at recombination and the leftover radiation at the current epoch can be restricted to a few-percent level in certain regions of our parameter space. However, it turns out that, at epochs close to the phase transition, these components of energy density can in fact be the dominant ones. Figure \ref{fig:ClosureTc} shows contours of closure fraction of early dark energy in our model at $T = T_c$. The contribution of early dark energy to the closure density at the critical temperature can be seen to vary across our parameter space from about $20\%$ to more than $99\%$.

Introducing this epoch of prodigious vacuum energy contribution at around $z \sim 100$ affects the time-evolution of the overall scale factor $a(t)$. Understandably, this effect is smaller in regions of the parameter space where the amount of vacuum domination, as well as the duration of the vacuum-dominant epoch, are relatively small.

\begin{figure}[t]
	\centering
	\includegraphics[width=0.5\textwidth]{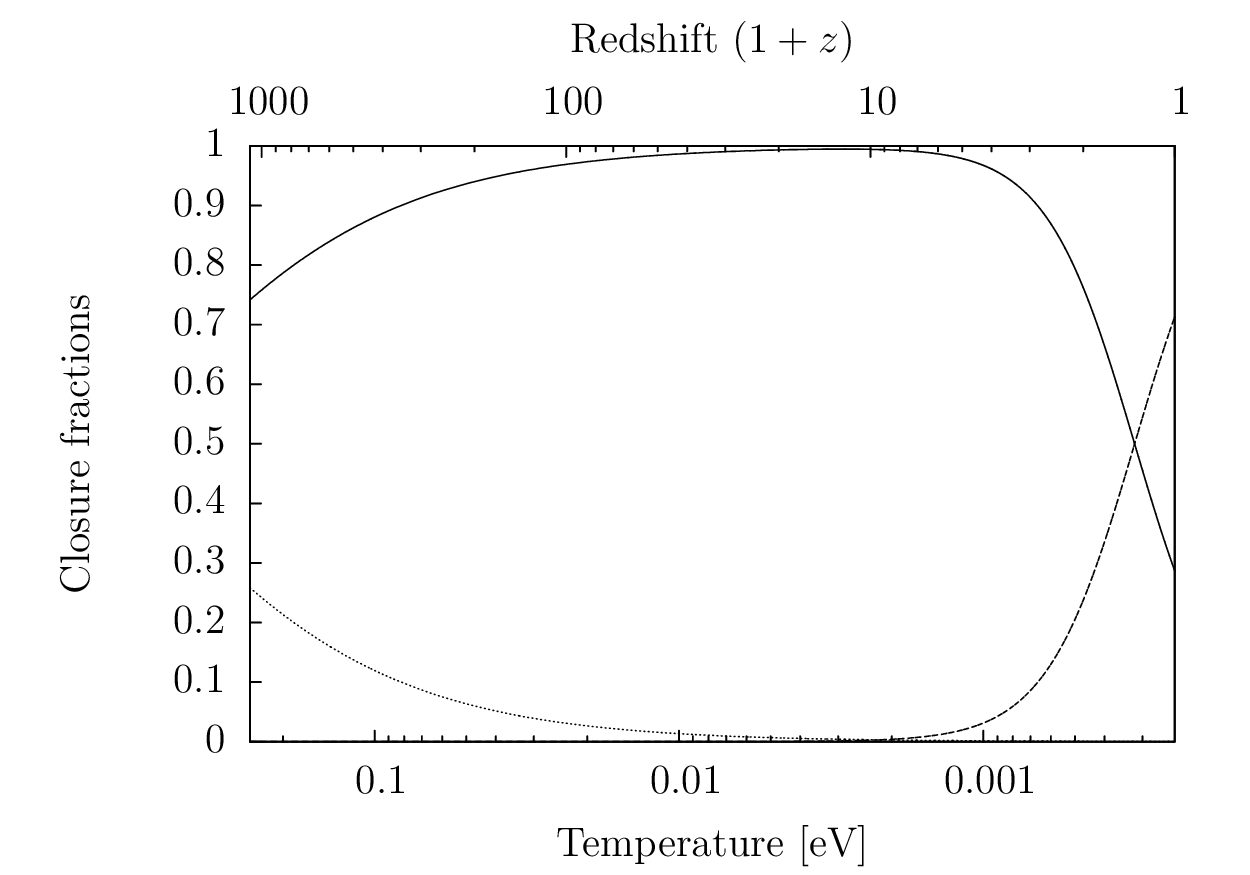}
	\caption{Evolution of closure fractions contributed by the various components of energy density with cosmic temperature in the standard $\Lambda$CDM cosmological model. The solid, dotted, and dashed lines represent nonrelativistic, relativistic, and vacuum energy densities, respectively. Neutrinos were approximated as being massless throughout the course of the evolution.}
	\label{fig:rhostd}
\end{figure}

\begin{figure}[t]
	\centering
	\subfloat[$T_c = 0.03$ eV, $\rho_v = 38$ GeV/cm$^3$]{\label{fig:rhomod3}\includegraphics[width=0.5\textwidth]{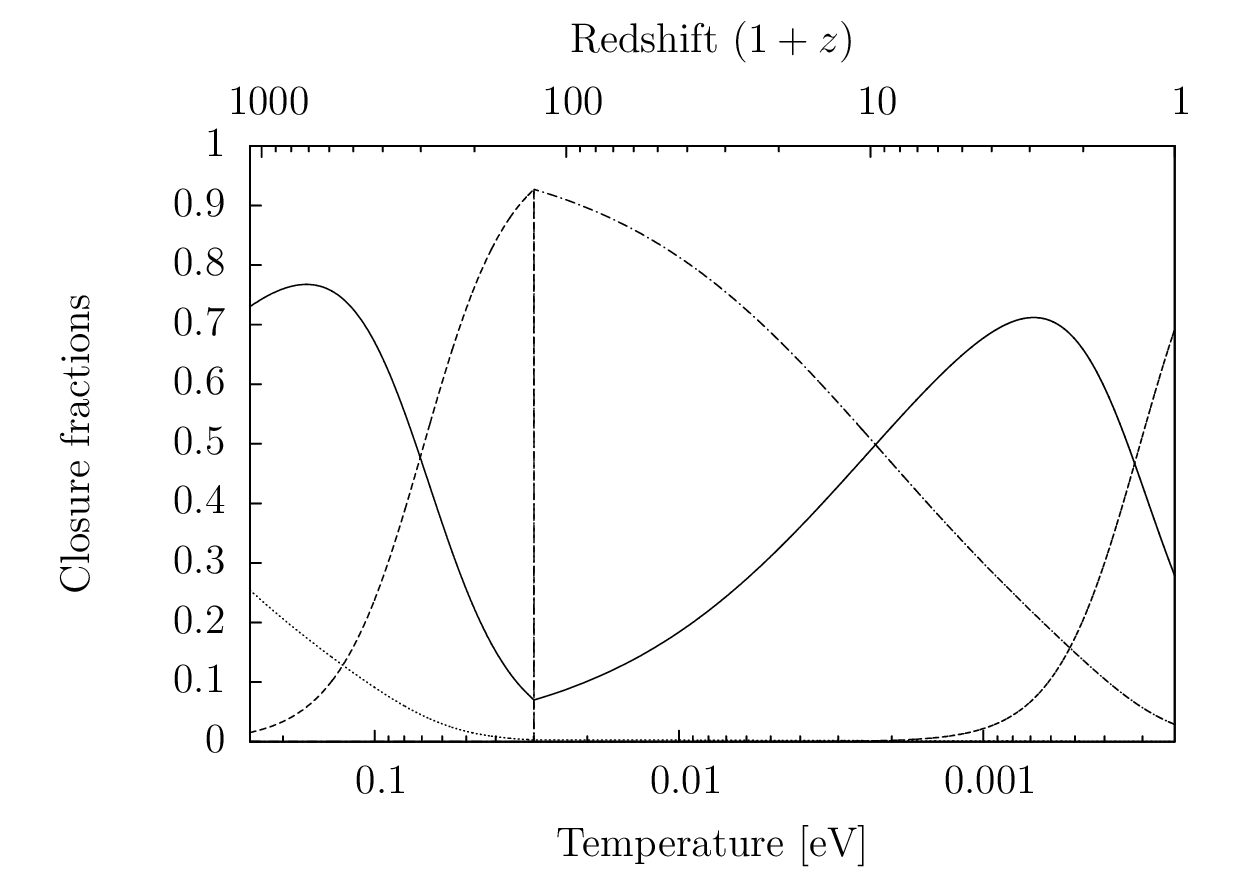}} \\
	\subfloat[$T_c = 0.05$ eV, $\rho_v = 76$ GeV/cm$^3$]{\label{fig:rhomod5}\includegraphics[width=0.5\textwidth]{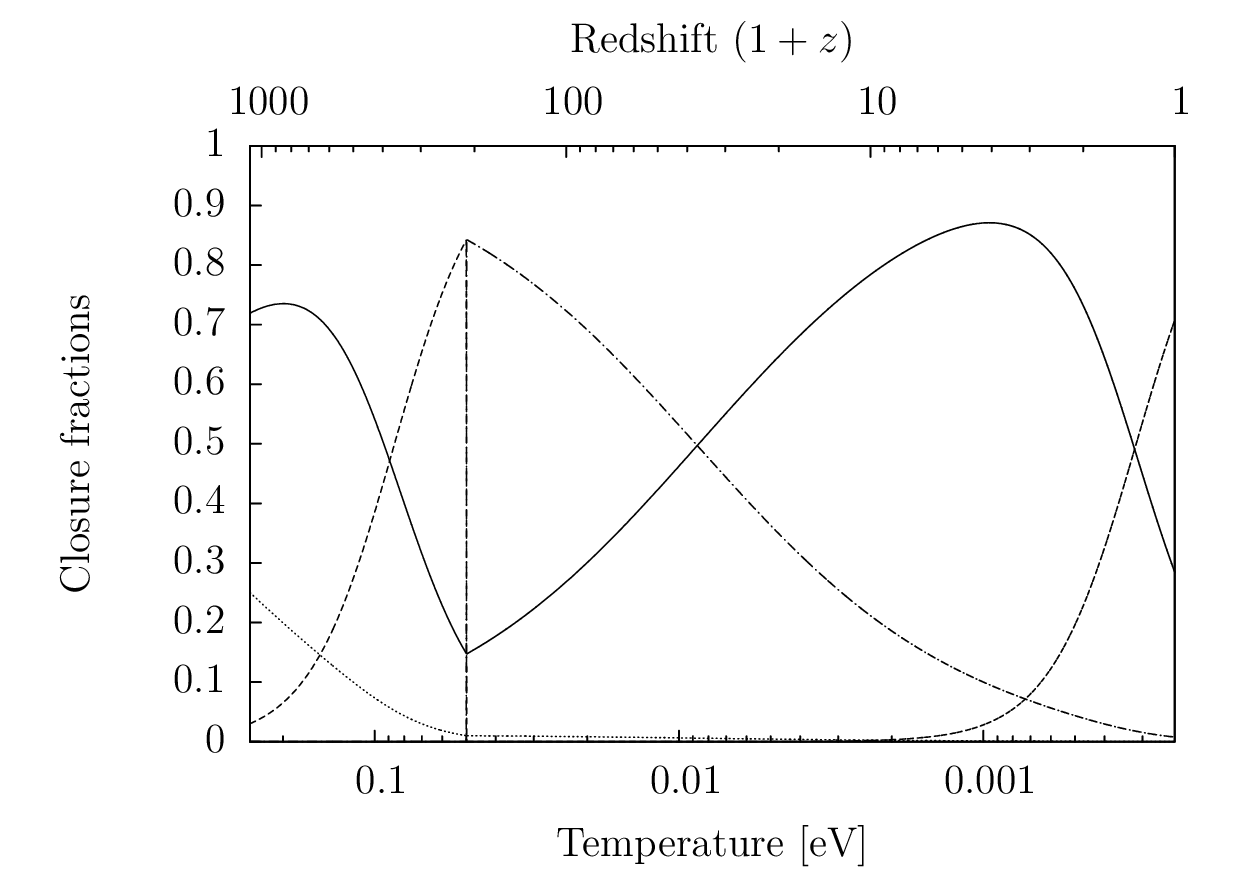}}
	\caption{Evolution of closure fractions contributed by the various components of energy density with cosmic temperature in our modified cosmological model, for different parameter values. The solid, dotted, and dashed lines represent nonrelativistic, primordial relativistic, and vacuum energy densities, respectively, whereas the dot-dashed line represents the leftover radiation from the phase transition. Notice how the vacuum energy density peaks at $T=T_c$, then suddenly plummets to a nearly zero closure fraction (as most of it gets converted to relativistic particles), before eventually becoming significant again at late times.}
	\label{fig:rhomod}
\end{figure}

Figure \ref{fig:rhostd} shows how the relative mix of the various components of energy density changes with time in the standard model, whereas Fig.~\ref{fig:rhomod} does the same in the context of our cosmological model. It is clear that a higher critical temperature (i.e., an earlier phase transition) is associated with a lower vacuum energy contribution at the epoch of the transition, and a smaller duration of vacuum energy domination.

\begin{figure}[t]
	\centering
	\includegraphics[width=0.5\textwidth]{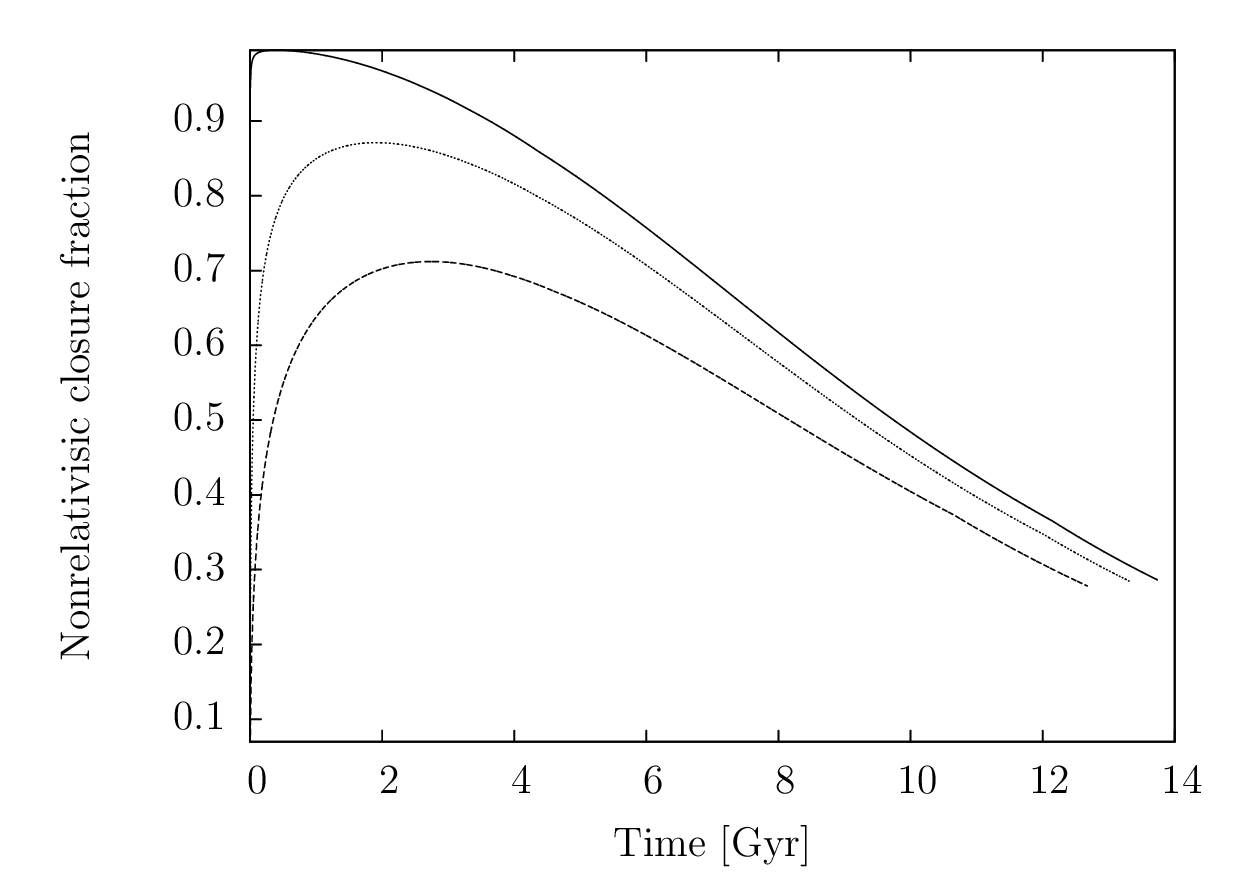}
	\caption{Evolution of the closure fraction contributed by the total nonrelativistic energy density ($\Omega_\text{NR}$) with time in our modified cosmological model, compared with the standard model picture. The solid line at the top represents the standard model, whereas the other two curves (from top to bottom) represent evolution in our cosmological model, with the following parameter values: $T_c = 0.05$ eV, $\rho_v = 76$ GeV/cm$^3$; and $T_c = 0.03$ eV, $\rho_v = 38$ GeV/cm$^3$, respectively.}
	\label{fig:nonrel}
\end{figure}

While it may seem that matter domination is quite substantially suppressed in our model as compared to the standard model, this effect appears much less dramatic when the curves representing the closure fractions are plotted against time, rather than temperature. Figure \ref{fig:nonrel} compares the $\Omega_\text{NR}$ vs $t$ curve in the standard model with the ones in our model, for different parameter values. While matter domination can be seen to be quite heavily suppressed early on, this effect tapers off as the leftover radiation from the phase transition redshifts away. It is not immediately clear how much of an impact this would have on structure growth at scales larger than the nucleation scale, e.g., galaxy formation. Large-scale numerical simulations that incorporate these effects may help address this question.

\begin{figure}[t]
	\centering
	\includegraphics[width=0.5\textwidth]{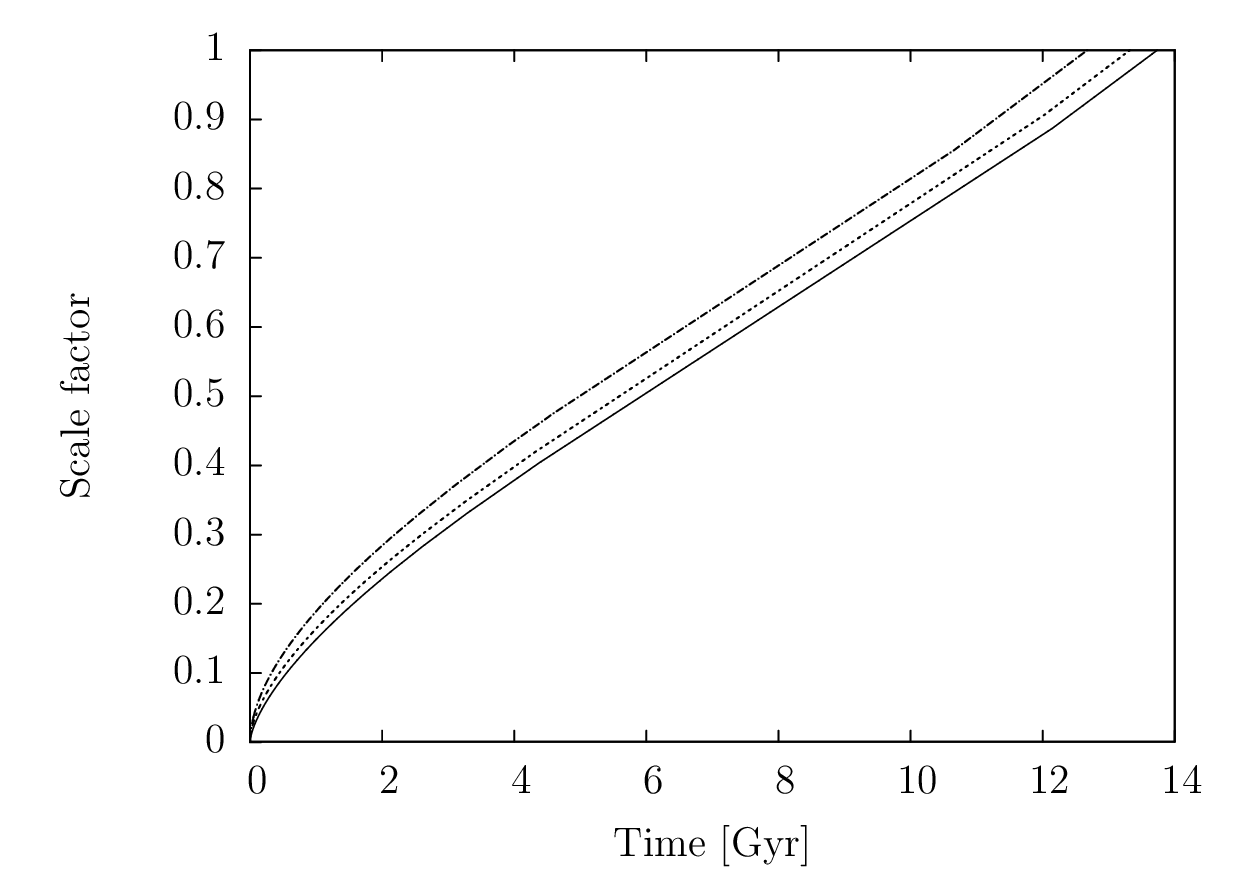}
	\caption{Evolution of the scale factor with time in our modified cosmological model, compared with the standard model picture. The solid line on the far right represents the standard model, whereas the other two curves (from left to right) represent evolution in our cosmological model, with the following parameter values: $T_c = 0.03$ eV, $\rho_v = 38$ GeV/cm$^3$; and $T_c = 0.05$ eV, $\rho_v = 76$ GeV/cm$^3$, respectively.}
	\label{fig:sclft}
\end{figure}

The impact of this new physics can also be analyzed by comparing the time-evolution of the universal scale factor $a(t)$ in our model (for various parameter values), with the standard $\Lambda$CDM cosmological model, as shown in Fig.~\ref{fig:sclft}. The scale factor has been rescaled in order to have $a = 1$ at the present epoch. It can be seen that for $T_c = 0.05$ eV, the deviation of the scale factor evolution curve from its standard model counterpart is a lot less, as compared to $T_c = 0.03$ eV. Apparently, a higher critical temperature serves to mitigate the effect of a large early vacuum energy density.

The plot in Fig.~\ref{fig:sclft} can be used to infer the age of the Universe in our model, by reading off the value of the time coordinate (x axis) when the scale factor becomes unity. For the case where $T_c = 0.03$ eV and $\rho_v = 38$ GeV/cm$^3$, the age can be inferred to be about $12.7 \times 10^9$ years, whereas for $T_c = 0.05$ eV and $\rho_v = 76$ GeV/cm$^3$, it increases to $13.3 \times 10^9$ years. The apparent disharmony between these numbers and the best-fit CMB prediction (i.e., $13.8$ Gyr), must not be taken too seriously, for the latter measurement is predicated on the assumption of the Universe having followed a standard $\Lambda$CDM-based evolutionary track throughout its history. The consistency of these calculated numbers with independent limits on the age of the Universe, derived using nucleocosmochronology \cite{Cowan:1999kx,Wanajo:2002yq}, main-sequence turnoff in globular clusters \cite{Krauss:2003vn}, and white dwarf cooling \cite{Hansen:2004rt,Hansen:2007ys}, may be noted. As of today, the tightest of these independent limits comes from observations of the oxygen-to-iron ratio in the ultra-metal-poor halo subgiant star HD 140283, coupled with stellar evolution theory. Bond \textit{et al.}, in Ref.~\cite{Bond:2013lr}, estimate the age of the star to be $14.5 \pm 0.8$ Gyr, where the uncertainty is part statistical and part systematic.

\subsection{Perturbations on the CMB temperature map}

The density fluctuations generated in our vacuum phase transition model can leave their imprint on the CMB temperature map, in the form of anisotropies arising via the integrated Sachs-Wolfe (ISW) effect. The ISW effect quantifies the differential redshift/blueshift experienced by photons falling into, and propagating out of, time-evolving potential wells, and is therefore proportional to the change in the gravitational potential in the time it takes for a photon to cross the density fluctuation. For a single fluctuation at a redshift $z$, we can write
\be
	\frac{\Delta T}{T}(z) \sim \Delta\phi(z) = \Delta[G\delta\rho R^2](z),
\ee
where $\delta\rho = \rho_\text{NR}' - \rho_\text{NR}$ (defined earlier in Sec.~\ref{sec:binding}) gives the strength of the density perturbation within the fluctuation region, and $R$ is a length scale representing the fluctuation size at the time of light crossing. Typically, a CMB photon will cross several such fluctuations along its trajectory before arriving at the detector, and the net effect can be expressed as a sum of ISW contributions from all the individual fluctuations.

It is possible to estimate the number of such fluctuations along the line of sight between an observer at the present time (i.e., $z = 0$) and a spatial slice at redshift $z = z_c$. This can be done by calculating the proper distance between the observer and the $z = z_c$ surface, at an epoch when a photon that is just arriving at the observer left the surface, and then dividing this distance by the typical fluctuation size. This distance, also known as the \lq\lq angular diameter distance,\rq\rq\ can be expressed as 
\be \label{eq:angdist}
	d_A(z) = \frac{1}{1+z}\int_{t(z)}^{t_0}\frac{dt}{a(t)},
\ee
where $t_0$ and $t(z)$ are cosmological times corresponding to the present epoch and redshift $z$, respectively. Knowing the fluctuation size $R_f$ from Eq. (\ref{eq:nusc}), we can then estimate the number of fluctuations along the photon trajectory as $N \sim d_A(z_c)/R_f$. Typically, for $z_c \sim 100$, we end up with $N \sim 10^4$. The net ISW effect experienced by a photon can then be bounded using
\be\label{eq:ISW}
	\left.\frac{\Delta T}{T}\right|_\text{total} = \sum_\text{i=1}^{N} \frac{\Delta T}{T}(z_i) \lesssim N \times \max_{0 \leq z \leq z_c}{\{\Delta\phi(z)\}}.
\ee

\begin{figure}[t]
	\centering
	\includegraphics[width=0.5\textwidth]{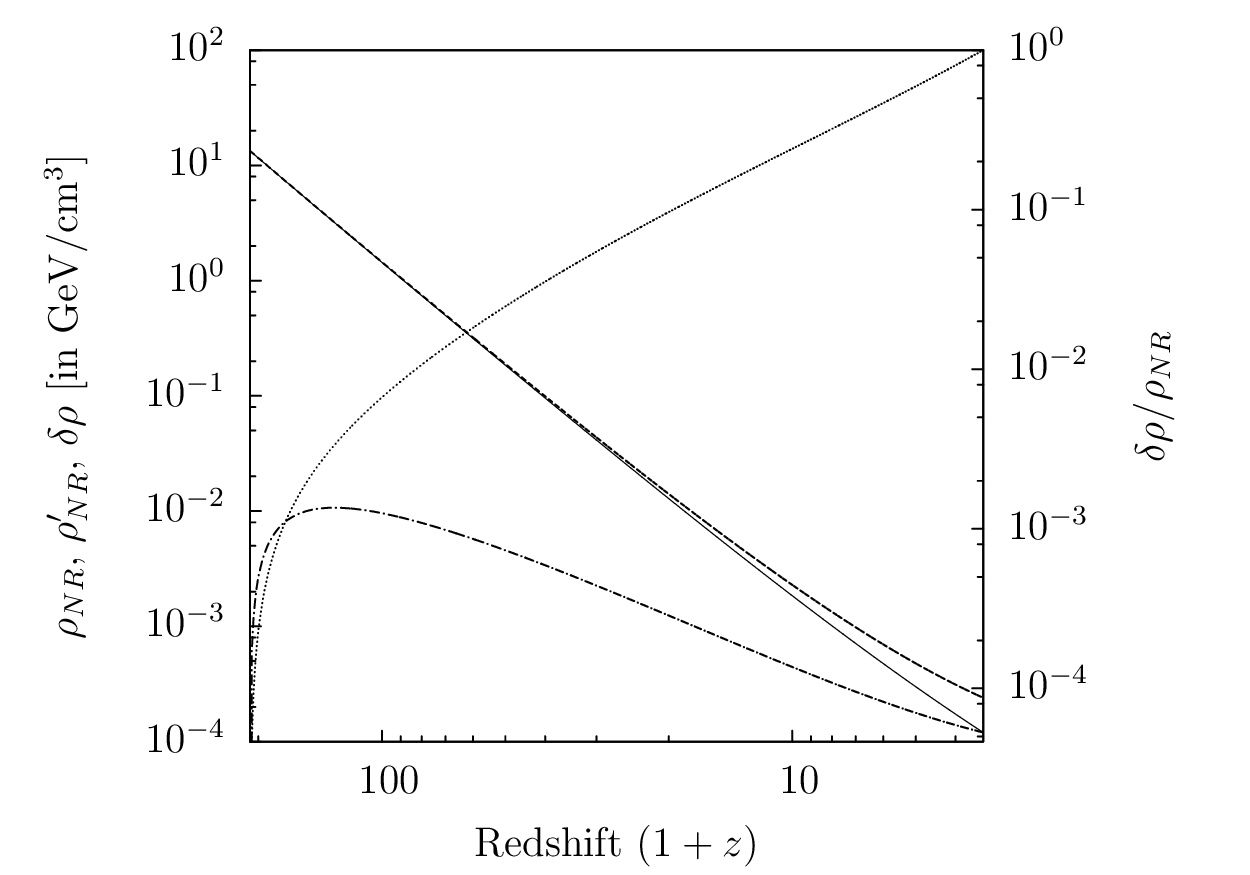}
	\caption{Plot showing the variation of nonrelativistic energy densities in our model (both the local and the average) as a function of redshift. The solid, dashed, dot-dashed and dotted curves correspond to $\rho_\text{NR}$, $\rho_\text{NR}'$, $\delta\rho$, and $\delta\rho/\rho_\text{NR}$, respectively. The parameters used for this calculation are $T_c = 0.05$ eV and $\rho_v = 76$ GeV/cm$^3$.}
	\label{fig:delrho}
\end{figure}

\begin{figure}[t]
	\centering
	\includegraphics[width=0.5\textwidth]{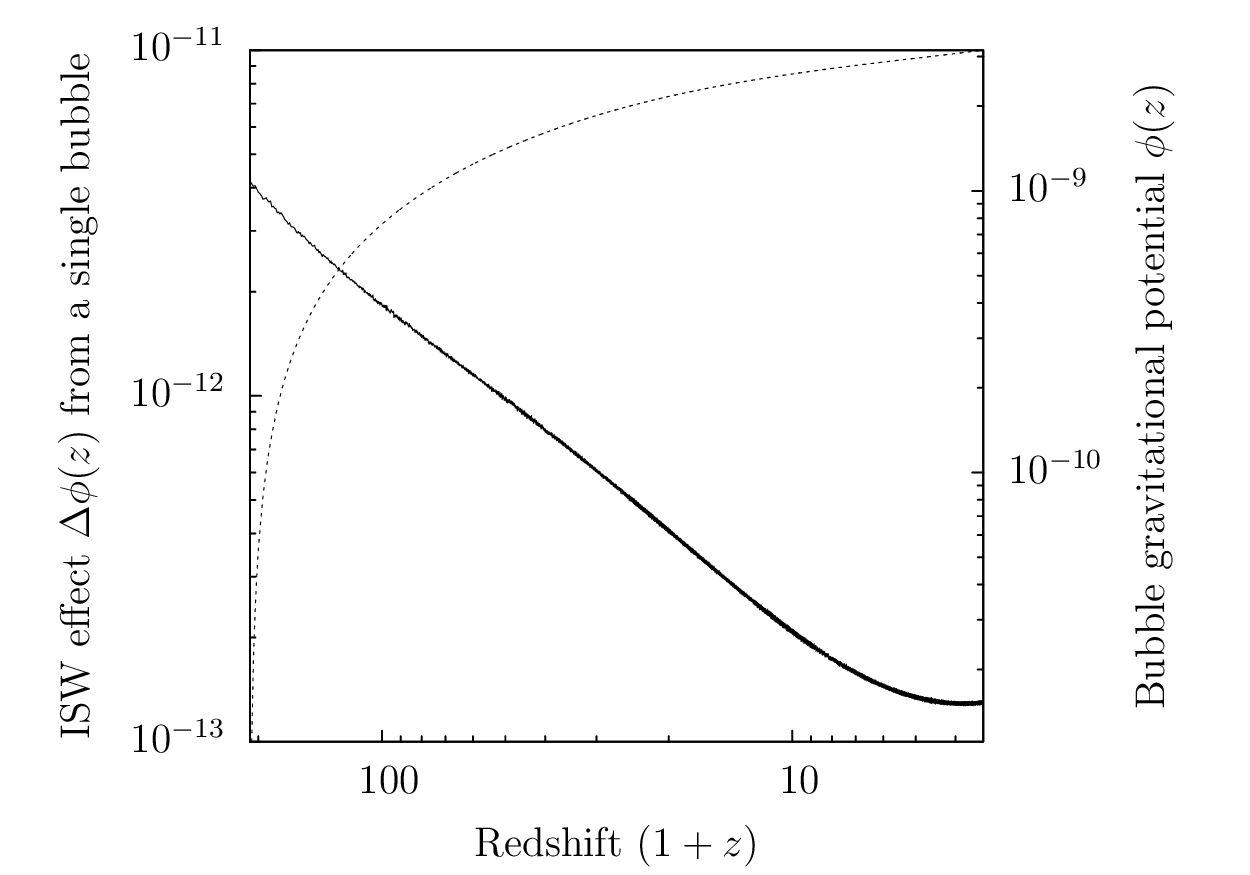}
	\caption{Plot showing the gravitational potentials $\phi(z)$ (dotted curve) of individual density fluctuations generated in our model at different redshifts along the line of sight, along with the integrated Sachs-Wolfe effect $\Delta\phi(z)$ (solid curve) that a photon passing through these fluctuations would experience. The parameters used for this calculation are $T_c = 0.05$ eV and $\rho_v = 76$ GeV/cm$^3$.}
	\label{fig:ISW}
\end{figure}

In order to estimate this upper bound, we attempted to compute the ISW effect $\Delta\phi(z)$ associated with individual fluctuations at different redshifts. Figures \ref{fig:delrho} and \ref{fig:ISW} summarize the results of one such computation, for a sample point in our parameter space, with parameter values $T_c = 0.05$ eV and $\rho_v = 76$ GeV/cm$^3$. Figure \ref{fig:delrho} shows the evolution of the local and average nonrelativistic matter densities ($\rho_\text{NR}'$ and $\rho_\text{NR}$) and the density contrast ($\delta\rho$ and $\delta\rho/\rho_\text{NR}$), whereas Fig.~\ref{fig:ISW} depicts how the gravitational potentials of the fluctuations [$\phi(z)$] evolve, along with the ISW effect [$\Delta\phi(z)$] as a function of redshift. The scale factor $a(t;r)$ at $r = 0.5 \, (\delta \, H_c^{-1})$ was used for evaluating the local nonrelativistic energy density $\rho_\text{NR}' = \rho_{\text{NR},n}/a^3(t;r)$, as well as the typical fluctuation size $R = R_f \,a(t;r)$ at different redshifts along the photon trajectory. The calculation was allowed to run, starting from a redshift $z = z_c$, down to a redshift corresponding to the fluctuations going nonlinear (we did not go all the way down to $z = 0$ so as to avoid operating in nonlinear regimes, and the effect would anyway be expected to drop at late redshifts as the extra radiation energy density becomes insignificant).

Looking at Fig.~\ref{fig:ISW}, we can conclude that, for our particular choice of parameter values, the ISW effect $\Delta\phi(z)$ from a single fluctuation at any redshift is $\lesssim$ $10^{-11}$. Multiplying this number by $N \sim 10^4$ gives an upper bound of $\left.\frac{\Delta T}{T}\right|_\text{total} \lesssim 10^{-7}$. The perturbations in the CMB temperature stemming from these small scale inhomogeneities would therefore fall well within the limits of about one part in $10^5$, set by present-day CMB experiments \cite{Banday:1997qy,Hu:2002fk,Hou:2012zr,Planck-Collaboration:2013zr,Hinshaw:2013ys,Bennett:2013fr,Sievers:2013lr}.

In general it is also possible to obtain an upper bound using a simple static Sachs-Wolfe explanation, wherein we argue that the relative amplitude of these perturbations is bounded by
\be\label{eq:SSW}
	\frac{\Delta T}{T} < \left.\frac{GE_f}{R_f}\right|_{T=T_c} = \frac{4\pi G}{3}\rho_c\, R_f^2,
\ee
where $R_f$ is the typical size of the density fluctuation, $E_f$ is the total enclosed mass-energy inside the fluctuation volume, and $\rho_c \equiv \rho(T = T_c)$ is the total energy density at the epoch of the phase transition. This works because $\delta\rho$ is always very small compared to $\rho_c$ (e.g., see Fig.~\ref{fig:delrho}), and in addition, $\Delta(\delta\rho R^2) \ll \delta\rho R^2$ on account of the smallness of the crossing time relative to the Hubble time (see Sec.~\ref{sec:nusc}). In summary, even after accounting for the large number of fluctuations $N$ along the photon trajectory, Eq. (\ref{eq:SSW}) can serve as a reasonably safe upper bound for the ISW effect calculated in Eq. (\ref{eq:ISW}), throughout the parameter space. Substituting the expression for $R_f$ from Eq. (\ref{eq:nusc}), we obtain
\be
\begin{split}
	\frac{\Delta T}{T} &< \frac{4\pi G}{3}\rho_c \, \[4B_1 \ln{\frac{m_P}{T_c}}\]^{-2} \[\frac{8\pi G}{3}\rho_c\]^{-1} \\
						&= \frac{1}{2}\[4B_1 \ln{\frac{m_P}{T_c}}\]^{-2}.
\end{split}
\ee

It is easy to see that this upper bound on the relative perturbation amplitude depends only logarithmically on the critical temperature, and following our calculation in Sec.~\ref{sec:nusc} we can write
\be
	\frac{\Delta T}{T} < \frac{1}{2}\[\frac{1}{300B_1}\]^2.
\ee

Since $B_1$ is always $\geq \mathcal{O}(1)$, this again is consistent with the observed upper bound of about one part in $10^5$, irrespective of the critical temperature and the early vacuum energy density.

The angular scales in today's sky that would correspond to the typical bubble size can be calculated using $\theta_f \sim R_f/d_A(z_c)$, where $d_A(z_c)$ is the angular diameter distance to the redshift $z_c$ of the phase transition. This happens to be just the inverse of the number of fluctuations $N$ along the photon trajectory, and can be expressed as
\be \label{eq:angscl}
	\theta_f \sim \frac{R_f}{d_A(z_c)} = \frac{\delta \, H_c^{-1}(1+z_c)}{\int_{t(z)}^{t_0}{dt/a(t)}}.
\ee

\begin{figure}[htb]
	\centering
	\includegraphics[width=0.5\textwidth]{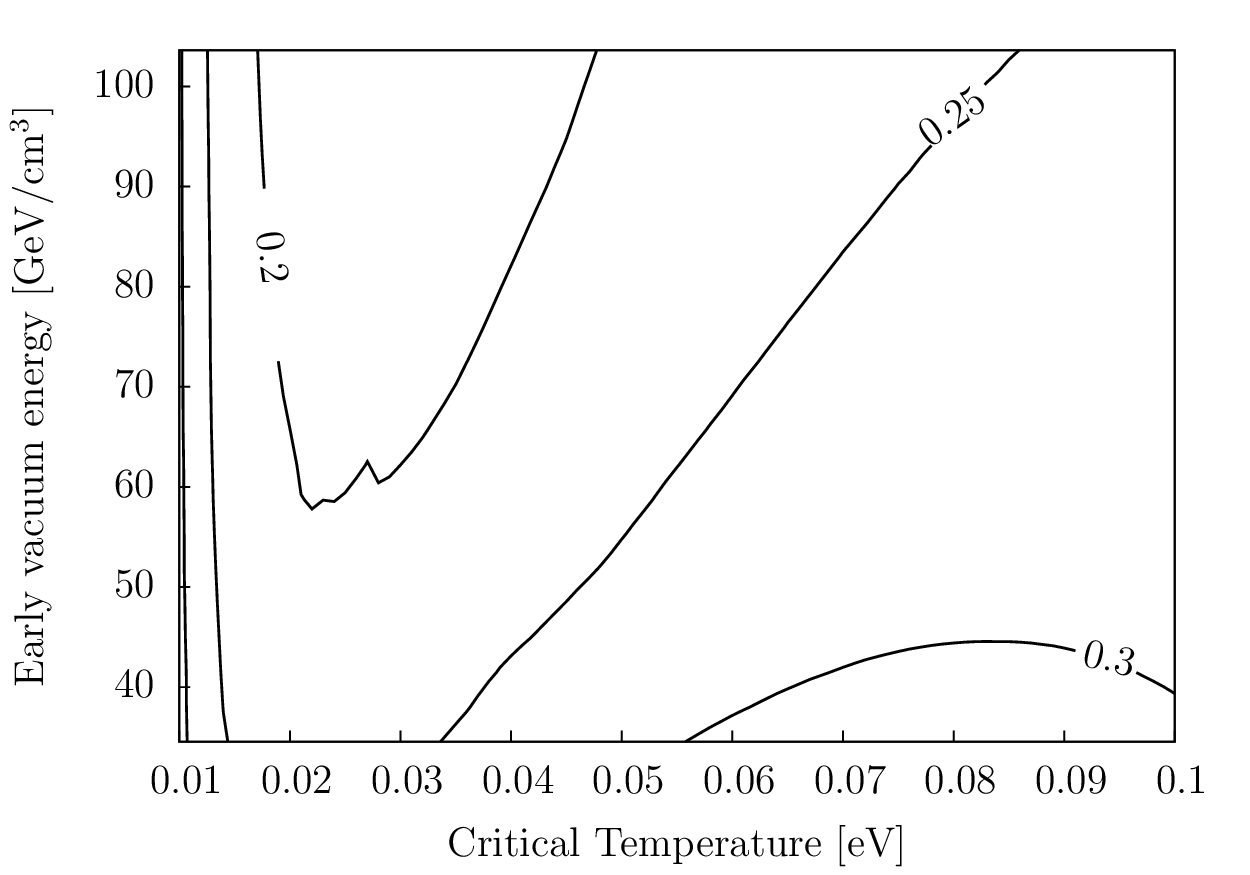}
	\caption{Contours representing the angular scales in today's sky (labeled in arc minutes) that would correspond to the nucleation scale. Plotted across a parameter space spanned by critical temperature in eV and early vacuum energy density in GeV/cm$^3$.}
	\label{fig:angsc}
\end{figure}

As we have seen, the evolution of the scale factor with time depends on the critical temperature and the early vacuum energy density, and therefore, so do the angular scales. Figure \ref{fig:angsc} shows contours of the angular size $\theta_f$ in the sky at the current epoch (in arc minutes), plotted across this parameter space.

These angular scales are a factor of few smaller compared to the angular resolution of state-of-the-art CMB experiments. Thus, any deviations from non-Gaussianity at these scales would have to be probed using other means (e.g., 21-cm observations, see Sec.~\ref{subsec:other}) or with better angular resolution in future CMB experiments.

\subsection{Other constraints} \label{subsec:other}

The late-time phase transitions we study here would result in moving around significant amounts of mass-energy. This potentially opens up other avenues for constraint, for example through observational probes of very low frequency gravitational radiation, or from the high-redshift distribution of baryons as inferred from future radio observations of redshifted 21-cm radiation.

Vacuum phase transition bubble wall dynamics and mergers (\lq\lq percolation\rq\rq) would generate gravitational radiation \cite{Kosowsky:1992uq,Kosowsky:1992fj,Kosowsky:1993fk,Kamionkowski:1994fj,Caprini:2008fr}. For the scenarios presented here, the characteristic frequency of this radiation would be of order the inverse nucleation scale, redshifted to the current epoch appropriately, i.e., $\omega_0 \sim \(1+z_c \)^{-1} (\delta \, H_c^{-1})^{-1}$, where $z_c$ is the redshift of the phase transition epoch. For $z_c \sim 100$, one would expect $\omega_0 \sim 10^{-14}\text{--}10^{-13}$ Hz. Using results from the above references, it is also possible to estimate the characteristic amplitude of the resulting gravitational wave spectrum at its typical frequency (see also Ref.~\cite{Maggiore:2000lr} for a general review). For a vacuum phase transition, the ratio of the liberated gravitational wave energy $E_{GW}$ to the total vacuum energy $E_\text{vac}$ is given by
\be
\frac{E_{GW}}{E_\text{vac}} \sim 0.06 \(\frac{H_c}{\beta}\)^2,
\ee
where $\beta \sim \delta^{-1}H_c$ is the inverse nucleation scale. With this, the closure fraction of gravitational waves at the time of the phase transition is $[\Omega_{GW}]_c = (E_{GW}/E_\text{vac})[\Omega_\text{vac}]_c \sim 0.06\,\delta^2[\Omega_\text{vac}]_c$, where the subscript \lq\lq$c$\rq\rq\ indicates that the quantities are being evaluated at $T = T_c$. This can be extrapolated to the current epoch to obtain
\be
\begin{split}
[\Omega_{GW}]_0 &= [\Omega_{GW}]_c\[\frac{H_c}{H_0}\]^2\[\frac{T_0}{T_c}\]^4\\
						&\sim 0.06\,\delta^2\[\frac{H_c}{H_0}\]^2\[\frac{T_0}{T_c}\]^4[\Omega_\text{vac}]_c.
\end{split}
\ee

The characteristic amplitude $\bar{h}(f)$ of the signal at a frequency $f$ can then be calculated using
\be
	\bar{h}(f) = \[\frac{3H_0^2}{2\pi^2}\,\frac{[\Omega_{GW}(f)]_0}{f^2}\]^{1/2},
\ee
where $\Omega_{GW}(f)$ is the gravitational wave energy contribution to closure per unit frequency octave at the frequency $f$. Using $f \sim \omega_0$, and approximating $\Omega_{GW}(\omega_0) \sim \Omega_{GW}$ (i.e., most of the energy density being concentrated around the peak frequency), we obtain a characteristic amplitude
\be
\bar{h}(\omega_0) \sim \frac{0.3}{\pi}\,\delta^2\,\frac{T_0}{T_c}\,[\Omega_\text{vac}]_c^{1/2} \sim 10^{-9}\text{--}10^{-8}.
\ee

Although this amplitude is seemingly quite substantial, the frequencies are too low for envisioned future gravitational radiation observatories like the Laser Interferometer Space Antenna (LISA). However, the nucleation event and bubble wall decay processes are complex, with a range of bubble sizes and percolation scales, so that a continuum of gravitational radiation extending to frequencies well above $\omega_0$ may be expected, albeit with low amplitude. The only conceivable probe of the higher frequency end of this radiation spectrum would come from precise timing of compact neutron star binary systems, or precision Doppler tracking of spacecraft \cite{Bertotti:1983lr,Armstrong:2003rt,Kramer:2006vn,Jenet:2006fk,Liu:2011yq,Hui:2013kx}, the former of which conceivably could push into the nano-Hertz frequency band \cite{Riles:2013uq}. Of course, direct collapse of massive nonlinear perturbations to black holes at redshift $z \sim 10$ and their subsequent mergers at lower redshift conceivably could be detectable in LISA-like experiments \cite{Berti:2005fk,Babak:2011qy}. Reference \cite{Backer:2004qy} gives a summary of the future methods for potentially detecting low frequency gravitational radiation, along with their expected sensitivities across the frequency spectrum.

Low frequency radio array observations targeting redshifted 21-cm radiation from the very early Universe are more promising as a means of constraining late phase transitions. These studies promise a direct, three-dimensional probe of structure formation from redshift $z \sim 200$ through the epoch of reionization at redshift $z \sim 6$ \cite{Scott:1990vn,Loeb:2004lr,Furlanetto:2006kx,Morales:2010qy,Pober:2013lr}. Extracting a high-redshift matter power spectrum from the low frequency radio data seems possible, if tricky because of foreground sources \cite{Zaldarriaga:2004rt,Morales:2012fj,Pritchard:2012uq}. Many of the late phase transition scenarios discussed here might distinguish themselves from standard $\Lambda$CDM structure formation through an earlier progression to nonlinearity on the relatively small scales that we are interested in.

\section{Conclusion and speculations on new neutrino sector physics} \label{sec:concl}

We have discussed how a first-order cosmological vacuum phase transition in the postrecombination era could potentially influence the growth of objects and structure at late times. We have examined various avenues through which the parameters of our cosmological model could be constrained using current and future observational data, especially from CMB observations, and potentially from future gravitational radiation experiments and redshifted 21-cm radiation observations. We conclude that perhaps the most sensitive constraints on, or probes of, these scenarios may come from the arguments about the content of radiation energy density at late epochs as derived from a comparison of CMB- and directly-derived cosmological parameters, especially the Hubble parameter $H_0$, and measures of overall large-scale structure power normalization, e.g., $\sigma_8$. That extra radiation energy density can effect such a reconciliation has been pointed out by Wyman \textit{et al.}~\cite{Wyman:2013mz}.

We have pointed out that such a process could result in an enhancement in power at relatively small scales ($\sim$$10^6\text{--}10^9\, M_\odot$). In addition to possibly boosting the growth of supermassive black holes at early redshifts, this could also have an influence on the evolution and distribution of dwarf galaxies within larger galactic halos. Discrepancies between the results of numerical simulations and the observed distribution of dwarf galaxies, such as the \lq\lq missing satellites\rq\rq\ and the \lq\lq too big to fail\rq\rq\ problems, have been well documented \cite{Boylan-Kolchin2012,Weinberg:2013fj}. While it may seem that, on the face of it, such a cosmological model would worsen the missing satellites problem, it is quite difficult 	to draw definitive conclusions until sophisticated numerical simulations are carried out with all of this speculative new physics built into the code. As mentioned earlier, fragmentation of the collapsing fluctuations might ensue, making it difficult to predict the exact scales at which the enhancement in power may occur. One could also potentially play around with different scenarios, combining late phase-transition dynamics with different dark matter models, such as warm or self-interacting dark matter.

Obviously, such a vacuum phase transition would have to be associated with new fundamental physics at sub-eV scales. A glance at Figs. \ref{fig:ctr} and \ref{fig:closure}--\ref{fig:Hz} would suggest that the sweet spot within our parameter space with associated collapse time scales of $\lesssim \mathcal{O}$(Gyr), and which is not yet ruled out by observations, lies in the region corresponding to $T_c \sim 0.03\text{--}0.05$ eV, numbers that are roughly in the same range as the likely neutrino absolute rest mass values. One could thus envisage that the leading suspect for something new at this energy scale would have to be the neutrino sector.

Though we know the neutrino mass-squared differences and three of the four parameters characterizing the vacuum unitary transformation between the neutrino mass states and the weak interaction (flavor) states, we do not know the neutrino absolute rest masses, and we do not even know the mass ordering of the neutrino rest mass eigenvalues, i.e., the neutrino mass hierarchy. Additionally, the very existence of nonzero neutrino rest masses invites speculation about \lq\lq sterile\rq\rq\ neutrinos, and these might not be sterile at all by virtue of their vacuum mixing with ordinary active neutrinos. Recent experiments have been interpreted as potentially bolstering the case for sterile neutrino species with rest masses in the $\sim$$1$ eV range \cite{Conrad:2013ys}. Reference \cite{de-Gouvea:2013fr} gives an overview of the current state of neutrino physics and the prospects for future laboratory and cosmological probes of this sector of particle physics.

Progress in probing neutrino mass/mixing and sterile neutrino physics from astrophysical considerations revolves around five current or coming developments \cite{Balantekin:2013zr,Gardner:2013lr}: (1) high precision baryon-to-photon ratio determinations from the CMB; (2) high precision determinations of the ratio of relativistic-to-nonrelativistic particle energy density at the epoch of photon decoupling, i.e., $N_\text{eff}$; (3) high precision CMB determinations of the primordial helium abundance; (4) high precision determinations of the primordial deuterium abundance; and (5) CMB- and large-scale structure-determined values of the sum of the light neutrino masses $\sum m_{\nu}$. Taken together, these data will provide constraints on relic neutrino number densities and energy spectra. This will be especially constraining for a putative sterile neutrino sector.

For example, sterile neutrinos with sufficiently large vacuum mixing with active species could have relic energy spectra and number densities that are comparable to those of ordinary active neutrino species, so long as the net lepton number in the Universe is small enough \cite{Abazajian:2005fk}. Given (1), their higher masses and the extra energy density they add in early epochs may cause a standard cosmology with these particles to run afoul of one or more of the observationally determined points (2)--(5) listed above \cite{Smith:2006qy}. We can speculate on what modifications to standard cosmology could reconcile (1)--(5) with such light sterile neutrinos with large vacuum mixing with active species. One way out might be a large net lepton number \cite{Abazajian:2005fk}, albeit one below the current bounds. These bounds currently come from the primordial helium abundance \cite{Kneller:2001uq,Smith:2006qy}. Future CMB polarization measurements are forecast to provide even more stringent bounds, e.g., see Ref.~\cite{Shimon:2010lr}.

Another way out could be a vacuum phase transition, occurring after photon decoupling, in which neutrinos acquire mass and flavor mixing. At earlier epochs the active and sterile neutrinos would not mix and, as a result, the sterile neutrino sea would not be populated by oscillations. This would lead to $N_\text{eff}$, light element abundances, and the active neutrino relic number densities and energy spectra (and hence $\sum m_\nu$), all being essentially identical to what would be expected in a standard cosmology without a light, large-mixing sterile neutrino species.

One might be tempted to associate such a low energy scale mass/mixing-generating phase transition with a symmetry breaking event. In that case, however, there would have to be some mechanism, perhaps large lepton number density, that suppresses restoration of that symmetry in the Sun (central temperature $\gg T_c$), where we know neutrino flavor mixing occurs. Likewise, detections of future core collapse supernova neutrino burst signatures might reveal neutrino flavor mixing processes and provide yet another probe of sub-eV scale physics in the neutrino sector. 

Obviously, such neutrino mass/mixing-generating vacuum phase transitions seem speculative and contrived at this point. However, future experimental and observational data may force us to take these models more seriously---or rule them out. As discussed in this paper, late-time phase transitions are wide open to constraint via current and near-future observations. There is much that remains mysterious about the origin of neutrino mass, the possible existence and mass/mixing scales of sterile neutrinos, many other issues in sub-eV scale particle physics, and the physics of the vacuum. Astrophysical considerations may be a key way to get at this physics.

\begin{acknowledgments}
This work was supported in part by NSF Grant No. PHY-1307372 at UCSD. We thank J. Carlstrom, J. F. Cherry, E. Grohs, W. Hu, C. Kishimoto, M. Millea, M. Morales, S. Profumo, and A. Vlasenko for useful conversations. G.M.F. would also like to thank the Kavli Institute for Cosmological Physics at The University of Chicago for hospitality and support.
\end{acknowledgments}

\bibliography{LPT_v11}

\begin{thebibliography}{97}%
\makeatletter
\providecommand \@ifxundefined [1]{%
 \@ifx{#1\undefined}
}%
\providecommand \@ifnum [1]{%
 \ifnum #1\expandafter \@firstoftwo
 \else \expandafter \@secondoftwo
 \fi
}%
\providecommand \@ifx [1]{%
 \ifx #1\expandafter \@firstoftwo
 \else \expandafter \@secondoftwo
 \fi
}%
\providecommand \natexlab [1]{#1}%
\providecommand \enquote  [1]{``#1''}%
\providecommand \bibnamefont  [1]{#1}%
\providecommand \bibfnamefont [1]{#1}%
\providecommand \citenamefont [1]{#1}%
\providecommand \href@noop [0]{\@secondoftwo}%
\providecommand \href [0]{\begingroup \@sanitize@url \@href}%
\providecommand \@href[1]{\@@startlink{#1}\@@href}%
\providecommand \@@href[1]{\endgroup#1\@@endlink}%
\providecommand \@sanitize@url [0]{\catcode `\\12\catcode `\$12\catcode
  `\&12\catcode `\#12\catcode `\^12\catcode `\_12\catcode `\%12\relax}%
\providecommand \@@startlink[1]{}%
\providecommand \@@endlink[0]{}%
\providecommand \url  [0]{\begingroup\@sanitize@url \@url }%
\providecommand \@url [1]{\endgroup\@href {#1}{\urlprefix }}%
\providecommand \urlprefix  [0]{URL }%
\providecommand \Eprint [0]{\href }%
\providecommand \doibase [0]{http://dx.doi.org/}%
\providecommand \selectlanguage [0]{\@gobble}%
\providecommand \bibinfo  [0]{\@secondoftwo}%
\providecommand \bibfield  [0]{\@secondoftwo}%
\providecommand \translation [1]{[#1]}%
\providecommand \BibitemOpen [0]{}%
\providecommand \bibitemStop [0]{}%
\providecommand \bibitemNoStop [0]{.\EOS\space}%
\providecommand \EOS [0]{\spacefactor3000\relax}%
\providecommand \BibitemShut  [1]{\csname bibitem#1\endcsname}%
\let\auto@bib@innerbib\@empty
\bibitem [{\citenamefont {{Raffelt}}\ and\ \citenamefont
  {{Silk}}(1987)}]{Raffelt:1987kx}%
  \BibitemOpen
  \bibfield  {author} {\bibinfo {author} {\bibfnamefont {G.}~\bibnamefont
  {{Raffelt}}}\ and\ \bibinfo {author} {\bibfnamefont {J.}~\bibnamefont
  {{Silk}}},\ }\href {\doibase 10.1016/0370-2693(87)91143-9} {\bibfield
  {journal} {\bibinfo  {journal} {Physics Letters B}\ }\textbf {\bibinfo
  {volume} {192}},\ \bibinfo {pages} {65} (\bibinfo {year} {1987})}\BibitemShut
  {NoStop}%
\bibitem [{\citenamefont {{Fuller}}\ and\ \citenamefont
  {{Schramm}}(1992)}]{Fuller:1992lr}%
  \BibitemOpen
  \bibfield  {author} {\bibinfo {author} {\bibfnamefont {G.~M.}\ \bibnamefont
  {{Fuller}}}\ and\ \bibinfo {author} {\bibfnamefont {D.~N.}\ \bibnamefont
  {{Schramm}}},\ }\href {\doibase 10.1103/PhysRevD.45.2595} {\bibfield
  {journal} {\bibinfo  {journal} {\prd}\ }\textbf {\bibinfo {volume} {45}},\
  \bibinfo {pages} {2595} (\bibinfo {year} {1992})}\BibitemShut {NoStop}%
\bibitem [{\citenamefont {{Kolb}}\ and\ \citenamefont
  {{Wang}}(1992)}]{Kolb:1992dq}%
  \BibitemOpen
  \bibfield  {author} {\bibinfo {author} {\bibfnamefont {E.~W.}\ \bibnamefont
  {{Kolb}}}\ and\ \bibinfo {author} {\bibfnamefont {Y.}~\bibnamefont
  {{Wang}}},\ }\href {\doibase 10.1103/PhysRevD.45.4421} {\bibfield  {journal}
  {\bibinfo  {journal} {\prd}\ }\textbf {\bibinfo {volume} {45}},\ \bibinfo
  {pages} {4421} (\bibinfo {year} {1992})}\BibitemShut {NoStop}%
\bibitem [{\citenamefont {{Frieman}}\ \emph {et~al.}(1992)\citenamefont
  {{Frieman}}, \citenamefont {{Hill}},\ and\ \citenamefont
  {{Watkins}}}]{Frieman:1992ly}%
  \BibitemOpen
  \bibfield  {author} {\bibinfo {author} {\bibfnamefont {J.~A.}\ \bibnamefont
  {{Frieman}}}, \bibinfo {author} {\bibfnamefont {C.~T.}\ \bibnamefont
  {{Hill}}}, \ and\ \bibinfo {author} {\bibfnamefont {R.}~\bibnamefont
  {{Watkins}}},\ }\href {\doibase 10.1103/PhysRevD.46.1226} {\bibfield
  {journal} {\bibinfo  {journal} {\prd}\ }\textbf {\bibinfo {volume} {46}},\
  \bibinfo {pages} {1226} (\bibinfo {year} {1992})}\BibitemShut {NoStop}%
\bibitem [{\citenamefont {{Barbieri}}\ \emph {et~al.}(2005)\citenamefont
  {{Barbieri}}, \citenamefont {{Hall}}, \citenamefont {{Oliver}},\ and\
  \citenamefont {{Strumia}}}]{Barbieri:2005qf}%
  \BibitemOpen
  \bibfield  {author} {\bibinfo {author} {\bibfnamefont {R.}~\bibnamefont
  {{Barbieri}}}, \bibinfo {author} {\bibfnamefont {L.~J.}\ \bibnamefont
  {{Hall}}}, \bibinfo {author} {\bibfnamefont {S.~J.}\ \bibnamefont
  {{Oliver}}}, \ and\ \bibinfo {author} {\bibfnamefont {A.}~\bibnamefont
  {{Strumia}}},\ }\href {\doibase 10.1016/j.physletb.2005.08.075} {\bibfield
  {journal} {\bibinfo  {journal} {Physics Letters B}\ }\textbf {\bibinfo
  {volume} {625}},\ \bibinfo {pages} {189} (\bibinfo {year} {2005})},\ \Eprint
  {http://arxiv.org/abs/arXiv:hep-ph/0505124} {arXiv:hep-ph/0505124}
  \BibitemShut {NoStop}%
\bibitem [{\citenamefont {{Bamba}}\ \emph {et~al.}(2008)\citenamefont
  {{Bamba}}, \citenamefont {{Geng}},\ and\ \citenamefont
  {{Ho}}}]{Bamba:2008pd}%
  \BibitemOpen
  \bibfield  {author} {\bibinfo {author} {\bibfnamefont {K.}~\bibnamefont
  {{Bamba}}}, \bibinfo {author} {\bibfnamefont {C.~Q.}\ \bibnamefont {{Geng}}},
  \ and\ \bibinfo {author} {\bibfnamefont {S.~H.}\ \bibnamefont {{Ho}}},\
  }\href {\doibase 10.1088/1475-7516/2008/09/001} {\bibfield  {journal}
  {\bibinfo  {journal} {Journal of Cosmology and Astroparticle Physics}\
  }\textbf {\bibinfo {volume} {09}},\ \bibinfo {eid} {001} (\bibinfo {year}
  {2008})},\ \Eprint {http://arxiv.org/abs/0806.0952} {arXiv:0806.0952
  [hep-ph]} \BibitemShut {NoStop}%
\bibitem [{\citenamefont {{Bhatt}}\ \emph {et~al.}(2010)\citenamefont
  {{Bhatt}}, \citenamefont {{Desai}}, \citenamefont {{Ma}}, \citenamefont
  {{Rajasekaran}},\ and\ \citenamefont {{Sarkar}}}]{Bhatt:2010ve}%
  \BibitemOpen
  \bibfield  {author} {\bibinfo {author} {\bibfnamefont {J.~R.}\ \bibnamefont
  {{Bhatt}}}, \bibinfo {author} {\bibfnamefont {B.~R.}\ \bibnamefont
  {{Desai}}}, \bibinfo {author} {\bibfnamefont {E.}~\bibnamefont {{Ma}}},
  \bibinfo {author} {\bibfnamefont {G.}~\bibnamefont {{Rajasekaran}}}, \ and\
  \bibinfo {author} {\bibfnamefont {U.}~\bibnamefont {{Sarkar}}},\ }\href
  {\doibase 10.1016/j.physletb.2010.02.079} {\bibfield  {journal} {\bibinfo
  {journal} {Physics Letters B}\ }\textbf {\bibinfo {volume} {687}},\ \bibinfo
  {pages} {75} (\bibinfo {year} {2010})},\ \Eprint
  {http://arxiv.org/abs/0911.5012} {arXiv:0911.5012 [hep-ph]} \BibitemShut
  {NoStop}%
\bibitem [{\citenamefont {{Primack}}\ and\ \citenamefont
  {{Sher}}(1980)}]{Primack:1980uq}%
  \BibitemOpen
  \bibfield  {author} {\bibinfo {author} {\bibfnamefont {J.~R.}\ \bibnamefont
  {{Primack}}}\ and\ \bibinfo {author} {\bibfnamefont {M.~A.}\ \bibnamefont
  {{Sher}}},\ }\href {\doibase 10.1038/288680a0} {\bibfield  {journal}
  {\bibinfo  {journal} {Nature}\ }\textbf {\bibinfo {volume} {288}},\ \bibinfo
  {pages} {680} (\bibinfo {year} {1980})}\BibitemShut {NoStop}%
\bibitem [{\citenamefont {{Wasserman}}(1986)}]{Wasserman:1986lr}%
  \BibitemOpen
  \bibfield  {author} {\bibinfo {author} {\bibfnamefont {I.}~\bibnamefont
  {{Wasserman}}},\ }\href {\doibase 10.1103/PhysRevLett.57.2234} {\bibfield
  {journal} {\bibinfo  {journal} {Physical Review Letters}\ }\textbf {\bibinfo
  {volume} {57}},\ \bibinfo {pages} {2234} (\bibinfo {year}
  {1986})}\BibitemShut {NoStop}%
\bibitem [{\citenamefont {{Hill}}\ \emph {et~al.}(1989)\citenamefont {{Hill}},
  \citenamefont {{Schramm}},\ and\ \citenamefont {{Fry}}}]{Hill:1989fj}%
  \BibitemOpen
  \bibfield  {author} {\bibinfo {author} {\bibfnamefont {C.~T.}\ \bibnamefont
  {{Hill}}}, \bibinfo {author} {\bibfnamefont {D.~N.}\ \bibnamefont
  {{Schramm}}}, \ and\ \bibinfo {author} {\bibfnamefont {J.~N.}\ \bibnamefont
  {{Fry}}},\ }\href@noop {} {\bibfield  {journal} {\bibinfo  {journal}
  {Comments Nucl.~Part.~Phys.}\ }\textbf {\bibinfo {volume} {19}},\ \bibinfo
  {pages} {25} (\bibinfo {year} {1989})}\BibitemShut {NoStop}%
\bibitem [{\citenamefont {{Press}}\ \emph {et~al.}(1990)\citenamefont
  {{Press}}, \citenamefont {{Ryden}},\ and\ \citenamefont
  {{Spergel}}}]{Press:1990vn}%
  \BibitemOpen
  \bibfield  {author} {\bibinfo {author} {\bibfnamefont {W.~H.}\ \bibnamefont
  {{Press}}}, \bibinfo {author} {\bibfnamefont {B.~S.}\ \bibnamefont
  {{Ryden}}}, \ and\ \bibinfo {author} {\bibfnamefont {D.~N.}\ \bibnamefont
  {{Spergel}}},\ }\href {\doibase 10.1103/PhysRevLett.64.1084} {\bibfield
  {journal} {\bibinfo  {journal} {Physical Review Letters}\ }\textbf {\bibinfo
  {volume} {64}},\ \bibinfo {pages} {1084} (\bibinfo {year}
  {1990})}\BibitemShut {NoStop}%
\bibitem [{\citenamefont {{Luo}}\ and\ \citenamefont
  {{Schramm}}(1994)}]{Luo:1994rt}%
  \BibitemOpen
  \bibfield  {author} {\bibinfo {author} {\bibfnamefont {X.}~\bibnamefont
  {{Luo}}}\ and\ \bibinfo {author} {\bibfnamefont {D.~N.}\ \bibnamefont
  {{Schramm}}},\ }\href {\doibase 10.1086/173658} {\bibfield  {journal}
  {\bibinfo  {journal} {\apj}\ }\textbf {\bibinfo {volume} {421}},\ \bibinfo
  {pages} {393} (\bibinfo {year} {1994})}\BibitemShut {NoStop}%
\bibitem [{\citenamefont {{Dutta}}\ \emph {et~al.}(2009)\citenamefont
  {{Dutta}}, \citenamefont {{Hsu}}, \citenamefont {{Reeb}},\ and\ \citenamefont
  {{Scherrer}}}]{Dutta:2009lr}%
  \BibitemOpen
  \bibfield  {author} {\bibinfo {author} {\bibfnamefont {S.}~\bibnamefont
  {{Dutta}}}, \bibinfo {author} {\bibfnamefont {S.~D.~H.}\ \bibnamefont
  {{Hsu}}}, \bibinfo {author} {\bibfnamefont {D.}~\bibnamefont {{Reeb}}}, \
  and\ \bibinfo {author} {\bibfnamefont {R.~J.}\ \bibnamefont {{Scherrer}}},\
  }\href {\doibase 10.1103/PhysRevD.79.103504} {\bibfield  {journal} {\bibinfo
  {journal} {\prd}\ }\textbf {\bibinfo {volume} {79}},\ \bibinfo {eid} {103504}
  (\bibinfo {year} {2009})},\ \Eprint {http://arxiv.org/abs/0902.4699}
  {arXiv:0902.4699 [astro-ph.CO]} \BibitemShut {NoStop}%
\bibitem [{\citenamefont {{Fan}}\ \emph {et~al.}(2001)\citenamefont {{Fan}},
  \citenamefont {{Narayanan}}, \citenamefont {{Lupton}}, \citenamefont
  {{Strauss}}, \citenamefont {{Knapp}}, \citenamefont {{Becker}}, \citenamefont
  {{White}}, \citenamefont {{Pentericci}}, \citenamefont {{Leggett}},
  \citenamefont {{Haiman}}, \citenamefont {{Gunn}}, \citenamefont
  {{Ivezi{\'c}}}, \citenamefont {{Schneider}}, \citenamefont {{Anderson}},
  \citenamefont {{Brinkmann}}, \citenamefont {{Bahcall}}, \citenamefont
  {{Connolly}}, \citenamefont {{Csabai}}, \citenamefont {{Doi}}, \citenamefont
  {{Fukugita}}, \citenamefont {{Geballe}}, \citenamefont {{Grebel}},
  \citenamefont {{Harbeck}}, \citenamefont {{Hennessy}}, \citenamefont
  {{Lamb}}, \citenamefont {{Miknaitis}}, \citenamefont {{Munn}}, \citenamefont
  {{Nichol}}, \citenamefont {{Okamura}}, \citenamefont {{Pier}}, \citenamefont
  {{Prada}}, \citenamefont {{Richards}}, \citenamefont {{Szalay}},\ and\
  \citenamefont {{York}}}]{Fan:2001qy}%
  \BibitemOpen
  \bibfield  {author} {\bibinfo {author} {\bibfnamefont {X.}~\bibnamefont
  {{Fan}}}, \bibinfo {author} {\bibfnamefont {V.~K.}\ \bibnamefont
  {{Narayanan}}}, \bibinfo {author} {\bibfnamefont {R.~H.}\ \bibnamefont
  {{Lupton}}}, \bibinfo {author} {\bibfnamefont {M.~A.}\ \bibnamefont
  {{Strauss}}}, \bibinfo {author} {\bibfnamefont {G.~R.}\ \bibnamefont
  {{Knapp}}}, \bibinfo {author} {\bibfnamefont {R.~H.}\ \bibnamefont
  {{Becker}}}, \bibinfo {author} {\bibfnamefont {R.~L.}\ \bibnamefont
  {{White}}}, \bibinfo {author} {\bibfnamefont {L.}~\bibnamefont
  {{Pentericci}}}, \bibinfo {author} {\bibfnamefont {S.~K.}\ \bibnamefont
  {{Leggett}}}, \bibinfo {author} {\bibfnamefont {Z.}~\bibnamefont {{Haiman}}},
  \bibinfo {author} {\bibfnamefont {J.~E.}\ \bibnamefont {{Gunn}}}, \bibinfo
  {author} {\bibfnamefont {{\v Z}.}~\bibnamefont {{Ivezi{\'c}}}}, \bibinfo
  {author} {\bibfnamefont {D.~P.}\ \bibnamefont {{Schneider}}}, \bibinfo
  {author} {\bibfnamefont {S.~F.}\ \bibnamefont {{Anderson}}}, \bibinfo
  {author} {\bibfnamefont {J.}~\bibnamefont {{Brinkmann}}}, \bibinfo {author}
  {\bibfnamefont {N.~A.}\ \bibnamefont {{Bahcall}}}, \bibinfo {author}
  {\bibfnamefont {A.~J.}\ \bibnamefont {{Connolly}}}, \bibinfo {author}
  {\bibfnamefont {I.}~\bibnamefont {{Csabai}}}, \bibinfo {author}
  {\bibfnamefont {M.}~\bibnamefont {{Doi}}}, \bibinfo {author} {\bibfnamefont
  {M.}~\bibnamefont {{Fukugita}}}, \bibinfo {author} {\bibfnamefont
  {T.}~\bibnamefont {{Geballe}}}, \bibinfo {author} {\bibfnamefont {E.~K.}\
  \bibnamefont {{Grebel}}}, \bibinfo {author} {\bibfnamefont {D.}~\bibnamefont
  {{Harbeck}}}, \bibinfo {author} {\bibfnamefont {G.}~\bibnamefont
  {{Hennessy}}}, \bibinfo {author} {\bibfnamefont {D.~Q.}\ \bibnamefont
  {{Lamb}}}, \bibinfo {author} {\bibfnamefont {G.}~\bibnamefont {{Miknaitis}}},
  \bibinfo {author} {\bibfnamefont {J.~A.}\ \bibnamefont {{Munn}}}, \bibinfo
  {author} {\bibfnamefont {R.}~\bibnamefont {{Nichol}}}, \bibinfo {author}
  {\bibfnamefont {S.}~\bibnamefont {{Okamura}}}, \bibinfo {author}
  {\bibfnamefont {J.~R.}\ \bibnamefont {{Pier}}}, \bibinfo {author}
  {\bibfnamefont {F.}~\bibnamefont {{Prada}}}, \bibinfo {author} {\bibfnamefont
  {G.~T.}\ \bibnamefont {{Richards}}}, \bibinfo {author} {\bibfnamefont
  {A.}~\bibnamefont {{Szalay}}}, \ and\ \bibinfo {author} {\bibfnamefont
  {D.~G.}\ \bibnamefont {{York}}},\ }\href {\doibase 10.1086/324111} {\bibfield
   {journal} {\bibinfo  {journal} {Astronomical Journal}\ }\textbf {\bibinfo
  {volume} {122}},\ \bibinfo {pages} {2833} (\bibinfo {year} {2001})},\ \Eprint
  {http://arxiv.org/abs/astro-ph/0108063} {astro-ph/0108063} \BibitemShut
  {NoStop}%
\bibitem [{\citenamefont {{Fan}}\ \emph {et~al.}(2003)\citenamefont {{Fan}},
  \citenamefont {{Strauss}}, \citenamefont {{Schneider}}, \citenamefont
  {{Becker}}, \citenamefont {{White}}, \citenamefont {{Haiman}}, \citenamefont
  {{Gregg}}, \citenamefont {{Pentericci}}, \citenamefont {{Grebel}},
  \citenamefont {{Narayanan}}, \citenamefont {{Loh}}, \citenamefont
  {{Richards}}, \citenamefont {{Gunn}}, \citenamefont {{Lupton}}, \citenamefont
  {{Knapp}}, \citenamefont {{Ivezi{\'c}}}, \citenamefont {{Brandt}},
  \citenamefont {{Collinge}}, \citenamefont {{Hao}}, \citenamefont {{Harbeck}},
  \citenamefont {{Prada}}, \citenamefont {{Schaye}}, \citenamefont
  {{Strateva}}, \citenamefont {{Zakamska}}, \citenamefont {{Anderson}},
  \citenamefont {{Brinkmann}}, \citenamefont {{Bahcall}}, \citenamefont
  {{Lamb}}, \citenamefont {{Okamura}}, \citenamefont {{Szalay}},\ and\
  \citenamefont {{York}}}]{Fan:2003lr}%
  \BibitemOpen
  \bibfield  {author} {\bibinfo {author} {\bibfnamefont {X.}~\bibnamefont
  {{Fan}}}, \bibinfo {author} {\bibfnamefont {M.~A.}\ \bibnamefont
  {{Strauss}}}, \bibinfo {author} {\bibfnamefont {D.~P.}\ \bibnamefont
  {{Schneider}}}, \bibinfo {author} {\bibfnamefont {R.~H.}\ \bibnamefont
  {{Becker}}}, \bibinfo {author} {\bibfnamefont {R.~L.}\ \bibnamefont
  {{White}}}, \bibinfo {author} {\bibfnamefont {Z.}~\bibnamefont {{Haiman}}},
  \bibinfo {author} {\bibfnamefont {M.}~\bibnamefont {{Gregg}}}, \bibinfo
  {author} {\bibfnamefont {L.}~\bibnamefont {{Pentericci}}}, \bibinfo {author}
  {\bibfnamefont {E.~K.}\ \bibnamefont {{Grebel}}}, \bibinfo {author}
  {\bibfnamefont {V.~K.}\ \bibnamefont {{Narayanan}}}, \bibinfo {author}
  {\bibfnamefont {Y.-S.}\ \bibnamefont {{Loh}}}, \bibinfo {author}
  {\bibfnamefont {G.~T.}\ \bibnamefont {{Richards}}}, \bibinfo {author}
  {\bibfnamefont {J.~E.}\ \bibnamefont {{Gunn}}}, \bibinfo {author}
  {\bibfnamefont {R.~H.}\ \bibnamefont {{Lupton}}}, \bibinfo {author}
  {\bibfnamefont {G.~R.}\ \bibnamefont {{Knapp}}}, \bibinfo {author}
  {\bibfnamefont {{\v Z}.}~\bibnamefont {{Ivezi{\'c}}}}, \bibinfo {author}
  {\bibfnamefont {W.~N.}\ \bibnamefont {{Brandt}}}, \bibinfo {author}
  {\bibfnamefont {M.}~\bibnamefont {{Collinge}}}, \bibinfo {author}
  {\bibfnamefont {L.}~\bibnamefont {{Hao}}}, \bibinfo {author} {\bibfnamefont
  {D.}~\bibnamefont {{Harbeck}}}, \bibinfo {author} {\bibfnamefont
  {F.}~\bibnamefont {{Prada}}}, \bibinfo {author} {\bibfnamefont
  {J.}~\bibnamefont {{Schaye}}}, \bibinfo {author} {\bibfnamefont
  {I.}~\bibnamefont {{Strateva}}}, \bibinfo {author} {\bibfnamefont
  {N.}~\bibnamefont {{Zakamska}}}, \bibinfo {author} {\bibfnamefont
  {S.}~\bibnamefont {{Anderson}}}, \bibinfo {author} {\bibfnamefont
  {J.}~\bibnamefont {{Brinkmann}}}, \bibinfo {author} {\bibfnamefont {N.~A.}\
  \bibnamefont {{Bahcall}}}, \bibinfo {author} {\bibfnamefont {D.~Q.}\
  \bibnamefont {{Lamb}}}, \bibinfo {author} {\bibfnamefont {S.}~\bibnamefont
  {{Okamura}}}, \bibinfo {author} {\bibfnamefont {A.}~\bibnamefont {{Szalay}}},
  \ and\ \bibinfo {author} {\bibfnamefont {D.~G.}\ \bibnamefont {{York}}},\
  }\href {\doibase 10.1086/368246} {\bibfield  {journal} {\bibinfo  {journal}
  {Astronomical Journal}\ }\textbf {\bibinfo {volume} {125}},\ \bibinfo {pages}
  {1649} (\bibinfo {year} {2003})},\ \Eprint
  {http://arxiv.org/abs/astro-ph/0301135} {astro-ph/0301135} \BibitemShut
  {NoStop}%
\bibitem [{\citenamefont {{Willott}}\ \emph {et~al.}(2003)\citenamefont
  {{Willott}}, \citenamefont {{McLure}},\ and\ \citenamefont
  {{Jarvis}}}]{Willott:2003fk}%
  \BibitemOpen
  \bibfield  {author} {\bibinfo {author} {\bibfnamefont {C.~J.}\ \bibnamefont
  {{Willott}}}, \bibinfo {author} {\bibfnamefont {R.~J.}\ \bibnamefont
  {{McLure}}}, \ and\ \bibinfo {author} {\bibfnamefont {M.~J.}\ \bibnamefont
  {{Jarvis}}},\ }\href {\doibase 10.1086/375126} {\bibfield  {journal}
  {\bibinfo  {journal} {Astrophys.~J.~(Letters)}\ }\textbf {\bibinfo {volume}
  {587}},\ \bibinfo {pages} {L15} (\bibinfo {year} {2003})},\ \Eprint
  {http://arxiv.org/abs/astro-ph/0303062} {astro-ph/0303062} \BibitemShut
  {NoStop}%
\bibitem [{\citenamefont {{Mortlock}}\ \emph {et~al.}(2011)\citenamefont
  {{Mortlock}}, \citenamefont {{Warren}}, \citenamefont {{Venemans}},
  \citenamefont {{Patel}}, \citenamefont {{Hewett}}, \citenamefont {{McMahon}},
  \citenamefont {{Simpson}}, \citenamefont {{Theuns}}, \citenamefont
  {{Gonz{\'a}les-Solares}}, \citenamefont {{Adamson}}, \citenamefont {{Dye}},
  \citenamefont {{Hambly}}, \citenamefont {{Hirst}}, \citenamefont {{Irwin}},
  \citenamefont {{Kuiper}}, \citenamefont {{Lawrence}},\ and\ \citenamefont
  {{R{\"o}ttgering}}}]{Mortlock:2011lr}%
  \BibitemOpen
  \bibfield  {author} {\bibinfo {author} {\bibfnamefont {D.~J.}\ \bibnamefont
  {{Mortlock}}}, \bibinfo {author} {\bibfnamefont {S.~J.}\ \bibnamefont
  {{Warren}}}, \bibinfo {author} {\bibfnamefont {B.~P.}\ \bibnamefont
  {{Venemans}}}, \bibinfo {author} {\bibfnamefont {M.}~\bibnamefont {{Patel}}},
  \bibinfo {author} {\bibfnamefont {P.~C.}\ \bibnamefont {{Hewett}}}, \bibinfo
  {author} {\bibfnamefont {R.~G.}\ \bibnamefont {{McMahon}}}, \bibinfo {author}
  {\bibfnamefont {C.}~\bibnamefont {{Simpson}}}, \bibinfo {author}
  {\bibfnamefont {T.}~\bibnamefont {{Theuns}}}, \bibinfo {author}
  {\bibfnamefont {E.~A.}\ \bibnamefont {{Gonz{\'a}les-Solares}}}, \bibinfo
  {author} {\bibfnamefont {A.}~\bibnamefont {{Adamson}}}, \bibinfo {author}
  {\bibfnamefont {S.}~\bibnamefont {{Dye}}}, \bibinfo {author} {\bibfnamefont
  {N.~C.}\ \bibnamefont {{Hambly}}}, \bibinfo {author} {\bibfnamefont
  {P.}~\bibnamefont {{Hirst}}}, \bibinfo {author} {\bibfnamefont {M.~J.}\
  \bibnamefont {{Irwin}}}, \bibinfo {author} {\bibfnamefont {E.}~\bibnamefont
  {{Kuiper}}}, \bibinfo {author} {\bibfnamefont {A.}~\bibnamefont
  {{Lawrence}}}, \ and\ \bibinfo {author} {\bibfnamefont {H.~J.~A.}\
  \bibnamefont {{R{\"o}ttgering}}},\ }\href {\doibase 10.1038/nature10159}
  {\bibfield  {journal} {\bibinfo  {journal} {\nat}\ }\textbf {\bibinfo
  {volume} {474}},\ \bibinfo {pages} {616} (\bibinfo {year} {2011})},\ \Eprint
  {http://arxiv.org/abs/1106.6088} {arXiv:1106.6088 [astro-ph.CO]} \BibitemShut
  {NoStop}%
\bibitem [{\citenamefont {{Efstathiou}}\ and\ \citenamefont
  {{Rees}}(1988)}]{Efstathiou:1988lr}%
  \BibitemOpen
  \bibfield  {author} {\bibinfo {author} {\bibfnamefont {G.}~\bibnamefont
  {{Efstathiou}}}\ and\ \bibinfo {author} {\bibfnamefont {M.~J.}\ \bibnamefont
  {{Rees}}},\ }\href@noop {} {\bibfield  {journal} {\bibinfo  {journal} {Mon.
  Not. Royal Astron. Soc.}\ }\textbf {\bibinfo {volume} {230}},\ \bibinfo
  {pages} {5P} (\bibinfo {year} {1988})}\BibitemShut {NoStop}%
\bibitem [{\citenamefont {{Haiman}}\ and\ \citenamefont
  {{Loeb}}(2001)}]{Haiman:2001fk}%
  \BibitemOpen
  \bibfield  {author} {\bibinfo {author} {\bibfnamefont {Z.}~\bibnamefont
  {{Haiman}}}\ and\ \bibinfo {author} {\bibfnamefont {A.}~\bibnamefont
  {{Loeb}}},\ }\href {\doibase 10.1086/320586} {\bibfield  {journal} {\bibinfo
  {journal} {\apj}\ }\textbf {\bibinfo {volume} {552}},\ \bibinfo {pages} {459}
  (\bibinfo {year} {2001})},\ \Eprint {http://arxiv.org/abs/astro-ph/0011529}
  {astro-ph/0011529} \BibitemShut {NoStop}%
\bibitem [{\citenamefont {{Yoo}}\ and\ \citenamefont
  {{Miralda-Escud{\'e}}}(2004)}]{Yoo:2004fk}%
  \BibitemOpen
  \bibfield  {author} {\bibinfo {author} {\bibfnamefont {J.}~\bibnamefont
  {{Yoo}}}\ and\ \bibinfo {author} {\bibfnamefont {J.}~\bibnamefont
  {{Miralda-Escud{\'e}}}},\ }\href {\doibase 10.1086/425416} {\bibfield
  {journal} {\bibinfo  {journal} {Astrophys. J. (Letters)}\ }\textbf {\bibinfo
  {volume} {614}},\ \bibinfo {pages} {L25} (\bibinfo {year} {2004})},\ \Eprint
  {http://arxiv.org/abs/astro-ph/0406217} {astro-ph/0406217} \BibitemShut
  {NoStop}%
\bibitem [{\citenamefont {{Shapiro}}(2005)}]{Shapiro:2005qy}%
  \BibitemOpen
  \bibfield  {author} {\bibinfo {author} {\bibfnamefont {S.~L.}\ \bibnamefont
  {{Shapiro}}},\ }\href {\doibase 10.1086/427065} {\bibfield  {journal}
  {\bibinfo  {journal} {\apj}\ }\textbf {\bibinfo {volume} {620}},\ \bibinfo
  {pages} {59} (\bibinfo {year} {2005})},\ \Eprint
  {http://arxiv.org/abs/astro-ph/0411156} {astro-ph/0411156} \BibitemShut
  {NoStop}%
\bibitem [{\citenamefont {{Khlopov}}\ \emph {et~al.}(2005)\citenamefont
  {{Khlopov}}, \citenamefont {{Rubin}},\ and\ \citenamefont
  {{Sakharov}}}]{Khlopov:2005uq}%
  \BibitemOpen
  \bibfield  {author} {\bibinfo {author} {\bibfnamefont {M.~Y.}\ \bibnamefont
  {{Khlopov}}}, \bibinfo {author} {\bibfnamefont {S.~G.}\ \bibnamefont
  {{Rubin}}}, \ and\ \bibinfo {author} {\bibfnamefont {A.~S.}\ \bibnamefont
  {{Sakharov}}},\ }\href {\doibase 10.1016/j.astropartphys.2004.12.002}
  {\bibfield  {journal} {\bibinfo  {journal} {Astroparticle Physics}\ }\textbf
  {\bibinfo {volume} {23}},\ \bibinfo {pages} {265} (\bibinfo {year} {2005})},\
  \Eprint {http://arxiv.org/abs/astro-ph/0401532} {astro-ph/0401532}
  \BibitemShut {NoStop}%
\bibitem [{\citenamefont {{Volonteri}}\ and\ \citenamefont
  {{Rees}}(2005)}]{Volonteri:2005uq}%
  \BibitemOpen
  \bibfield  {author} {\bibinfo {author} {\bibfnamefont {M.}~\bibnamefont
  {{Volonteri}}}\ and\ \bibinfo {author} {\bibfnamefont {M.~J.}\ \bibnamefont
  {{Rees}}},\ }\href {\doibase 10.1086/466521} {\bibfield  {journal} {\bibinfo
  {journal} {\apj}\ }\textbf {\bibinfo {volume} {633}},\ \bibinfo {pages} {624}
  (\bibinfo {year} {2005})},\ \Eprint {http://arxiv.org/abs/astro-ph/0506040}
  {astro-ph/0506040} \BibitemShut {NoStop}%
\bibitem [{\citenamefont {{Volonteri}}\ and\ \citenamefont
  {{Rees}}(2006)}]{Volonteri:2006fj}%
  \BibitemOpen
  \bibfield  {author} {\bibinfo {author} {\bibfnamefont {M.}~\bibnamefont
  {{Volonteri}}}\ and\ \bibinfo {author} {\bibfnamefont {M.~J.}\ \bibnamefont
  {{Rees}}},\ }\href {\doibase 10.1086/507444} {\bibfield  {journal} {\bibinfo
  {journal} {\apj}\ }\textbf {\bibinfo {volume} {650}},\ \bibinfo {pages} {669}
  (\bibinfo {year} {2006})},\ \Eprint {http://arxiv.org/abs/astro-ph/0607093}
  {astro-ph/0607093} \BibitemShut {NoStop}%
\bibitem [{\citenamefont {{King}}\ and\ \citenamefont
  {{Pringle}}(2006)}]{King:2006kx}%
  \BibitemOpen
  \bibfield  {author} {\bibinfo {author} {\bibfnamefont {A.~R.}\ \bibnamefont
  {{King}}}\ and\ \bibinfo {author} {\bibfnamefont {J.~E.}\ \bibnamefont
  {{Pringle}}},\ }\href {\doibase 10.1111/j.1745-3933.2006.00249.x} {\bibfield
  {journal} {\bibinfo  {journal} {Mon. Not. Royal Astron. Soc.}\ }\textbf
  {\bibinfo {volume} {373}},\ \bibinfo {pages} {L90} (\bibinfo {year}
  {2006})},\ \Eprint {http://arxiv.org/abs/astro-ph/0609598} {astro-ph/0609598}
  \BibitemShut {NoStop}%
\bibitem [{\citenamefont {{Li}}\ \emph {et~al.}(2007)\citenamefont {{Li}},
  \citenamefont {{Hernquist}}, \citenamefont {{Robertson}}, \citenamefont
  {{Cox}}, \citenamefont {{Hopkins}}, \citenamefont {{Springel}}, \citenamefont
  {{Gao}}, \citenamefont {{Di Matteo}}, \citenamefont {{Zentner}},
  \citenamefont {{Jenkins}},\ and\ \citenamefont {{Yoshida}}}]{Li:2007yq}%
  \BibitemOpen
  \bibfield  {author} {\bibinfo {author} {\bibfnamefont {Y.}~\bibnamefont
  {{Li}}}, \bibinfo {author} {\bibfnamefont {L.}~\bibnamefont {{Hernquist}}},
  \bibinfo {author} {\bibfnamefont {B.}~\bibnamefont {{Robertson}}}, \bibinfo
  {author} {\bibfnamefont {T.~J.}\ \bibnamefont {{Cox}}}, \bibinfo {author}
  {\bibfnamefont {P.~F.}\ \bibnamefont {{Hopkins}}}, \bibinfo {author}
  {\bibfnamefont {V.}~\bibnamefont {{Springel}}}, \bibinfo {author}
  {\bibfnamefont {L.}~\bibnamefont {{Gao}}}, \bibinfo {author} {\bibfnamefont
  {T.}~\bibnamefont {{Di Matteo}}}, \bibinfo {author} {\bibfnamefont {A.~R.}\
  \bibnamefont {{Zentner}}}, \bibinfo {author} {\bibfnamefont {A.}~\bibnamefont
  {{Jenkins}}}, \ and\ \bibinfo {author} {\bibfnamefont {N.}~\bibnamefont
  {{Yoshida}}},\ }\href {\doibase 10.1086/519297} {\bibfield  {journal}
  {\bibinfo  {journal} {\apj}\ }\textbf {\bibinfo {volume} {665}},\ \bibinfo
  {pages} {187} (\bibinfo {year} {2007})},\ \Eprint
  {http://arxiv.org/abs/astro-ph/0608190} {astro-ph/0608190} \BibitemShut
  {NoStop}%
\bibitem [{\citenamefont {{Kawakatu}}\ and\ \citenamefont
  {{Wada}}(2009)}]{Kawakatu:2009vn}%
  \BibitemOpen
  \bibfield  {author} {\bibinfo {author} {\bibfnamefont {N.}~\bibnamefont
  {{Kawakatu}}}\ and\ \bibinfo {author} {\bibfnamefont {K.}~\bibnamefont
  {{Wada}}},\ }\href {\doibase 10.1088/0004-637X/706/1/676} {\bibfield
  {journal} {\bibinfo  {journal} {\apj}\ }\textbf {\bibinfo {volume} {706}},\
  \bibinfo {pages} {676} (\bibinfo {year} {2009})},\ \Eprint
  {http://arxiv.org/abs/0910.1379} {arXiv:0910.1379 [astro-ph.CO]} \BibitemShut
  {NoStop}%
\bibitem [{\citenamefont {{Sijacki}}\ \emph {et~al.}(2009)\citenamefont
  {{Sijacki}}, \citenamefont {{Springel}},\ and\ \citenamefont
  {{Haehnelt}}}]{Sijacki:2009rt}%
  \BibitemOpen
  \bibfield  {author} {\bibinfo {author} {\bibfnamefont {D.}~\bibnamefont
  {{Sijacki}}}, \bibinfo {author} {\bibfnamefont {V.}~\bibnamefont
  {{Springel}}}, \ and\ \bibinfo {author} {\bibfnamefont {M.~G.}\ \bibnamefont
  {{Haehnelt}}},\ }\href {\doibase 10.1111/j.1365-2966.2009.15452.x} {\bibfield
   {journal} {\bibinfo  {journal} {Mon. Not. Royal Astron. Soc.}\ }\textbf
  {\bibinfo {volume} {400}},\ \bibinfo {pages} {100} (\bibinfo {year}
  {2009})},\ \Eprint {http://arxiv.org/abs/0905.1689} {arXiv:0905.1689
  [astro-ph.CO]} \BibitemShut {NoStop}%
\bibitem [{\citenamefont {{Hou}}\ \emph {et~al.}(2014)\citenamefont {{Hou}},
  \citenamefont {{Reichardt}}, \citenamefont {{Story}}, \citenamefont
  {{Follin}}, \citenamefont {{Keisler}}, \citenamefont {{Aird}}, \citenamefont
  {{Benson}}, \citenamefont {{Bleem}}, \citenamefont {{Carlstrom}},
  \citenamefont {{Chang}}, \citenamefont {{Cho}}, \citenamefont {{Crawford}},
  \citenamefont {{Crites}}, \citenamefont {{de Haan}}, \citenamefont {{de
  Putter}}, \citenamefont {{Dobbs}}, \citenamefont {{Dodelson}}, \citenamefont
  {{Dudley}}, \citenamefont {{George}}, \citenamefont {{Halverson}},
  \citenamefont {{Holder}}, \citenamefont {{Holzapfel}}, \citenamefont
  {{Hoover}}, \citenamefont {{Hrubes}}, \citenamefont {{Joy}}, \citenamefont
  {{Knox}}, \citenamefont {{Lee}}, \citenamefont {{Leitch}}, \citenamefont
  {{Lueker}}, \citenamefont {{Luong-Van}}, \citenamefont {{McMahon}},
  \citenamefont {{Mehl}}, \citenamefont {{Meyer}}, \citenamefont {{Millea}},
  \citenamefont {{Mohr}}, \citenamefont {{Montroy}}, \citenamefont {{Padin}},
  \citenamefont {{Plagge}}, \citenamefont {{Pryke}}, \citenamefont {{Ruhl}},
  \citenamefont {{Sayre}}, \citenamefont {{Schaffer}}, \citenamefont {{Shaw}},
  \citenamefont {{Shirokoff}}, \citenamefont {{Spieler}}, \citenamefont
  {{Staniszewski}}, \citenamefont {{Stark}}, \citenamefont {{van Engelen}},
  \citenamefont {{Vanderlinde}}, \citenamefont {{Vieira}}, \citenamefont
  {{Williamson}},\ and\ \citenamefont {{Zahn}}}]{Hou:2012zr}%
  \BibitemOpen
  \bibfield  {author} {\bibinfo {author} {\bibfnamefont {Z.}~\bibnamefont
  {{Hou}}}, \bibinfo {author} {\bibfnamefont {C.~L.}\ \bibnamefont
  {{Reichardt}}}, \bibinfo {author} {\bibfnamefont {K.~T.}\ \bibnamefont
  {{Story}}}, \bibinfo {author} {\bibfnamefont {B.}~\bibnamefont {{Follin}}},
  \bibinfo {author} {\bibfnamefont {R.}~\bibnamefont {{Keisler}}}, \bibinfo
  {author} {\bibfnamefont {K.~A.}\ \bibnamefont {{Aird}}}, \bibinfo {author}
  {\bibfnamefont {B.~A.}\ \bibnamefont {{Benson}}}, \bibinfo {author}
  {\bibfnamefont {L.~E.}\ \bibnamefont {{Bleem}}}, \bibinfo {author}
  {\bibfnamefont {J.~E.}\ \bibnamefont {{Carlstrom}}}, \bibinfo {author}
  {\bibfnamefont {C.~L.}\ \bibnamefont {{Chang}}}, \bibinfo {author}
  {\bibfnamefont {H.-M.}\ \bibnamefont {{Cho}}}, \bibinfo {author}
  {\bibfnamefont {T.~M.}\ \bibnamefont {{Crawford}}}, \bibinfo {author}
  {\bibfnamefont {A.~T.}\ \bibnamefont {{Crites}}}, \bibinfo {author}
  {\bibfnamefont {T.}~\bibnamefont {{de Haan}}}, \bibinfo {author}
  {\bibfnamefont {R.}~\bibnamefont {{de Putter}}}, \bibinfo {author}
  {\bibfnamefont {M.~A.}\ \bibnamefont {{Dobbs}}}, \bibinfo {author}
  {\bibfnamefont {S.}~\bibnamefont {{Dodelson}}}, \bibinfo {author}
  {\bibfnamefont {J.}~\bibnamefont {{Dudley}}}, \bibinfo {author}
  {\bibfnamefont {E.~M.}\ \bibnamefont {{George}}}, \bibinfo {author}
  {\bibfnamefont {N.~W.}\ \bibnamefont {{Halverson}}}, \bibinfo {author}
  {\bibfnamefont {G.~P.}\ \bibnamefont {{Holder}}}, \bibinfo {author}
  {\bibfnamefont {W.~L.}\ \bibnamefont {{Holzapfel}}}, \bibinfo {author}
  {\bibfnamefont {S.}~\bibnamefont {{Hoover}}}, \bibinfo {author}
  {\bibfnamefont {J.~D.}\ \bibnamefont {{Hrubes}}}, \bibinfo {author}
  {\bibfnamefont {M.}~\bibnamefont {{Joy}}}, \bibinfo {author} {\bibfnamefont
  {L.}~\bibnamefont {{Knox}}}, \bibinfo {author} {\bibfnamefont {A.~T.}\
  \bibnamefont {{Lee}}}, \bibinfo {author} {\bibfnamefont {E.~M.}\ \bibnamefont
  {{Leitch}}}, \bibinfo {author} {\bibfnamefont {M.}~\bibnamefont {{Lueker}}},
  \bibinfo {author} {\bibfnamefont {D.}~\bibnamefont {{Luong-Van}}}, \bibinfo
  {author} {\bibfnamefont {J.~J.}\ \bibnamefont {{McMahon}}}, \bibinfo {author}
  {\bibfnamefont {J.}~\bibnamefont {{Mehl}}}, \bibinfo {author} {\bibfnamefont
  {S.~S.}\ \bibnamefont {{Meyer}}}, \bibinfo {author} {\bibfnamefont
  {M.}~\bibnamefont {{Millea}}}, \bibinfo {author} {\bibfnamefont {J.~J.}\
  \bibnamefont {{Mohr}}}, \bibinfo {author} {\bibfnamefont {T.~E.}\
  \bibnamefont {{Montroy}}}, \bibinfo {author} {\bibfnamefont {S.}~\bibnamefont
  {{Padin}}}, \bibinfo {author} {\bibfnamefont {T.}~\bibnamefont {{Plagge}}},
  \bibinfo {author} {\bibfnamefont {C.}~\bibnamefont {{Pryke}}}, \bibinfo
  {author} {\bibfnamefont {J.~E.}\ \bibnamefont {{Ruhl}}}, \bibinfo {author}
  {\bibfnamefont {J.~T.}\ \bibnamefont {{Sayre}}}, \bibinfo {author}
  {\bibfnamefont {K.~K.}\ \bibnamefont {{Schaffer}}}, \bibinfo {author}
  {\bibfnamefont {L.}~\bibnamefont {{Shaw}}}, \bibinfo {author} {\bibfnamefont
  {E.}~\bibnamefont {{Shirokoff}}}, \bibinfo {author} {\bibfnamefont {H.~G.}\
  \bibnamefont {{Spieler}}}, \bibinfo {author} {\bibfnamefont {Z.}~\bibnamefont
  {{Staniszewski}}}, \bibinfo {author} {\bibfnamefont {A.~A.}\ \bibnamefont
  {{Stark}}}, \bibinfo {author} {\bibfnamefont {A.}~\bibnamefont {{van
  Engelen}}}, \bibinfo {author} {\bibfnamefont {K.}~\bibnamefont
  {{Vanderlinde}}}, \bibinfo {author} {\bibfnamefont {J.~D.}\ \bibnamefont
  {{Vieira}}}, \bibinfo {author} {\bibfnamefont {R.}~\bibnamefont
  {{Williamson}}}, \ and\ \bibinfo {author} {\bibfnamefont {O.}~\bibnamefont
  {{Zahn}}},\ }\href {\doibase 10.1088/0004-637X/782/2/74} {\bibfield
  {journal} {\bibinfo  {journal} {\apj}\ }\textbf {\bibinfo {volume} {782}},\
  \bibinfo {eid} {74} (\bibinfo {year} {2014})},\ \Eprint
  {http://arxiv.org/abs/1212.6267} {arXiv:1212.6267 [astro-ph.CO]} \BibitemShut
  {NoStop}%
\bibitem [{\citenamefont {{Planck Collaboration}}\ \emph
  {et~al.}(2013)\citenamefont {{Planck Collaboration}}, \citenamefont {{Ade}},
  \citenamefont {{Aghanim}}, \citenamefont {{Armitage-Caplan}}, \citenamefont
  {{Arnaud}}, \citenamefont {{Ashdown}}, \citenamefont {{Atrio-Barandela}},
  \citenamefont {{Aumont}}, \citenamefont {{Baccigalupi}}, \citenamefont
  {{Banday}},\ and\ \citenamefont {et~al.}}]{Planck-Collaboration:2013zr}%
  \BibitemOpen
  \bibfield  {author} {\bibinfo {author} {\bibnamefont {{Planck
  Collaboration}}}, \bibinfo {author} {\bibfnamefont {P.~A.~R.}\ \bibnamefont
  {{Ade}}}, \bibinfo {author} {\bibfnamefont {N.}~\bibnamefont {{Aghanim}}},
  \bibinfo {author} {\bibfnamefont {C.}~\bibnamefont {{Armitage-Caplan}}},
  \bibinfo {author} {\bibfnamefont {M.}~\bibnamefont {{Arnaud}}}, \bibinfo
  {author} {\bibfnamefont {M.}~\bibnamefont {{Ashdown}}}, \bibinfo {author}
  {\bibfnamefont {F.}~\bibnamefont {{Atrio-Barandela}}}, \bibinfo {author}
  {\bibfnamefont {J.}~\bibnamefont {{Aumont}}}, \bibinfo {author}
  {\bibfnamefont {C.}~\bibnamefont {{Baccigalupi}}}, \bibinfo {author}
  {\bibfnamefont {A.~J.}\ \bibnamefont {{Banday}}}, \ and\ \bibinfo {author}
  {\bibnamefont {et~al.}},\ }\href@noop {} {\bibfield  {journal} {\bibinfo
  {journal} {ArXiv e-prints}\ } (\bibinfo {year} {2013})},\ \Eprint
  {http://arxiv.org/abs/1303.5076} {arXiv:1303.5076 [astro-ph.CO]} \BibitemShut
  {NoStop}%
\bibitem [{\citenamefont {{Hinshaw}}\ \emph {et~al.}(2013)\citenamefont
  {{Hinshaw}}, \citenamefont {{Larson}}, \citenamefont {{Komatsu}},
  \citenamefont {{Spergel}}, \citenamefont {{Bennett}}, \citenamefont
  {{Dunkley}}, \citenamefont {{Nolta}}, \citenamefont {{Halpern}},
  \citenamefont {{Hill}}, \citenamefont {{Odegard}}, \citenamefont {{Page}},
  \citenamefont {{Smith}}, \citenamefont {{Weiland}}, \citenamefont {{Gold}},
  \citenamefont {{Jarosik}}, \citenamefont {{Kogut}}, \citenamefont {{Limon}},
  \citenamefont {{Meyer}}, \citenamefont {{Tucker}}, \citenamefont
  {{Wollack}},\ and\ \citenamefont {{Wright}}}]{Hinshaw:2013ys}%
  \BibitemOpen
  \bibfield  {author} {\bibinfo {author} {\bibfnamefont {G.}~\bibnamefont
  {{Hinshaw}}}, \bibinfo {author} {\bibfnamefont {D.}~\bibnamefont {{Larson}}},
  \bibinfo {author} {\bibfnamefont {E.}~\bibnamefont {{Komatsu}}}, \bibinfo
  {author} {\bibfnamefont {D.~N.}\ \bibnamefont {{Spergel}}}, \bibinfo {author}
  {\bibfnamefont {C.~L.}\ \bibnamefont {{Bennett}}}, \bibinfo {author}
  {\bibfnamefont {J.}~\bibnamefont {{Dunkley}}}, \bibinfo {author}
  {\bibfnamefont {M.~R.}\ \bibnamefont {{Nolta}}}, \bibinfo {author}
  {\bibfnamefont {M.}~\bibnamefont {{Halpern}}}, \bibinfo {author}
  {\bibfnamefont {R.~S.}\ \bibnamefont {{Hill}}}, \bibinfo {author}
  {\bibfnamefont {N.}~\bibnamefont {{Odegard}}}, \bibinfo {author}
  {\bibfnamefont {L.}~\bibnamefont {{Page}}}, \bibinfo {author} {\bibfnamefont
  {K.~M.}\ \bibnamefont {{Smith}}}, \bibinfo {author} {\bibfnamefont {J.~L.}\
  \bibnamefont {{Weiland}}}, \bibinfo {author} {\bibfnamefont {B.}~\bibnamefont
  {{Gold}}}, \bibinfo {author} {\bibfnamefont {N.}~\bibnamefont {{Jarosik}}},
  \bibinfo {author} {\bibfnamefont {A.}~\bibnamefont {{Kogut}}}, \bibinfo
  {author} {\bibfnamefont {M.}~\bibnamefont {{Limon}}}, \bibinfo {author}
  {\bibfnamefont {S.~S.}\ \bibnamefont {{Meyer}}}, \bibinfo {author}
  {\bibfnamefont {G.~S.}\ \bibnamefont {{Tucker}}}, \bibinfo {author}
  {\bibfnamefont {E.}~\bibnamefont {{Wollack}}}, \ and\ \bibinfo {author}
  {\bibfnamefont {E.~L.}\ \bibnamefont {{Wright}}},\ }\href {\doibase
  10.1088/0067-0049/208/2/19} {\bibfield  {journal} {\bibinfo  {journal}
  {Astrophys.~J.~(Supplement)}\ }\textbf {\bibinfo {volume} {208}},\ \bibinfo
  {eid} {19} (\bibinfo {year} {2013})},\ \Eprint
  {http://arxiv.org/abs/1212.5226} {arXiv:1212.5226 [astro-ph.CO]} \BibitemShut
  {NoStop}%
\bibitem [{\citenamefont {{Bennett}}\ \emph {et~al.}(2013)\citenamefont
  {{Bennett}}, \citenamefont {{Larson}}, \citenamefont {{Weiland}},
  \citenamefont {{Jarosik}}, \citenamefont {{Hinshaw}}, \citenamefont
  {{Odegard}}, \citenamefont {{Smith}}, \citenamefont {{Hill}}, \citenamefont
  {{Gold}}, \citenamefont {{Halpern}}, \citenamefont {{Komatsu}}, \citenamefont
  {{Nolta}}, \citenamefont {{Page}}, \citenamefont {{Spergel}}, \citenamefont
  {{Wollack}}, \citenamefont {{Dunkley}}, \citenamefont {{Kogut}},
  \citenamefont {{Limon}}, \citenamefont {{Meyer}}, \citenamefont {{Tucker}},\
  and\ \citenamefont {{Wright}}}]{Bennett:2013fr}%
  \BibitemOpen
  \bibfield  {author} {\bibinfo {author} {\bibfnamefont {C.~L.}\ \bibnamefont
  {{Bennett}}}, \bibinfo {author} {\bibfnamefont {D.}~\bibnamefont {{Larson}}},
  \bibinfo {author} {\bibfnamefont {J.~L.}\ \bibnamefont {{Weiland}}}, \bibinfo
  {author} {\bibfnamefont {N.}~\bibnamefont {{Jarosik}}}, \bibinfo {author}
  {\bibfnamefont {G.}~\bibnamefont {{Hinshaw}}}, \bibinfo {author}
  {\bibfnamefont {N.}~\bibnamefont {{Odegard}}}, \bibinfo {author}
  {\bibfnamefont {K.~M.}\ \bibnamefont {{Smith}}}, \bibinfo {author}
  {\bibfnamefont {R.~S.}\ \bibnamefont {{Hill}}}, \bibinfo {author}
  {\bibfnamefont {B.}~\bibnamefont {{Gold}}}, \bibinfo {author} {\bibfnamefont
  {M.}~\bibnamefont {{Halpern}}}, \bibinfo {author} {\bibfnamefont
  {E.}~\bibnamefont {{Komatsu}}}, \bibinfo {author} {\bibfnamefont {M.~R.}\
  \bibnamefont {{Nolta}}}, \bibinfo {author} {\bibfnamefont {L.}~\bibnamefont
  {{Page}}}, \bibinfo {author} {\bibfnamefont {D.~N.}\ \bibnamefont
  {{Spergel}}}, \bibinfo {author} {\bibfnamefont {E.}~\bibnamefont
  {{Wollack}}}, \bibinfo {author} {\bibfnamefont {J.}~\bibnamefont
  {{Dunkley}}}, \bibinfo {author} {\bibfnamefont {A.}~\bibnamefont {{Kogut}}},
  \bibinfo {author} {\bibfnamefont {M.}~\bibnamefont {{Limon}}}, \bibinfo
  {author} {\bibfnamefont {S.~S.}\ \bibnamefont {{Meyer}}}, \bibinfo {author}
  {\bibfnamefont {G.~S.}\ \bibnamefont {{Tucker}}}, \ and\ \bibinfo {author}
  {\bibfnamefont {E.~L.}\ \bibnamefont {{Wright}}},\ }\href {\doibase
  10.1088/0067-0049/208/2/20} {\bibfield  {journal} {\bibinfo  {journal}
  {Astrophys.~J.~(Supplement)}\ }\textbf {\bibinfo {volume} {208}},\ \bibinfo
  {eid} {20} (\bibinfo {year} {2013})},\ \Eprint
  {http://arxiv.org/abs/1212.5225} {arXiv:1212.5225 [astro-ph.CO]} \BibitemShut
  {NoStop}%
\bibitem [{\citenamefont {{Sievers}}\ \emph {et~al.}(2013)\citenamefont
  {{Sievers}}, \citenamefont {{Hlozek}}, \citenamefont {{Nolta}}, \citenamefont
  {{Acquaviva}}, \citenamefont {{Addison}}, \citenamefont {{Ade}},
  \citenamefont {{Aguirre}}, \citenamefont {{Amiri}}, \citenamefont {{Appel}},
  \citenamefont {{Barrientos}}, \citenamefont {{Battistelli}}, \citenamefont
  {{Battaglia}}, \citenamefont {{Bond}}, \citenamefont {{Brown}}, \citenamefont
  {{Burger}}, \citenamefont {{Calabrese}}, \citenamefont {{Chervenak}},
  \citenamefont {{Crichton}}, \citenamefont {{Das}}, \citenamefont {{Devlin}},
  \citenamefont {{Dicker}}, \citenamefont {{Bertrand Doriese}}, \citenamefont
  {{Dunkley}}, \citenamefont {{D{\"u}nner}}, \citenamefont
  {{Essinger-Hileman}}, \citenamefont {{Faber}}, \citenamefont {{Fisher}},
  \citenamefont {{Fowler}}, \citenamefont {{Gallardo}}, \citenamefont
  {{Gordon}}, \citenamefont {{Gralla}}, \citenamefont {{Hajian}}, \citenamefont
  {{Halpern}}, \citenamefont {{Hasselfield}}, \citenamefont
  {{Hern{\'a}ndez-Monteagudo}}, \citenamefont {{Hill}}, \citenamefont
  {{Hilton}}, \citenamefont {{Hilton}}, \citenamefont {{Hincks}}, \citenamefont
  {{Holtz}}, \citenamefont {{Huffenberger}}, \citenamefont {{Hughes}},
  \citenamefont {{Hughes}}, \citenamefont {{Infante}}, \citenamefont {{Irwin}},
  \citenamefont {{Jacobson}}, \citenamefont {{Johnstone}}, \citenamefont
  {{Baptiste Juin}}, \citenamefont {{Kaul}}, \citenamefont {{Klein}},
  \citenamefont {{Kosowsky}}, \citenamefont {{Lau}}, \citenamefont {{Limon}},
  \citenamefont {{Lin}}, \citenamefont {{Louis}}, \citenamefont {{Lupton}},
  \citenamefont {{Marriage}}, \citenamefont {{Marsden}}, \citenamefont
  {{Martocci}}, \citenamefont {{Mauskopf}}, \citenamefont {{McLaren}},
  \citenamefont {{Menanteau}}, \citenamefont {{Moodley}}, \citenamefont
  {{Moseley}}, \citenamefont {{Netterfield}}, \citenamefont {{Niemack}},
  \citenamefont {{Page}}, \citenamefont {{Page}}, \citenamefont {{Parker}},
  \citenamefont {{Partridge}}, \citenamefont {{Plimpton}}, \citenamefont
  {{Quintana}}, \citenamefont {{Reese}}, \citenamefont {{Reid}}, \citenamefont
  {{Rojas}}, \citenamefont {{Sehgal}}, \citenamefont {{Sherwin}}, \citenamefont
  {{Schmitt}}, \citenamefont {{Spergel}}, \citenamefont {{Staggs}},
  \citenamefont {{Stryzak}}, \citenamefont {{Swetz}}, \citenamefont
  {{Switzer}}, \citenamefont {{Thornton}}, \citenamefont {{Trac}},
  \citenamefont {{Tucker}}, \citenamefont {{Uehara}}, \citenamefont
  {{Visnjic}}, \citenamefont {{Warne}}, \citenamefont {{Wilson}}, \citenamefont
  {{Wollack}}, \citenamefont {{Zhao}},\ and\ \citenamefont
  {{Zunckel}}}]{Sievers:2013lr}%
  \BibitemOpen
  \bibfield  {author} {\bibinfo {author} {\bibfnamefont {J.~L.}\ \bibnamefont
  {{Sievers}}}, \bibinfo {author} {\bibfnamefont {R.~A.}\ \bibnamefont
  {{Hlozek}}}, \bibinfo {author} {\bibfnamefont {M.~R.}\ \bibnamefont
  {{Nolta}}}, \bibinfo {author} {\bibfnamefont {V.}~\bibnamefont
  {{Acquaviva}}}, \bibinfo {author} {\bibfnamefont {G.~E.}\ \bibnamefont
  {{Addison}}}, \bibinfo {author} {\bibfnamefont {P.~A.~R.}\ \bibnamefont
  {{Ade}}}, \bibinfo {author} {\bibfnamefont {P.}~\bibnamefont {{Aguirre}}},
  \bibinfo {author} {\bibfnamefont {M.}~\bibnamefont {{Amiri}}}, \bibinfo
  {author} {\bibfnamefont {J.~W.}\ \bibnamefont {{Appel}}}, \bibinfo {author}
  {\bibfnamefont {L.~F.}\ \bibnamefont {{Barrientos}}}, \bibinfo {author}
  {\bibfnamefont {E.~S.}\ \bibnamefont {{Battistelli}}}, \bibinfo {author}
  {\bibfnamefont {N.}~\bibnamefont {{Battaglia}}}, \bibinfo {author}
  {\bibfnamefont {J.~R.}\ \bibnamefont {{Bond}}}, \bibinfo {author}
  {\bibfnamefont {B.}~\bibnamefont {{Brown}}}, \bibinfo {author} {\bibfnamefont
  {B.}~\bibnamefont {{Burger}}}, \bibinfo {author} {\bibfnamefont
  {E.}~\bibnamefont {{Calabrese}}}, \bibinfo {author} {\bibfnamefont
  {J.}~\bibnamefont {{Chervenak}}}, \bibinfo {author} {\bibfnamefont
  {D.}~\bibnamefont {{Crichton}}}, \bibinfo {author} {\bibfnamefont
  {S.}~\bibnamefont {{Das}}}, \bibinfo {author} {\bibfnamefont {M.~J.}\
  \bibnamefont {{Devlin}}}, \bibinfo {author} {\bibfnamefont {S.~R.}\
  \bibnamefont {{Dicker}}}, \bibinfo {author} {\bibfnamefont {W.}~\bibnamefont
  {{Bertrand Doriese}}}, \bibinfo {author} {\bibfnamefont {J.}~\bibnamefont
  {{Dunkley}}}, \bibinfo {author} {\bibfnamefont {R.}~\bibnamefont
  {{D{\"u}nner}}}, \bibinfo {author} {\bibfnamefont {T.}~\bibnamefont
  {{Essinger-Hileman}}}, \bibinfo {author} {\bibfnamefont {D.}~\bibnamefont
  {{Faber}}}, \bibinfo {author} {\bibfnamefont {R.~P.}\ \bibnamefont
  {{Fisher}}}, \bibinfo {author} {\bibfnamefont {J.~W.}\ \bibnamefont
  {{Fowler}}}, \bibinfo {author} {\bibfnamefont {P.}~\bibnamefont
  {{Gallardo}}}, \bibinfo {author} {\bibfnamefont {M.~S.}\ \bibnamefont
  {{Gordon}}}, \bibinfo {author} {\bibfnamefont {M.~B.}\ \bibnamefont
  {{Gralla}}}, \bibinfo {author} {\bibfnamefont {A.}~\bibnamefont {{Hajian}}},
  \bibinfo {author} {\bibfnamefont {M.}~\bibnamefont {{Halpern}}}, \bibinfo
  {author} {\bibfnamefont {M.}~\bibnamefont {{Hasselfield}}}, \bibinfo {author}
  {\bibfnamefont {C.}~\bibnamefont {{Hern{\'a}ndez-Monteagudo}}}, \bibinfo
  {author} {\bibfnamefont {J.~C.}\ \bibnamefont {{Hill}}}, \bibinfo {author}
  {\bibfnamefont {G.~C.}\ \bibnamefont {{Hilton}}}, \bibinfo {author}
  {\bibfnamefont {M.}~\bibnamefont {{Hilton}}}, \bibinfo {author}
  {\bibfnamefont {A.~D.}\ \bibnamefont {{Hincks}}}, \bibinfo {author}
  {\bibfnamefont {D.}~\bibnamefont {{Holtz}}}, \bibinfo {author} {\bibfnamefont
  {K.~M.}\ \bibnamefont {{Huffenberger}}}, \bibinfo {author} {\bibfnamefont
  {D.~H.}\ \bibnamefont {{Hughes}}}, \bibinfo {author} {\bibfnamefont {J.~P.}\
  \bibnamefont {{Hughes}}}, \bibinfo {author} {\bibfnamefont {L.}~\bibnamefont
  {{Infante}}}, \bibinfo {author} {\bibfnamefont {K.~D.}\ \bibnamefont
  {{Irwin}}}, \bibinfo {author} {\bibfnamefont {D.~R.}\ \bibnamefont
  {{Jacobson}}}, \bibinfo {author} {\bibfnamefont {B.}~\bibnamefont
  {{Johnstone}}}, \bibinfo {author} {\bibfnamefont {J.}~\bibnamefont {{Baptiste
  Juin}}}, \bibinfo {author} {\bibfnamefont {M.}~\bibnamefont {{Kaul}}},
  \bibinfo {author} {\bibfnamefont {J.}~\bibnamefont {{Klein}}}, \bibinfo
  {author} {\bibfnamefont {A.}~\bibnamefont {{Kosowsky}}}, \bibinfo {author}
  {\bibfnamefont {J.~M.}\ \bibnamefont {{Lau}}}, \bibinfo {author}
  {\bibfnamefont {M.}~\bibnamefont {{Limon}}}, \bibinfo {author} {\bibfnamefont
  {Y.-T.}\ \bibnamefont {{Lin}}}, \bibinfo {author} {\bibfnamefont
  {T.}~\bibnamefont {{Louis}}}, \bibinfo {author} {\bibfnamefont {R.~H.}\
  \bibnamefont {{Lupton}}}, \bibinfo {author} {\bibfnamefont {T.~A.}\
  \bibnamefont {{Marriage}}}, \bibinfo {author} {\bibfnamefont
  {D.}~\bibnamefont {{Marsden}}}, \bibinfo {author} {\bibfnamefont
  {K.}~\bibnamefont {{Martocci}}}, \bibinfo {author} {\bibfnamefont
  {P.}~\bibnamefont {{Mauskopf}}}, \bibinfo {author} {\bibfnamefont
  {M.}~\bibnamefont {{McLaren}}}, \bibinfo {author} {\bibfnamefont
  {F.}~\bibnamefont {{Menanteau}}}, \bibinfo {author} {\bibfnamefont
  {K.}~\bibnamefont {{Moodley}}}, \bibinfo {author} {\bibfnamefont
  {H.}~\bibnamefont {{Moseley}}}, \bibinfo {author} {\bibfnamefont {C.~B.}\
  \bibnamefont {{Netterfield}}}, \bibinfo {author} {\bibfnamefont {M.~D.}\
  \bibnamefont {{Niemack}}}, \bibinfo {author} {\bibfnamefont {L.~A.}\
  \bibnamefont {{Page}}}, \bibinfo {author} {\bibfnamefont {W.~A.}\
  \bibnamefont {{Page}}}, \bibinfo {author} {\bibfnamefont {L.}~\bibnamefont
  {{Parker}}}, \bibinfo {author} {\bibfnamefont {B.}~\bibnamefont
  {{Partridge}}}, \bibinfo {author} {\bibfnamefont {R.}~\bibnamefont
  {{Plimpton}}}, \bibinfo {author} {\bibfnamefont {H.}~\bibnamefont
  {{Quintana}}}, \bibinfo {author} {\bibfnamefont {E.~D.}\ \bibnamefont
  {{Reese}}}, \bibinfo {author} {\bibfnamefont {B.}~\bibnamefont {{Reid}}},
  \bibinfo {author} {\bibfnamefont {F.}~\bibnamefont {{Rojas}}}, \bibinfo
  {author} {\bibfnamefont {N.}~\bibnamefont {{Sehgal}}}, \bibinfo {author}
  {\bibfnamefont {B.~D.}\ \bibnamefont {{Sherwin}}}, \bibinfo {author}
  {\bibfnamefont {B.~L.}\ \bibnamefont {{Schmitt}}}, \bibinfo {author}
  {\bibfnamefont {D.~N.}\ \bibnamefont {{Spergel}}}, \bibinfo {author}
  {\bibfnamefont {S.~T.}\ \bibnamefont {{Staggs}}}, \bibinfo {author}
  {\bibfnamefont {O.}~\bibnamefont {{Stryzak}}}, \bibinfo {author}
  {\bibfnamefont {D.~S.}\ \bibnamefont {{Swetz}}}, \bibinfo {author}
  {\bibfnamefont {E.~R.}\ \bibnamefont {{Switzer}}}, \bibinfo {author}
  {\bibfnamefont {R.}~\bibnamefont {{Thornton}}}, \bibinfo {author}
  {\bibfnamefont {H.}~\bibnamefont {{Trac}}}, \bibinfo {author} {\bibfnamefont
  {C.}~\bibnamefont {{Tucker}}}, \bibinfo {author} {\bibfnamefont
  {M.}~\bibnamefont {{Uehara}}}, \bibinfo {author} {\bibfnamefont
  {K.}~\bibnamefont {{Visnjic}}}, \bibinfo {author} {\bibfnamefont
  {R.}~\bibnamefont {{Warne}}}, \bibinfo {author} {\bibfnamefont
  {G.}~\bibnamefont {{Wilson}}}, \bibinfo {author} {\bibfnamefont
  {E.}~\bibnamefont {{Wollack}}}, \bibinfo {author} {\bibfnamefont
  {Y.}~\bibnamefont {{Zhao}}}, \ and\ \bibinfo {author} {\bibfnamefont
  {C.}~\bibnamefont {{Zunckel}}},\ }\href {\doibase
  10.1088/1475-7516/2013/10/060} {\bibfield  {journal} {\bibinfo  {journal}
  {Journal of Cosmology and Astroparticle Physics}\ }\textbf {\bibinfo {volume}
  {10}},\ \bibinfo {eid} {060} (\bibinfo {year} {2013})},\ \Eprint
  {http://arxiv.org/abs/1301.0824} {arXiv:1301.0824 [astro-ph.CO]} \BibitemShut
  {NoStop}%
\bibitem [{\citenamefont {{Coleman}}(1977{\natexlab{a}})}]{Coleman:1977lr}%
  \BibitemOpen
  \bibfield  {author} {\bibinfo {author} {\bibfnamefont {S.}~\bibnamefont
  {{Coleman}}},\ }\href {\doibase 10.1103/PhysRevD.15.2929} {\bibfield
  {journal} {\bibinfo  {journal} {\prd}\ }\textbf {\bibinfo {volume} {15}},\
  \bibinfo {pages} {2929} (\bibinfo {year} {1977}{\natexlab{a}})}\BibitemShut
  {NoStop}%
\bibitem [{\citenamefont {{Coleman}}(1977{\natexlab{b}})}]{Coleman:1977qy}%
  \BibitemOpen
  \bibfield  {author} {\bibinfo {author} {\bibfnamefont {S.}~\bibnamefont
  {{Coleman}}},\ }\href {\doibase 10.1103/PhysRevD.16.1248} {\bibfield
  {journal} {\bibinfo  {journal} {\prd}\ }\textbf {\bibinfo {volume} {16}},\
  \bibinfo {pages} {1248} (\bibinfo {year} {1977}{\natexlab{b}})}\BibitemShut
  {NoStop}%
\bibitem [{\citenamefont {{Callan}}\ and\ \citenamefont
  {{Coleman}}(1977)}]{Callan:1977fk}%
  \BibitemOpen
  \bibfield  {author} {\bibinfo {author} {\bibfnamefont {C.~G.}\ \bibnamefont
  {{Callan}}, \bibfnamefont {Jr.}}\ and\ \bibinfo {author} {\bibfnamefont
  {S.}~\bibnamefont {{Coleman}}},\ }\href {\doibase 10.1103/PhysRevD.16.1762}
  {\bibfield  {journal} {\bibinfo  {journal} {\prd}\ }\textbf {\bibinfo
  {volume} {16}},\ \bibinfo {pages} {1762} (\bibinfo {year}
  {1977})}\BibitemShut {NoStop}%
\bibitem [{\citenamefont {{Ignatius}}\ \emph {et~al.}(1994)\citenamefont
  {{Ignatius}}, \citenamefont {{Kajantie}}, \citenamefont {{Kurki-Suonio}},\
  and\ \citenamefont {{Laine}}}]{Ignatius:1994uq}%
  \BibitemOpen
  \bibfield  {author} {\bibinfo {author} {\bibfnamefont {J.}~\bibnamefont
  {{Ignatius}}}, \bibinfo {author} {\bibfnamefont {K.}~\bibnamefont
  {{Kajantie}}}, \bibinfo {author} {\bibfnamefont {H.}~\bibnamefont
  {{Kurki-Suonio}}}, \ and\ \bibinfo {author} {\bibfnamefont {M.}~\bibnamefont
  {{Laine}}},\ }\href {\doibase 10.1103/PhysRevD.49.3854} {\bibfield  {journal}
  {\bibinfo  {journal} {\prd}\ }\textbf {\bibinfo {volume} {49}},\ \bibinfo
  {pages} {3854} (\bibinfo {year} {1994})},\ \Eprint
  {http://arxiv.org/abs/arXiv:astro-ph/9309059} {arXiv:astro-ph/9309059}
  \BibitemShut {NoStop}%
\bibitem [{\citenamefont {{Giblin}}\ \emph {et~al.}(2010)\citenamefont
  {{Giblin}}, \citenamefont {{Hui}}, \citenamefont {{Lim}},\ and\ \citenamefont
  {{Yang}}}]{Giblin:2010lr}%
  \BibitemOpen
  \bibfield  {author} {\bibinfo {author} {\bibfnamefont {J.~T.}\ \bibnamefont
  {{Giblin}}, \bibfnamefont {Jr.}}, \bibinfo {author} {\bibfnamefont
  {L.}~\bibnamefont {{Hui}}}, \bibinfo {author} {\bibfnamefont {E.~A.}\
  \bibnamefont {{Lim}}}, \ and\ \bibinfo {author} {\bibfnamefont {I.-S.}\
  \bibnamefont {{Yang}}},\ }\href {\doibase 10.1103/PhysRevD.82.045019}
  {\bibfield  {journal} {\bibinfo  {journal} {\prd}\ }\textbf {\bibinfo
  {volume} {82}},\ \bibinfo {eid} {045019} (\bibinfo {year} {2010})},\ \Eprint
  {http://arxiv.org/abs/1005.3493} {arXiv:1005.3493 [hep-th]} \BibitemShut
  {NoStop}%
\bibitem [{\citenamefont {{Johnson}}\ and\ \citenamefont
  {{Yang}}(2010)}]{Johnson:2010fk}%
  \BibitemOpen
  \bibfield  {author} {\bibinfo {author} {\bibfnamefont {M.~C.}\ \bibnamefont
  {{Johnson}}}\ and\ \bibinfo {author} {\bibfnamefont {I.-S.}\ \bibnamefont
  {{Yang}}},\ }\href {\doibase 10.1103/PhysRevD.82.065023} {\bibfield
  {journal} {\bibinfo  {journal} {\prd}\ }\textbf {\bibinfo {volume} {82}},\
  \bibinfo {eid} {065023} (\bibinfo {year} {2010})},\ \Eprint
  {http://arxiv.org/abs/1005.3506} {arXiv:1005.3506 [hep-th]} \BibitemShut
  {NoStop}%
\bibitem [{\citenamefont {{Giblin}}\ and\ \citenamefont
  {{Mertens}}(2013)}]{Giblin:2013uq}%
  \BibitemOpen
  \bibfield  {author} {\bibinfo {author} {\bibfnamefont {J.~T.}\ \bibnamefont
  {{Giblin}}}\ and\ \bibinfo {author} {\bibfnamefont {J.~B.}\ \bibnamefont
  {{Mertens}}},\ }\href {\doibase 10.1007/JHEP12(2013)042} {\bibfield
  {journal} {\bibinfo  {journal} {Journal of High Energy Physics}\ }\textbf
  {\bibinfo {volume} {12}},\ \bibinfo {pages} {042} (\bibinfo {year} {2013})},\
  \Eprint {http://arxiv.org/abs/1310.2948} {arXiv:1310.2948 [hep-th]}
  \BibitemShut {NoStop}%
\bibitem [{\citenamefont {{Hogan}}(1983)}]{Hogan:1983fk}%
  \BibitemOpen
  \bibfield  {author} {\bibinfo {author} {\bibfnamefont {C.~J.}\ \bibnamefont
  {{Hogan}}},\ }\href {\doibase 10.1016/0370-2693(83)90553-1} {\bibfield
  {journal} {\bibinfo  {journal} {Physics Letters B}\ }\textbf {\bibinfo
  {volume} {133}},\ \bibinfo {pages} {172} (\bibinfo {year}
  {1983})}\BibitemShut {NoStop}%
\bibitem [{\citenamefont {{Doran}}\ and\ \citenamefont
  {{Robbers}}(2006)}]{Doran:2006mz}%
  \BibitemOpen
  \bibfield  {author} {\bibinfo {author} {\bibfnamefont {M.}~\bibnamefont
  {{Doran}}}\ and\ \bibinfo {author} {\bibfnamefont {G.}~\bibnamefont
  {{Robbers}}},\ }\href {\doibase 10.1088/1475-7516/2006/06/026} {\bibfield
  {journal} {\bibinfo  {journal} {Journal of Cosmology and Astroparticle
  Physics}\ }\textbf {\bibinfo {volume} {06}},\ \bibinfo {eid} {026} (\bibinfo
  {year} {2006})},\ \Eprint {http://arxiv.org/abs/astro-ph/0601544}
  {astro-ph/0601544} \BibitemShut {NoStop}%
\bibitem [{\citenamefont {{Calabrese}}\ \emph
  {et~al.}(2011{\natexlab{a}})\citenamefont {{Calabrese}}, \citenamefont {{de
  Putter}}, \citenamefont {{Huterer}}, \citenamefont {{Linder}},\ and\
  \citenamefont {{Melchiorri}}}]{Calabrese:2011gf}%
  \BibitemOpen
  \bibfield  {author} {\bibinfo {author} {\bibfnamefont {E.}~\bibnamefont
  {{Calabrese}}}, \bibinfo {author} {\bibfnamefont {R.}~\bibnamefont {{de
  Putter}}}, \bibinfo {author} {\bibfnamefont {D.}~\bibnamefont {{Huterer}}},
  \bibinfo {author} {\bibfnamefont {E.~V.}\ \bibnamefont {{Linder}}}, \ and\
  \bibinfo {author} {\bibfnamefont {A.}~\bibnamefont {{Melchiorri}}},\ }\href
  {\doibase 10.1103/PhysRevD.83.023011} {\bibfield  {journal} {\bibinfo
  {journal} {\prd}\ }\textbf {\bibinfo {volume} {83}},\ \bibinfo {eid} {023011}
  (\bibinfo {year} {2011}{\natexlab{a}})},\ \Eprint
  {http://arxiv.org/abs/1010.5612} {arXiv:1010.5612 [astro-ph.CO]} \BibitemShut
  {NoStop}%
\bibitem [{\citenamefont {{Calabrese}}\ \emph
  {et~al.}(2011{\natexlab{b}})\citenamefont {{Calabrese}}, \citenamefont
  {{Huterer}}, \citenamefont {{Linder}}, \citenamefont {{Melchiorri}},\ and\
  \citenamefont {{Pagano}}}]{Calabrese:2011ly}%
  \BibitemOpen
  \bibfield  {author} {\bibinfo {author} {\bibfnamefont {E.}~\bibnamefont
  {{Calabrese}}}, \bibinfo {author} {\bibfnamefont {D.}~\bibnamefont
  {{Huterer}}}, \bibinfo {author} {\bibfnamefont {E.~V.}\ \bibnamefont
  {{Linder}}}, \bibinfo {author} {\bibfnamefont {A.}~\bibnamefont
  {{Melchiorri}}}, \ and\ \bibinfo {author} {\bibfnamefont {L.}~\bibnamefont
  {{Pagano}}},\ }\href {\doibase 10.1103/PhysRevD.83.123504} {\bibfield
  {journal} {\bibinfo  {journal} {\prd}\ }\textbf {\bibinfo {volume} {83}},\
  \bibinfo {eid} {123504} (\bibinfo {year} {2011}{\natexlab{b}})},\ \Eprint
  {http://arxiv.org/abs/1103.4132} {arXiv:1103.4132 [astro-ph.CO]} \BibitemShut
  {NoStop}%
\bibitem [{\citenamefont {{Reichardt}}\ \emph {et~al.}(2012)\citenamefont
  {{Reichardt}}, \citenamefont {{de Putter}}, \citenamefont {{Zahn}},\ and\
  \citenamefont {{Hou}}}]{Reichardt:2012ys}%
  \BibitemOpen
  \bibfield  {author} {\bibinfo {author} {\bibfnamefont {C.~L.}\ \bibnamefont
  {{Reichardt}}}, \bibinfo {author} {\bibfnamefont {R.}~\bibnamefont {{de
  Putter}}}, \bibinfo {author} {\bibfnamefont {O.}~\bibnamefont {{Zahn}}}, \
  and\ \bibinfo {author} {\bibfnamefont {Z.}~\bibnamefont {{Hou}}},\ }\href
  {\doibase 10.1088/2041-8205/749/1/L9} {\bibfield  {journal} {\bibinfo
  {journal} {Astrophys.~J.~(Letters)}\ }\textbf {\bibinfo {volume} {749}},\
  \bibinfo {eid} {L9} (\bibinfo {year} {2012})},\ \Eprint
  {http://arxiv.org/abs/1110.5328} {arXiv:1110.5328 [astro-ph.CO]} \BibitemShut
  {NoStop}%
\bibitem [{\citenamefont {{Pettorino}}\ \emph {et~al.}(2013)\citenamefont
  {{Pettorino}}, \citenamefont {{Amendola}},\ and\ \citenamefont
  {{Wetterich}}}]{Pettorino:2013ve}%
  \BibitemOpen
  \bibfield  {author} {\bibinfo {author} {\bibfnamefont {V.}~\bibnamefont
  {{Pettorino}}}, \bibinfo {author} {\bibfnamefont {L.}~\bibnamefont
  {{Amendola}}}, \ and\ \bibinfo {author} {\bibfnamefont {C.}~\bibnamefont
  {{Wetterich}}},\ }\href {\doibase 10.1103/PhysRevD.87.083009} {\bibfield
  {journal} {\bibinfo  {journal} {\prd}\ }\textbf {\bibinfo {volume} {87}},\
  \bibinfo {eid} {083009} (\bibinfo {year} {2013})},\ \Eprint
  {http://arxiv.org/abs/1301.5279} {arXiv:1301.5279 [astro-ph.CO]} \BibitemShut
  {NoStop}%
\bibitem [{\citenamefont {{Bielefeld}}\ \emph {et~al.}(2013)\citenamefont
  {{Bielefeld}}, \citenamefont {{Wu}}, \citenamefont {{Caldwell}},\ and\
  \citenamefont {{Dor{\'e}}}}]{Bielefeld:2013yq}%
  \BibitemOpen
  \bibfield  {author} {\bibinfo {author} {\bibfnamefont {J.}~\bibnamefont
  {{Bielefeld}}}, \bibinfo {author} {\bibfnamefont {W.~L.~K.}\ \bibnamefont
  {{Wu}}}, \bibinfo {author} {\bibfnamefont {R.~R.}\ \bibnamefont
  {{Caldwell}}}, \ and\ \bibinfo {author} {\bibfnamefont {O.}~\bibnamefont
  {{Dor{\'e}}}},\ }\href {\doibase 10.1103/PhysRevD.88.103004} {\bibfield
  {journal} {\bibinfo  {journal} {\prd}\ }\textbf {\bibinfo {volume} {88}},\
  \bibinfo {eid} {103004} (\bibinfo {year} {2013})},\ \Eprint
  {http://arxiv.org/abs/1305.2209} {arXiv:1305.2209 [astro-ph.CO]} \BibitemShut
  {NoStop}%
\bibitem [{\citenamefont {{Fixsen}}(2009)}]{Fixsen:2009qy}%
  \BibitemOpen
  \bibfield  {author} {\bibinfo {author} {\bibfnamefont {D.~J.}\ \bibnamefont
  {{Fixsen}}},\ }\href {\doibase 10.1088/0004-637X/707/2/916} {\bibfield
  {journal} {\bibinfo  {journal} {\apj}\ }\textbf {\bibinfo {volume} {707}},\
  \bibinfo {pages} {916} (\bibinfo {year} {2009})},\ \Eprint
  {http://arxiv.org/abs/0911.1955} {arXiv:0911.1955 [astro-ph.CO]} \BibitemShut
  {NoStop}%
\bibitem [{\citenamefont {{Wyman}}\ \emph {et~al.}(2014)\citenamefont
  {{Wyman}}, \citenamefont {{Rudd}}, \citenamefont {{Vanderveld}},\ and\
  \citenamefont {{Hu}}}]{Wyman:2013mz}%
  \BibitemOpen
  \bibfield  {author} {\bibinfo {author} {\bibfnamefont {M.}~\bibnamefont
  {{Wyman}}}, \bibinfo {author} {\bibfnamefont {D.~H.}\ \bibnamefont {{Rudd}}},
  \bibinfo {author} {\bibfnamefont {R.~A.}\ \bibnamefont {{Vanderveld}}}, \
  and\ \bibinfo {author} {\bibfnamefont {W.}~\bibnamefont {{Hu}}},\ }\href
  {\doibase 10.1103/PhysRevLett.112.051302} {\bibfield  {journal} {\bibinfo
  {journal} {Physical Review Letters}\ }\textbf {\bibinfo {volume} {112}},\
  \bibinfo {eid} {051302} (\bibinfo {year} {2014})},\ \Eprint
  {http://arxiv.org/abs/1307.7715} {arXiv:1307.7715 [astro-ph.CO]} \BibitemShut
  {NoStop}%
\bibitem [{\citenamefont {{Calabrese}}\ \emph {et~al.}(2013)\citenamefont
  {{Calabrese}}, \citenamefont {{Hlozek}}, \citenamefont {{Battaglia}},
  \citenamefont {{Battistelli}}, \citenamefont {{Bond}}, \citenamefont
  {{Chluba}}, \citenamefont {{Crichton}}, \citenamefont {{Das}}, \citenamefont
  {{Devlin}}, \citenamefont {{Dunkley}}, \citenamefont {{D{\"u}nner}},
  \citenamefont {{Farhang}}, \citenamefont {{Gralla}}, \citenamefont
  {{Hajian}}, \citenamefont {{Halpern}}, \citenamefont {{Hasselfield}},
  \citenamefont {{Hincks}}, \citenamefont {{Irwin}}, \citenamefont
  {{Kosowsky}}, \citenamefont {{Louis}}, \citenamefont {{Marriage}},
  \citenamefont {{Moodley}}, \citenamefont {{Newburgh}}, \citenamefont
  {{Niemack}}, \citenamefont {{Nolta}}, \citenamefont {{Page}}, \citenamefont
  {{Sehgal}}, \citenamefont {{Sherwin}}, \citenamefont {{Sievers}},
  \citenamefont {{Sif{\'o}n}}, \citenamefont {{Spergel}}, \citenamefont
  {{Staggs}}, \citenamefont {{Switzer}},\ and\ \citenamefont
  {{Wollack}}}]{Calabrese:2013fr}%
  \BibitemOpen
  \bibfield  {author} {\bibinfo {author} {\bibfnamefont {E.}~\bibnamefont
  {{Calabrese}}}, \bibinfo {author} {\bibfnamefont {R.~A.}\ \bibnamefont
  {{Hlozek}}}, \bibinfo {author} {\bibfnamefont {N.}~\bibnamefont
  {{Battaglia}}}, \bibinfo {author} {\bibfnamefont {E.~S.}\ \bibnamefont
  {{Battistelli}}}, \bibinfo {author} {\bibfnamefont {J.~R.}\ \bibnamefont
  {{Bond}}}, \bibinfo {author} {\bibfnamefont {J.}~\bibnamefont {{Chluba}}},
  \bibinfo {author} {\bibfnamefont {D.}~\bibnamefont {{Crichton}}}, \bibinfo
  {author} {\bibfnamefont {S.}~\bibnamefont {{Das}}}, \bibinfo {author}
  {\bibfnamefont {M.~J.}\ \bibnamefont {{Devlin}}}, \bibinfo {author}
  {\bibfnamefont {J.}~\bibnamefont {{Dunkley}}}, \bibinfo {author}
  {\bibfnamefont {R.}~\bibnamefont {{D{\"u}nner}}}, \bibinfo {author}
  {\bibfnamefont {M.}~\bibnamefont {{Farhang}}}, \bibinfo {author}
  {\bibfnamefont {M.~B.}\ \bibnamefont {{Gralla}}}, \bibinfo {author}
  {\bibfnamefont {A.}~\bibnamefont {{Hajian}}}, \bibinfo {author}
  {\bibfnamefont {M.}~\bibnamefont {{Halpern}}}, \bibinfo {author}
  {\bibfnamefont {M.}~\bibnamefont {{Hasselfield}}}, \bibinfo {author}
  {\bibfnamefont {A.~D.}\ \bibnamefont {{Hincks}}}, \bibinfo {author}
  {\bibfnamefont {K.~D.}\ \bibnamefont {{Irwin}}}, \bibinfo {author}
  {\bibfnamefont {A.}~\bibnamefont {{Kosowsky}}}, \bibinfo {author}
  {\bibfnamefont {T.}~\bibnamefont {{Louis}}}, \bibinfo {author} {\bibfnamefont
  {T.~A.}\ \bibnamefont {{Marriage}}}, \bibinfo {author} {\bibfnamefont
  {K.}~\bibnamefont {{Moodley}}}, \bibinfo {author} {\bibfnamefont
  {L.}~\bibnamefont {{Newburgh}}}, \bibinfo {author} {\bibfnamefont {M.~D.}\
  \bibnamefont {{Niemack}}}, \bibinfo {author} {\bibfnamefont {M.~R.}\
  \bibnamefont {{Nolta}}}, \bibinfo {author} {\bibfnamefont {L.~A.}\
  \bibnamefont {{Page}}}, \bibinfo {author} {\bibfnamefont {N.}~\bibnamefont
  {{Sehgal}}}, \bibinfo {author} {\bibfnamefont {B.~D.}\ \bibnamefont
  {{Sherwin}}}, \bibinfo {author} {\bibfnamefont {J.~L.}\ \bibnamefont
  {{Sievers}}}, \bibinfo {author} {\bibfnamefont {C.}~\bibnamefont
  {{Sif{\'o}n}}}, \bibinfo {author} {\bibfnamefont {D.~N.}\ \bibnamefont
  {{Spergel}}}, \bibinfo {author} {\bibfnamefont {S.~T.}\ \bibnamefont
  {{Staggs}}}, \bibinfo {author} {\bibfnamefont {E.~R.}\ \bibnamefont
  {{Switzer}}}, \ and\ \bibinfo {author} {\bibfnamefont {E.~J.}\ \bibnamefont
  {{Wollack}}},\ }\href {\doibase 10.1103/PhysRevD.87.103012} {\bibfield
  {journal} {\bibinfo  {journal} {\prd}\ }\textbf {\bibinfo {volume} {87}},\
  \bibinfo {eid} {103012} (\bibinfo {year} {2013})},\ \Eprint
  {http://arxiv.org/abs/1302.1841} {arXiv:1302.1841 [astro-ph.CO]} \BibitemShut
  {NoStop}%
\bibitem [{\citenamefont {{Riess}}\ \emph
  {et~al.}(2011{\natexlab{a}})\citenamefont {{Riess}}, \citenamefont {{Macri}},
  \citenamefont {{Casertano}}, \citenamefont {{Lampeitl}}, \citenamefont
  {{Ferguson}}, \citenamefont {{Filippenko}}, \citenamefont {{Jha}},
  \citenamefont {{Li}},\ and\ \citenamefont {{Chornock}}}]{Riess:2011lr}%
  \BibitemOpen
  \bibfield  {author} {\bibinfo {author} {\bibfnamefont {A.~G.}\ \bibnamefont
  {{Riess}}}, \bibinfo {author} {\bibfnamefont {L.}~\bibnamefont {{Macri}}},
  \bibinfo {author} {\bibfnamefont {S.}~\bibnamefont {{Casertano}}}, \bibinfo
  {author} {\bibfnamefont {H.}~\bibnamefont {{Lampeitl}}}, \bibinfo {author}
  {\bibfnamefont {H.~C.}\ \bibnamefont {{Ferguson}}}, \bibinfo {author}
  {\bibfnamefont {A.~V.}\ \bibnamefont {{Filippenko}}}, \bibinfo {author}
  {\bibfnamefont {S.~W.}\ \bibnamefont {{Jha}}}, \bibinfo {author}
  {\bibfnamefont {W.}~\bibnamefont {{Li}}}, \ and\ \bibinfo {author}
  {\bibfnamefont {R.}~\bibnamefont {{Chornock}}},\ }\href {\doibase
  10.1088/0004-637X/730/2/119} {\bibfield  {journal} {\bibinfo  {journal}
  {\apj}\ }\textbf {\bibinfo {volume} {730}},\ \bibinfo {eid} {119} (\bibinfo
  {year} {2011}{\natexlab{a}})},\ \Eprint {http://arxiv.org/abs/1103.2976}
  {arXiv:1103.2976 [astro-ph.CO]} \BibitemShut {NoStop}%
\bibitem [{\citenamefont {{Riess}}\ \emph
  {et~al.}(2011{\natexlab{b}})\citenamefont {{Riess}}, \citenamefont {{Macri}},
  \citenamefont {{Casertano}}, \citenamefont {{Lampeit}}, \citenamefont
  {{Ferguson}}, \citenamefont {{Filippenko}}, \citenamefont {{Jha}},
  \citenamefont {{Li}}, \citenamefont {{Chornock}},\ and\ \citenamefont
  {{Silverman}}}]{Riess:2011lrE}%
  \BibitemOpen
  \bibfield  {author} {\bibinfo {author} {\bibfnamefont {A.~G.}\ \bibnamefont
  {{Riess}}}, \bibinfo {author} {\bibfnamefont {L.}~\bibnamefont {{Macri}}},
  \bibinfo {author} {\bibfnamefont {S.}~\bibnamefont {{Casertano}}}, \bibinfo
  {author} {\bibfnamefont {H.}~\bibnamefont {{Lampeit}}}, \bibinfo {author}
  {\bibfnamefont {H.~C.}\ \bibnamefont {{Ferguson}}}, \bibinfo {author}
  {\bibfnamefont {A.~V.}\ \bibnamefont {{Filippenko}}}, \bibinfo {author}
  {\bibfnamefont {S.~W.}\ \bibnamefont {{Jha}}}, \bibinfo {author}
  {\bibfnamefont {W.}~\bibnamefont {{Li}}}, \bibinfo {author} {\bibfnamefont
  {R.}~\bibnamefont {{Chornock}}}, \ and\ \bibinfo {author} {\bibfnamefont
  {J.~M.}\ \bibnamefont {{Silverman}}},\ }\href {\doibase
  10.1088/0004-637X/732/2/129} {\bibfield  {journal} {\bibinfo  {journal}
  {\apj}\ }\textbf {\bibinfo {volume} {732}},\ \bibinfo {eid} {129(E)}
  (\bibinfo {year} {2011}{\natexlab{b}})}\BibitemShut {NoStop}%
\bibitem [{\citenamefont {{Freedman}}\ \emph {et~al.}(2012)\citenamefont
  {{Freedman}}, \citenamefont {{Madore}}, \citenamefont {{Scowcroft}},
  \citenamefont {{Burns}}, \citenamefont {{Monson}}, \citenamefont {{Persson}},
  \citenamefont {{Seibert}},\ and\ \citenamefont {{Rigby}}}]{Freedman:2012fk}%
  \BibitemOpen
  \bibfield  {author} {\bibinfo {author} {\bibfnamefont {W.~L.}\ \bibnamefont
  {{Freedman}}}, \bibinfo {author} {\bibfnamefont {B.~F.}\ \bibnamefont
  {{Madore}}}, \bibinfo {author} {\bibfnamefont {V.}~\bibnamefont
  {{Scowcroft}}}, \bibinfo {author} {\bibfnamefont {C.}~\bibnamefont
  {{Burns}}}, \bibinfo {author} {\bibfnamefont {A.}~\bibnamefont {{Monson}}},
  \bibinfo {author} {\bibfnamefont {S.~E.}\ \bibnamefont {{Persson}}}, \bibinfo
  {author} {\bibfnamefont {M.}~\bibnamefont {{Seibert}}}, \ and\ \bibinfo
  {author} {\bibfnamefont {J.}~\bibnamefont {{Rigby}}},\ }\href {\doibase
  10.1088/0004-637X/758/1/24} {\bibfield  {journal} {\bibinfo  {journal}
  {\apj}\ }\textbf {\bibinfo {volume} {758}},\ \bibinfo {eid} {24} (\bibinfo
  {year} {2012})},\ \Eprint {http://arxiv.org/abs/1208.3281} {arXiv:1208.3281
  [astro-ph.CO]} \BibitemShut {NoStop}%
\bibitem [{\citenamefont {{Font-Ribera}}\ \emph {et~al.}(2014)\citenamefont
  {{Font-Ribera}}, \citenamefont {{Kirkby}}, \citenamefont {{Busca}},
  \citenamefont {{Miralda-Escud{\'e}}}, \citenamefont {{Ross}}, \citenamefont
  {{Slosar}}, \citenamefont {{Rich}}, \citenamefont {{Aubourg}}, \citenamefont
  {{Bailey}}, \citenamefont {{Bhardwaj}}, \citenamefont {{Bautista}},
  \citenamefont {{Beutler}}, \citenamefont {{Bizyaev}}, \citenamefont
  {{Blomqvist}}, \citenamefont {{Brewington}}, \citenamefont {{Brinkmann}},
  \citenamefont {{Brownstein}}, \citenamefont {{Carithers}}, \citenamefont
  {{Dawson}}, \citenamefont {{Delubac}}, \citenamefont {{Ebelke}},
  \citenamefont {{Eisenstein}}, \citenamefont {{Ge}}, \citenamefont
  {{Kinemuchi}}, \citenamefont {{Lee}}, \citenamefont {{Malanushenko}},
  \citenamefont {{Malanushenko}}, \citenamefont {{Marchante}}, \citenamefont
  {{Margala}}, \citenamefont {{Muna}}, \citenamefont {{Myers}}, \citenamefont
  {{Noterdaeme}}, \citenamefont {{Oravetz}}, \citenamefont
  {{Palanque-Delabrouille}}, \citenamefont {{P{\^a}ris}}, \citenamefont
  {{Petitjean}}, \citenamefont {{Pieri}}, \citenamefont {{Rossi}},
  \citenamefont {{Schneider}}, \citenamefont {{Simmons}}, \citenamefont
  {{Viel}}, \citenamefont {{Yeche}},\ and\ \citenamefont
  {{York}}}]{Font-Ribera:2013uq}%
  \BibitemOpen
  \bibfield  {author} {\bibinfo {author} {\bibfnamefont {A.}~\bibnamefont
  {{Font-Ribera}}}, \bibinfo {author} {\bibfnamefont {D.}~\bibnamefont
  {{Kirkby}}}, \bibinfo {author} {\bibfnamefont {N.}~\bibnamefont {{Busca}}},
  \bibinfo {author} {\bibfnamefont {J.}~\bibnamefont {{Miralda-Escud{\'e}}}},
  \bibinfo {author} {\bibfnamefont {N.~P.}\ \bibnamefont {{Ross}}}, \bibinfo
  {author} {\bibfnamefont {A.}~\bibnamefont {{Slosar}}}, \bibinfo {author}
  {\bibfnamefont {J.}~\bibnamefont {{Rich}}}, \bibinfo {author} {\bibfnamefont
  {{\'E}.}~\bibnamefont {{Aubourg}}}, \bibinfo {author} {\bibfnamefont
  {S.}~\bibnamefont {{Bailey}}}, \bibinfo {author} {\bibfnamefont
  {V.}~\bibnamefont {{Bhardwaj}}}, \bibinfo {author} {\bibfnamefont
  {J.}~\bibnamefont {{Bautista}}}, \bibinfo {author} {\bibfnamefont
  {F.}~\bibnamefont {{Beutler}}}, \bibinfo {author} {\bibfnamefont
  {D.}~\bibnamefont {{Bizyaev}}}, \bibinfo {author} {\bibfnamefont
  {M.}~\bibnamefont {{Blomqvist}}}, \bibinfo {author} {\bibfnamefont
  {H.}~\bibnamefont {{Brewington}}}, \bibinfo {author} {\bibfnamefont
  {J.}~\bibnamefont {{Brinkmann}}}, \bibinfo {author} {\bibfnamefont {J.~R.}\
  \bibnamefont {{Brownstein}}}, \bibinfo {author} {\bibfnamefont
  {B.}~\bibnamefont {{Carithers}}}, \bibinfo {author} {\bibfnamefont {K.~S.}\
  \bibnamefont {{Dawson}}}, \bibinfo {author} {\bibfnamefont {T.}~\bibnamefont
  {{Delubac}}}, \bibinfo {author} {\bibfnamefont {G.}~\bibnamefont {{Ebelke}}},
  \bibinfo {author} {\bibfnamefont {D.~J.}\ \bibnamefont {{Eisenstein}}},
  \bibinfo {author} {\bibfnamefont {J.}~\bibnamefont {{Ge}}}, \bibinfo {author}
  {\bibfnamefont {K.}~\bibnamefont {{Kinemuchi}}}, \bibinfo {author}
  {\bibfnamefont {K.-G.}\ \bibnamefont {{Lee}}}, \bibinfo {author}
  {\bibfnamefont {V.}~\bibnamefont {{Malanushenko}}}, \bibinfo {author}
  {\bibfnamefont {E.}~\bibnamefont {{Malanushenko}}}, \bibinfo {author}
  {\bibfnamefont {M.}~\bibnamefont {{Marchante}}}, \bibinfo {author}
  {\bibfnamefont {D.}~\bibnamefont {{Margala}}}, \bibinfo {author}
  {\bibfnamefont {D.}~\bibnamefont {{Muna}}}, \bibinfo {author} {\bibfnamefont
  {A.~D.}\ \bibnamefont {{Myers}}}, \bibinfo {author} {\bibfnamefont
  {P.}~\bibnamefont {{Noterdaeme}}}, \bibinfo {author} {\bibfnamefont
  {D.}~\bibnamefont {{Oravetz}}}, \bibinfo {author} {\bibfnamefont
  {N.}~\bibnamefont {{Palanque-Delabrouille}}}, \bibinfo {author}
  {\bibfnamefont {I.}~\bibnamefont {{P{\^a}ris}}}, \bibinfo {author}
  {\bibfnamefont {P.}~\bibnamefont {{Petitjean}}}, \bibinfo {author}
  {\bibfnamefont {M.~M.}\ \bibnamefont {{Pieri}}}, \bibinfo {author}
  {\bibfnamefont {G.}~\bibnamefont {{Rossi}}}, \bibinfo {author} {\bibfnamefont
  {D.~P.}\ \bibnamefont {{Schneider}}}, \bibinfo {author} {\bibfnamefont
  {A.}~\bibnamefont {{Simmons}}}, \bibinfo {author} {\bibfnamefont
  {M.}~\bibnamefont {{Viel}}}, \bibinfo {author} {\bibfnamefont
  {C.}~\bibnamefont {{Yeche}}}, \ and\ \bibinfo {author} {\bibfnamefont
  {D.~G.}\ \bibnamefont {{York}}},\ }\href {\doibase
  10.1088/1475-7516/2014/05/027} {\bibfield  {journal} {\bibinfo  {journal}
  {Journal of Cosmology and Astroparticle Physics}\ }\textbf {\bibinfo {volume}
  {05}},\ \bibinfo {eid} {027} (\bibinfo {year} {2014})},\ \Eprint
  {http://arxiv.org/abs/1311.1767} {arXiv:1311.1767} \BibitemShut {NoStop}%
\bibitem [{\citenamefont {{Yuan}}\ \emph {et~al.}(2013)\citenamefont {{Yuan}},
  \citenamefont {{Liu}},\ and\ \citenamefont {{Zhang}}}]{Yuan:2013qy}%
  \BibitemOpen
  \bibfield  {author} {\bibinfo {author} {\bibfnamefont {S.}~\bibnamefont
  {{Yuan}}}, \bibinfo {author} {\bibfnamefont {S.}~\bibnamefont {{Liu}}}, \
  and\ \bibinfo {author} {\bibfnamefont {T.-J.}\ \bibnamefont {{Zhang}}},\
  }\href@noop {} {\bibfield  {journal} {\bibinfo  {journal} {ArXiv e-prints}\ }
  (\bibinfo {year} {2013})},\ \Eprint {http://arxiv.org/abs/1311.1583}
  {arXiv:1311.1583 [astro-ph.CO]} \BibitemShut {NoStop}%
\bibitem [{\citenamefont {{Cowan}}\ \emph {et~al.}(1999)\citenamefont
  {{Cowan}}, \citenamefont {{Pfeiffer}}, \citenamefont {{Kratz}}, \citenamefont
  {{Thielemann}}, \citenamefont {{Sneden}}, \citenamefont {{Burles}},
  \citenamefont {{Tytler}},\ and\ \citenamefont {{Beers}}}]{Cowan:1999kx}%
  \BibitemOpen
  \bibfield  {author} {\bibinfo {author} {\bibfnamefont {J.~J.}\ \bibnamefont
  {{Cowan}}}, \bibinfo {author} {\bibfnamefont {B.}~\bibnamefont {{Pfeiffer}}},
  \bibinfo {author} {\bibfnamefont {K.-L.}\ \bibnamefont {{Kratz}}}, \bibinfo
  {author} {\bibfnamefont {F.-K.}\ \bibnamefont {{Thielemann}}}, \bibinfo
  {author} {\bibfnamefont {C.}~\bibnamefont {{Sneden}}}, \bibinfo {author}
  {\bibfnamefont {S.}~\bibnamefont {{Burles}}}, \bibinfo {author}
  {\bibfnamefont {D.}~\bibnamefont {{Tytler}}}, \ and\ \bibinfo {author}
  {\bibfnamefont {T.~C.}\ \bibnamefont {{Beers}}},\ }\href {\doibase
  10.1086/307512} {\bibfield  {journal} {\bibinfo  {journal} {\apj}\ }\textbf
  {\bibinfo {volume} {521}},\ \bibinfo {pages} {194} (\bibinfo {year}
  {1999})},\ \Eprint {http://arxiv.org/abs/astro-ph/9808272} {astro-ph/9808272}
  \BibitemShut {NoStop}%
\bibitem [{\citenamefont {{Wanajo}}\ \emph {et~al.}(2002)\citenamefont
  {{Wanajo}}, \citenamefont {{Itoh}}, \citenamefont {{Ishimaru}}, \citenamefont
  {{Nozawa}},\ and\ \citenamefont {{Beers}}}]{Wanajo:2002yq}%
  \BibitemOpen
  \bibfield  {author} {\bibinfo {author} {\bibfnamefont {S.}~\bibnamefont
  {{Wanajo}}}, \bibinfo {author} {\bibfnamefont {N.}~\bibnamefont {{Itoh}}},
  \bibinfo {author} {\bibfnamefont {Y.}~\bibnamefont {{Ishimaru}}}, \bibinfo
  {author} {\bibfnamefont {S.}~\bibnamefont {{Nozawa}}}, \ and\ \bibinfo
  {author} {\bibfnamefont {T.~C.}\ \bibnamefont {{Beers}}},\ }\href {\doibase
  10.1086/342230} {\bibfield  {journal} {\bibinfo  {journal} {\apj}\ }\textbf
  {\bibinfo {volume} {577}},\ \bibinfo {pages} {853} (\bibinfo {year}
  {2002})},\ \Eprint {http://arxiv.org/abs/astro-ph/0206133} {astro-ph/0206133}
  \BibitemShut {NoStop}%
\bibitem [{\citenamefont {{Krauss}}\ and\ \citenamefont
  {{Chaboyer}}(2003)}]{Krauss:2003vn}%
  \BibitemOpen
  \bibfield  {author} {\bibinfo {author} {\bibfnamefont {L.~M.}\ \bibnamefont
  {{Krauss}}}\ and\ \bibinfo {author} {\bibfnamefont {B.}~\bibnamefont
  {{Chaboyer}}},\ }\href {\doibase 10.1126/science.1075631} {\bibfield
  {journal} {\bibinfo  {journal} {Science}\ }\textbf {\bibinfo {volume}
  {299}},\ \bibinfo {pages} {65} (\bibinfo {year} {2003})}\BibitemShut
  {NoStop}%
\bibitem [{\citenamefont {{Hansen}}\ \emph {et~al.}(2004)\citenamefont
  {{Hansen}}, \citenamefont {{Richer}}, \citenamefont {{Fahlman}},
  \citenamefont {{Stetson}}, \citenamefont {{Brewer}}, \citenamefont
  {{Currie}}, \citenamefont {{Gibson}}, \citenamefont {{Ibata}}, \citenamefont
  {{Rich}},\ and\ \citenamefont {{Shara}}}]{Hansen:2004rt}%
  \BibitemOpen
  \bibfield  {author} {\bibinfo {author} {\bibfnamefont {B.~M.~S.}\
  \bibnamefont {{Hansen}}}, \bibinfo {author} {\bibfnamefont {H.~B.}\
  \bibnamefont {{Richer}}}, \bibinfo {author} {\bibfnamefont {G.~G.}\
  \bibnamefont {{Fahlman}}}, \bibinfo {author} {\bibfnamefont {P.~B.}\
  \bibnamefont {{Stetson}}}, \bibinfo {author} {\bibfnamefont {J.}~\bibnamefont
  {{Brewer}}}, \bibinfo {author} {\bibfnamefont {T.}~\bibnamefont {{Currie}}},
  \bibinfo {author} {\bibfnamefont {B.~K.}\ \bibnamefont {{Gibson}}}, \bibinfo
  {author} {\bibfnamefont {R.}~\bibnamefont {{Ibata}}}, \bibinfo {author}
  {\bibfnamefont {R.~M.}\ \bibnamefont {{Rich}}}, \ and\ \bibinfo {author}
  {\bibfnamefont {M.~M.}\ \bibnamefont {{Shara}}},\ }\href {\doibase
  10.1086/424832} {\bibfield  {journal} {\bibinfo  {journal}
  {Astrophys.~J.~(Supplement)}\ }\textbf {\bibinfo {volume} {155}},\ \bibinfo
  {pages} {551} (\bibinfo {year} {2004})},\ \Eprint
  {http://arxiv.org/abs/astro-ph/0401443} {astro-ph/0401443} \BibitemShut
  {NoStop}%
\bibitem [{\citenamefont {{Hansen}}\ \emph {et~al.}(2007)\citenamefont
  {{Hansen}}, \citenamefont {{Anderson}}, \citenamefont {{Brewer}},
  \citenamefont {{Dotter}}, \citenamefont {{Fahlman}}, \citenamefont
  {{Hurley}}, \citenamefont {{Kalirai}}, \citenamefont {{King}}, \citenamefont
  {{Reitzel}}, \citenamefont {{Richer}}, \citenamefont {{Rich}}, \citenamefont
  {{Shara}},\ and\ \citenamefont {{Stetson}}}]{Hansen:2007ys}%
  \BibitemOpen
  \bibfield  {author} {\bibinfo {author} {\bibfnamefont {B.~M.~S.}\
  \bibnamefont {{Hansen}}}, \bibinfo {author} {\bibfnamefont {J.}~\bibnamefont
  {{Anderson}}}, \bibinfo {author} {\bibfnamefont {J.}~\bibnamefont
  {{Brewer}}}, \bibinfo {author} {\bibfnamefont {A.}~\bibnamefont {{Dotter}}},
  \bibinfo {author} {\bibfnamefont {G.~G.}\ \bibnamefont {{Fahlman}}}, \bibinfo
  {author} {\bibfnamefont {J.}~\bibnamefont {{Hurley}}}, \bibinfo {author}
  {\bibfnamefont {J.}~\bibnamefont {{Kalirai}}}, \bibinfo {author}
  {\bibfnamefont {I.}~\bibnamefont {{King}}}, \bibinfo {author} {\bibfnamefont
  {D.}~\bibnamefont {{Reitzel}}}, \bibinfo {author} {\bibfnamefont {H.~B.}\
  \bibnamefont {{Richer}}}, \bibinfo {author} {\bibfnamefont {R.~M.}\
  \bibnamefont {{Rich}}}, \bibinfo {author} {\bibfnamefont {M.~M.}\
  \bibnamefont {{Shara}}}, \ and\ \bibinfo {author} {\bibfnamefont {P.~B.}\
  \bibnamefont {{Stetson}}},\ }\href {\doibase 10.1086/522567} {\bibfield
  {journal} {\bibinfo  {journal} {\apj}\ }\textbf {\bibinfo {volume} {671}},\
  \bibinfo {pages} {380} (\bibinfo {year} {2007})},\ \Eprint
  {http://arxiv.org/abs/astro-ph/0701738} {astro-ph/0701738} \BibitemShut
  {NoStop}%
\bibitem [{\citenamefont {{Bond}}\ \emph {et~al.}(2013)\citenamefont {{Bond}},
  \citenamefont {{Nelan}}, \citenamefont {{VandenBerg}}, \citenamefont
  {{Schaefer}},\ and\ \citenamefont {{Harmer}}}]{Bond:2013lr}%
  \BibitemOpen
  \bibfield  {author} {\bibinfo {author} {\bibfnamefont {H.~E.}\ \bibnamefont
  {{Bond}}}, \bibinfo {author} {\bibfnamefont {E.~P.}\ \bibnamefont {{Nelan}}},
  \bibinfo {author} {\bibfnamefont {D.~A.}\ \bibnamefont {{VandenBerg}}},
  \bibinfo {author} {\bibfnamefont {G.~H.}\ \bibnamefont {{Schaefer}}}, \ and\
  \bibinfo {author} {\bibfnamefont {D.}~\bibnamefont {{Harmer}}},\ }\href
  {\doibase 10.1088/2041-8205/765/1/L12} {\bibfield  {journal} {\bibinfo
  {journal} {Astrophys. J. (Letters)}\ }\textbf {\bibinfo {volume} {765}},\
  \bibinfo {eid} {L12} (\bibinfo {year} {2013})},\ \Eprint
  {http://arxiv.org/abs/1302.3180} {arXiv:1302.3180 [astro-ph.SR]} \BibitemShut
  {NoStop}%
\bibitem [{\citenamefont {{Banday}}\ \emph {et~al.}(1997)\citenamefont
  {{Banday}}, \citenamefont {{Gorski}}, \citenamefont {{Bennett}},
  \citenamefont {{Hinshaw}}, \citenamefont {{Kogut}}, \citenamefont
  {{Lineweaver}}, \citenamefont {{Smoot}},\ and\ \citenamefont
  {{Tenorio}}}]{Banday:1997qy}%
  \BibitemOpen
  \bibfield  {author} {\bibinfo {author} {\bibfnamefont {A.~J.}\ \bibnamefont
  {{Banday}}}, \bibinfo {author} {\bibfnamefont {K.~M.}\ \bibnamefont
  {{Gorski}}}, \bibinfo {author} {\bibfnamefont {C.~L.}\ \bibnamefont
  {{Bennett}}}, \bibinfo {author} {\bibfnamefont {G.}~\bibnamefont
  {{Hinshaw}}}, \bibinfo {author} {\bibfnamefont {A.}~\bibnamefont {{Kogut}}},
  \bibinfo {author} {\bibfnamefont {C.}~\bibnamefont {{Lineweaver}}}, \bibinfo
  {author} {\bibfnamefont {G.~F.}\ \bibnamefont {{Smoot}}}, \ and\ \bibinfo
  {author} {\bibfnamefont {L.}~\bibnamefont {{Tenorio}}},\ }\href {\doibase
  10.1086/303585} {\bibfield  {journal} {\bibinfo  {journal} {\apj}\ }\textbf
  {\bibinfo {volume} {475}},\ \bibinfo {pages} {393} (\bibinfo {year}
  {1997})},\ \Eprint {http://arxiv.org/abs/astro-ph/9601065} {astro-ph/9601065}
  \BibitemShut {NoStop}%
\bibitem [{\citenamefont {{Hu}}\ and\ \citenamefont
  {{Dodelson}}(2002)}]{Hu:2002fk}%
  \BibitemOpen
  \bibfield  {author} {\bibinfo {author} {\bibfnamefont {W.}~\bibnamefont
  {{Hu}}}\ and\ \bibinfo {author} {\bibfnamefont {S.}~\bibnamefont
  {{Dodelson}}},\ }\href {\doibase 10.1146/annurev.astro.40.060401.093926}
  {\bibfield  {journal} {\bibinfo  {journal} {Ann.~Rev.~Astron.~Astrophys.}\
  }\textbf {\bibinfo {volume} {40}},\ \bibinfo {pages} {171} (\bibinfo {year}
  {2002})},\ \Eprint {http://arxiv.org/abs/astro-ph/0110414} {astro-ph/0110414}
  \BibitemShut {NoStop}%
\bibitem [{\citenamefont {{Kosowsky}}\ \emph
  {et~al.}(1992{\natexlab{a}})\citenamefont {{Kosowsky}}, \citenamefont
  {{Turner}},\ and\ \citenamefont {{Watkins}}}]{Kosowsky:1992uq}%
  \BibitemOpen
  \bibfield  {author} {\bibinfo {author} {\bibfnamefont {A.}~\bibnamefont
  {{Kosowsky}}}, \bibinfo {author} {\bibfnamefont {M.~S.}\ \bibnamefont
  {{Turner}}}, \ and\ \bibinfo {author} {\bibfnamefont {R.}~\bibnamefont
  {{Watkins}}},\ }\href {\doibase 10.1103/PhysRevD.45.4514} {\bibfield
  {journal} {\bibinfo  {journal} {\prd}\ }\textbf {\bibinfo {volume} {45}},\
  \bibinfo {pages} {4514} (\bibinfo {year} {1992}{\natexlab{a}})}\BibitemShut
  {NoStop}%
\bibitem [{\citenamefont {{Kosowsky}}\ \emph
  {et~al.}(1992{\natexlab{b}})\citenamefont {{Kosowsky}}, \citenamefont
  {{Turner}},\ and\ \citenamefont {{Watkins}}}]{Kosowsky:1992fj}%
  \BibitemOpen
  \bibfield  {author} {\bibinfo {author} {\bibfnamefont {A.}~\bibnamefont
  {{Kosowsky}}}, \bibinfo {author} {\bibfnamefont {M.~S.}\ \bibnamefont
  {{Turner}}}, \ and\ \bibinfo {author} {\bibfnamefont {R.}~\bibnamefont
  {{Watkins}}},\ }\href {\doibase 10.1103/PhysRevLett.69.2026} {\bibfield
  {journal} {\bibinfo  {journal} {Physical Review Letters}\ }\textbf {\bibinfo
  {volume} {69}},\ \bibinfo {pages} {2026} (\bibinfo {year}
  {1992}{\natexlab{b}})}\BibitemShut {NoStop}%
\bibitem [{\citenamefont {{Kosowsky}}\ and\ \citenamefont
  {{Turner}}(1993)}]{Kosowsky:1993fk}%
  \BibitemOpen
  \bibfield  {author} {\bibinfo {author} {\bibfnamefont {A.}~\bibnamefont
  {{Kosowsky}}}\ and\ \bibinfo {author} {\bibfnamefont {M.~S.}\ \bibnamefont
  {{Turner}}},\ }\href {\doibase 10.1103/PhysRevD.47.4372} {\bibfield
  {journal} {\bibinfo  {journal} {\prd}\ }\textbf {\bibinfo {volume} {47}},\
  \bibinfo {pages} {4372} (\bibinfo {year} {1993})},\ \Eprint
  {http://arxiv.org/abs/astro-ph/9211004} {astro-ph/9211004} \BibitemShut
  {NoStop}%
\bibitem [{\citenamefont {{Kamionkowski}}\ \emph {et~al.}(1994)\citenamefont
  {{Kamionkowski}}, \citenamefont {{Kosowsky}},\ and\ \citenamefont
  {{Turner}}}]{Kamionkowski:1994fj}%
  \BibitemOpen
  \bibfield  {author} {\bibinfo {author} {\bibfnamefont {M.}~\bibnamefont
  {{Kamionkowski}}}, \bibinfo {author} {\bibfnamefont {A.}~\bibnamefont
  {{Kosowsky}}}, \ and\ \bibinfo {author} {\bibfnamefont {M.~S.}\ \bibnamefont
  {{Turner}}},\ }\href {\doibase 10.1103/PhysRevD.49.2837} {\bibfield
  {journal} {\bibinfo  {journal} {\prd}\ }\textbf {\bibinfo {volume} {49}},\
  \bibinfo {pages} {2837} (\bibinfo {year} {1994})},\ \Eprint
  {http://arxiv.org/abs/arXiv:astro-ph/9310044} {arXiv:astro-ph/9310044}
  \BibitemShut {NoStop}%
\bibitem [{\citenamefont {{Caprini}}\ \emph {et~al.}(2008)\citenamefont
  {{Caprini}}, \citenamefont {{Durrer}},\ and\ \citenamefont
  {{Servant}}}]{Caprini:2008fr}%
  \BibitemOpen
  \bibfield  {author} {\bibinfo {author} {\bibfnamefont {C.}~\bibnamefont
  {{Caprini}}}, \bibinfo {author} {\bibfnamefont {R.}~\bibnamefont {{Durrer}}},
  \ and\ \bibinfo {author} {\bibfnamefont {G.}~\bibnamefont {{Servant}}},\
  }\href {\doibase 10.1103/PhysRevD.77.124015} {\bibfield  {journal} {\bibinfo
  {journal} {\prd}\ }\textbf {\bibinfo {volume} {77}},\ \bibinfo {eid} {124015}
  (\bibinfo {year} {2008})},\ \Eprint {http://arxiv.org/abs/0711.2593}
  {arXiv:0711.2593} \BibitemShut {NoStop}%
\bibitem [{\citenamefont {{Maggiore}}(2000)}]{Maggiore:2000lr}%
  \BibitemOpen
  \bibfield  {author} {\bibinfo {author} {\bibfnamefont {M.}~\bibnamefont
  {{Maggiore}}},\ }\href {\doibase 10.1016/S0370-1573(99)00102-7} {\bibfield
  {journal} {\bibinfo  {journal} {Physics Reports}\ }\textbf {\bibinfo {volume}
  {331}},\ \bibinfo {pages} {283} (\bibinfo {year} {2000})},\ \Eprint
  {http://arxiv.org/abs/gr-qc/9909001} {gr-qc/9909001} \BibitemShut {NoStop}%
\bibitem [{\citenamefont {{Bertotti}}\ \emph {et~al.}(1983)\citenamefont
  {{Bertotti}}, \citenamefont {{Carr}},\ and\ \citenamefont
  {{Rees}}}]{Bertotti:1983lr}%
  \BibitemOpen
  \bibfield  {author} {\bibinfo {author} {\bibfnamefont {B.}~\bibnamefont
  {{Bertotti}}}, \bibinfo {author} {\bibfnamefont {B.~J.}\ \bibnamefont
  {{Carr}}}, \ and\ \bibinfo {author} {\bibfnamefont {M.~J.}\ \bibnamefont
  {{Rees}}},\ }\href@noop {} {\bibfield  {journal} {\bibinfo  {journal} {Mon.
  Not. Royal Astron. Soc.}\ }\textbf {\bibinfo {volume} {203}},\ \bibinfo
  {pages} {945} (\bibinfo {year} {1983})}\BibitemShut {NoStop}%
\bibitem [{\citenamefont {{Armstrong}}\ \emph {et~al.}(2003)\citenamefont
  {{Armstrong}}, \citenamefont {{Iess}}, \citenamefont {{Tortora}},\ and\
  \citenamefont {{Bertotti}}}]{Armstrong:2003rt}%
  \BibitemOpen
  \bibfield  {author} {\bibinfo {author} {\bibfnamefont {J.~W.}\ \bibnamefont
  {{Armstrong}}}, \bibinfo {author} {\bibfnamefont {L.}~\bibnamefont {{Iess}}},
  \bibinfo {author} {\bibfnamefont {P.}~\bibnamefont {{Tortora}}}, \ and\
  \bibinfo {author} {\bibfnamefont {B.}~\bibnamefont {{Bertotti}}},\ }\href
  {\doibase 10.1086/379505} {\bibfield  {journal} {\bibinfo  {journal} {\apj}\
  }\textbf {\bibinfo {volume} {599}},\ \bibinfo {pages} {806} (\bibinfo {year}
  {2003})}\BibitemShut {NoStop}%
\bibitem [{\citenamefont {{Kramer}}\ \emph {et~al.}(2006)\citenamefont
  {{Kramer}}, \citenamefont {{Stairs}}, \citenamefont {{Manchester}},
  \citenamefont {{McLaughlin}}, \citenamefont {{Lyne}}, \citenamefont
  {{Ferdman}}, \citenamefont {{Burgay}}, \citenamefont {{Lorimer}},
  \citenamefont {{Possenti}}, \citenamefont {{D'Amico}}, \citenamefont
  {{Sarkissian}}, \citenamefont {{Hobbs}}, \citenamefont {{Reynolds}},
  \citenamefont {{Freire}},\ and\ \citenamefont {{Camilo}}}]{Kramer:2006vn}%
  \BibitemOpen
  \bibfield  {author} {\bibinfo {author} {\bibfnamefont {M.}~\bibnamefont
  {{Kramer}}}, \bibinfo {author} {\bibfnamefont {I.~H.}\ \bibnamefont
  {{Stairs}}}, \bibinfo {author} {\bibfnamefont {R.~N.}\ \bibnamefont
  {{Manchester}}}, \bibinfo {author} {\bibfnamefont {M.~A.}\ \bibnamefont
  {{McLaughlin}}}, \bibinfo {author} {\bibfnamefont {A.~G.}\ \bibnamefont
  {{Lyne}}}, \bibinfo {author} {\bibfnamefont {R.~D.}\ \bibnamefont
  {{Ferdman}}}, \bibinfo {author} {\bibfnamefont {M.}~\bibnamefont {{Burgay}}},
  \bibinfo {author} {\bibfnamefont {D.~R.}\ \bibnamefont {{Lorimer}}}, \bibinfo
  {author} {\bibfnamefont {A.}~\bibnamefont {{Possenti}}}, \bibinfo {author}
  {\bibfnamefont {N.}~\bibnamefont {{D'Amico}}}, \bibinfo {author}
  {\bibfnamefont {J.~M.}\ \bibnamefont {{Sarkissian}}}, \bibinfo {author}
  {\bibfnamefont {G.~B.}\ \bibnamefont {{Hobbs}}}, \bibinfo {author}
  {\bibfnamefont {J.~E.}\ \bibnamefont {{Reynolds}}}, \bibinfo {author}
  {\bibfnamefont {P.~C.~C.}\ \bibnamefont {{Freire}}}, \ and\ \bibinfo {author}
  {\bibfnamefont {F.}~\bibnamefont {{Camilo}}},\ }\href {\doibase
  10.1126/science.1132305} {\bibfield  {journal} {\bibinfo  {journal}
  {Science}\ }\textbf {\bibinfo {volume} {314}},\ \bibinfo {pages} {97}
  (\bibinfo {year} {2006})},\ \Eprint
  {http://arxiv.org/abs/arXiv:astro-ph/0609417} {arXiv:astro-ph/0609417}
  \BibitemShut {NoStop}%
\bibitem [{\citenamefont {{Jenet}}\ \emph {et~al.}(2006)\citenamefont
  {{Jenet}}, \citenamefont {{Hobbs}}, \citenamefont {{van Straten}},
  \citenamefont {{Manchester}}, \citenamefont {{Bailes}}, \citenamefont
  {{Verbiest}}, \citenamefont {{Edwards}}, \citenamefont {{Hotan}},
  \citenamefont {{Sarkissian}},\ and\ \citenamefont {{Ord}}}]{Jenet:2006fk}%
  \BibitemOpen
  \bibfield  {author} {\bibinfo {author} {\bibfnamefont {F.~A.}\ \bibnamefont
  {{Jenet}}}, \bibinfo {author} {\bibfnamefont {G.~B.}\ \bibnamefont
  {{Hobbs}}}, \bibinfo {author} {\bibfnamefont {W.}~\bibnamefont {{van
  Straten}}}, \bibinfo {author} {\bibfnamefont {R.~N.}\ \bibnamefont
  {{Manchester}}}, \bibinfo {author} {\bibfnamefont {M.}~\bibnamefont
  {{Bailes}}}, \bibinfo {author} {\bibfnamefont {J.~P.~W.}\ \bibnamefont
  {{Verbiest}}}, \bibinfo {author} {\bibfnamefont {R.~T.}\ \bibnamefont
  {{Edwards}}}, \bibinfo {author} {\bibfnamefont {A.~W.}\ \bibnamefont
  {{Hotan}}}, \bibinfo {author} {\bibfnamefont {J.~M.}\ \bibnamefont
  {{Sarkissian}}}, \ and\ \bibinfo {author} {\bibfnamefont {S.~M.}\
  \bibnamefont {{Ord}}},\ }\href {\doibase 10.1086/508702} {\bibfield
  {journal} {\bibinfo  {journal} {\apj}\ }\textbf {\bibinfo {volume} {653}},\
  \bibinfo {pages} {1571} (\bibinfo {year} {2006})},\ \Eprint
  {http://arxiv.org/abs/astro-ph/0609013} {astro-ph/0609013} \BibitemShut
  {NoStop}%
\bibitem [{\citenamefont {{Liu}}\ \emph {et~al.}(2011)\citenamefont {{Liu}},
  \citenamefont {{Verbiest}}, \citenamefont {{Kramer}}, \citenamefont
  {{Stappers}}, \citenamefont {{van Straten}},\ and\ \citenamefont
  {{Cordes}}}]{Liu:2011yq}%
  \BibitemOpen
  \bibfield  {author} {\bibinfo {author} {\bibfnamefont {K.}~\bibnamefont
  {{Liu}}}, \bibinfo {author} {\bibfnamefont {J.~P.~W.}\ \bibnamefont
  {{Verbiest}}}, \bibinfo {author} {\bibfnamefont {M.}~\bibnamefont
  {{Kramer}}}, \bibinfo {author} {\bibfnamefont {B.~W.}\ \bibnamefont
  {{Stappers}}}, \bibinfo {author} {\bibfnamefont {W.}~\bibnamefont {{van
  Straten}}}, \ and\ \bibinfo {author} {\bibfnamefont {J.~M.}\ \bibnamefont
  {{Cordes}}},\ }\href {\doibase 10.1111/j.1365-2966.2011.19452.x} {\bibfield
  {journal} {\bibinfo  {journal} {Mon. Not. Royal Astron. Soc.}\ }\textbf
  {\bibinfo {volume} {417}},\ \bibinfo {pages} {2916} (\bibinfo {year}
  {2011})},\ \Eprint {http://arxiv.org/abs/1107.3086} {arXiv:1107.3086
  [astro-ph.HE]} \BibitemShut {NoStop}%
\bibitem [{\citenamefont {{Hui}}\ \emph {et~al.}(2013)\citenamefont {{Hui}},
  \citenamefont {{McWilliams}},\ and\ \citenamefont {{Yang}}}]{Hui:2013kx}%
  \BibitemOpen
  \bibfield  {author} {\bibinfo {author} {\bibfnamefont {L.}~\bibnamefont
  {{Hui}}}, \bibinfo {author} {\bibfnamefont {S.~T.}\ \bibnamefont
  {{McWilliams}}}, \ and\ \bibinfo {author} {\bibfnamefont {I.-S.}\
  \bibnamefont {{Yang}}},\ }\href {\doibase 10.1103/PhysRevD.87.084009}
  {\bibfield  {journal} {\bibinfo  {journal} {\prd}\ }\textbf {\bibinfo
  {volume} {87}},\ \bibinfo {eid} {084009} (\bibinfo {year} {2013})},\ \Eprint
  {http://arxiv.org/abs/1212.2623} {arXiv:1212.2623 [gr-qc]} \BibitemShut
  {NoStop}%
\bibitem [{\citenamefont {{Riles}}(2013)}]{Riles:2013uq}%
  \BibitemOpen
  \bibfield  {author} {\bibinfo {author} {\bibfnamefont {K.}~\bibnamefont
  {{Riles}}},\ }\href {\doibase 10.1016/j.ppnp.2012.08.001} {\bibfield
  {journal} {\bibinfo  {journal} {Progress in Particle and Nuclear Physics}\
  }\textbf {\bibinfo {volume} {68}},\ \bibinfo {pages} {1} (\bibinfo {year}
  {2013})},\ \Eprint {http://arxiv.org/abs/1209.0667} {arXiv:1209.0667
  [hep-ex]} \BibitemShut {NoStop}%
\bibitem [{\citenamefont {{Berti}}\ \emph {et~al.}(2005)\citenamefont
  {{Berti}}, \citenamefont {{Buonanno}},\ and\ \citenamefont
  {{Will}}}]{Berti:2005fk}%
  \BibitemOpen
  \bibfield  {author} {\bibinfo {author} {\bibfnamefont {E.}~\bibnamefont
  {{Berti}}}, \bibinfo {author} {\bibfnamefont {A.}~\bibnamefont {{Buonanno}}},
  \ and\ \bibinfo {author} {\bibfnamefont {C.~M.}\ \bibnamefont {{Will}}},\
  }\href {\doibase 10.1088/0264-9381/22/18/S08} {\bibfield  {journal} {\bibinfo
   {journal} {Classical and Quantum Gravity}\ }\textbf {\bibinfo {volume}
  {22}},\ \bibinfo {pages} {S943} (\bibinfo {year} {2005})},\ \Eprint
  {http://arxiv.org/abs/gr-qc/0504017} {gr-qc/0504017} \BibitemShut {NoStop}%
\bibitem [{\citenamefont {{Babak}}\ \emph {et~al.}(2011)\citenamefont
  {{Babak}}, \citenamefont {{Gair}}, \citenamefont {{Petiteau}},\ and\
  \citenamefont {{Sesana}}}]{Babak:2011qy}%
  \BibitemOpen
  \bibfield  {author} {\bibinfo {author} {\bibfnamefont {S.}~\bibnamefont
  {{Babak}}}, \bibinfo {author} {\bibfnamefont {J.~R.}\ \bibnamefont {{Gair}}},
  \bibinfo {author} {\bibfnamefont {A.}~\bibnamefont {{Petiteau}}}, \ and\
  \bibinfo {author} {\bibfnamefont {A.}~\bibnamefont {{Sesana}}},\ }\href
  {\doibase 10.1088/0264-9381/28/11/114001} {\bibfield  {journal} {\bibinfo
  {journal} {Classical and Quantum Gravity}\ }\textbf {\bibinfo {volume}
  {28}},\ \bibinfo {eid} {114001} (\bibinfo {year} {2011})},\ \Eprint
  {http://arxiv.org/abs/1011.2062} {arXiv:1011.2062 [gr-qc]} \BibitemShut
  {NoStop}%
\bibitem [{\citenamefont {{Backer}}\ \emph {et~al.}(2004)\citenamefont
  {{Backer}}, \citenamefont {{Jaffe}},\ and\ \citenamefont
  {{Lommen}}}]{Backer:2004qy}%
  \BibitemOpen
  \bibfield  {author} {\bibinfo {author} {\bibfnamefont {D.~C.}\ \bibnamefont
  {{Backer}}}, \bibinfo {author} {\bibfnamefont {A.~H.}\ \bibnamefont
  {{Jaffe}}}, \ and\ \bibinfo {author} {\bibfnamefont {A.~N.}\ \bibnamefont
  {{Lommen}}},\ }\enquote {\bibinfo {title} {{Massive Black Holes,
  Gravitational Waves and Pulsars}},}\ in\ \href@noop {} {\emph {\bibinfo
  {booktitle} {{Coevolution of Black Holes and Galaxies}}}},\ \bibinfo {series}
  {Carnegie Observatories Astrophysics Series}, Vol.~\bibinfo {volume} {I},\
  \bibinfo {editor} {edited by\ \bibinfo {editor} {\bibfnamefont {L.~C.}\
  \bibnamefont {{Ho}}}}\ (\bibinfo  {publisher} {Cambridge University Press,
  Cambridge, England},\ \bibinfo {year} {2004})\ p.~\bibinfo {pages}
  {1}\BibitemShut {NoStop}%
\bibitem [{\citenamefont {{Scott}}\ and\ \citenamefont
  {{Rees}}(1990)}]{Scott:1990vn}%
  \BibitemOpen
  \bibfield  {author} {\bibinfo {author} {\bibfnamefont {D.}~\bibnamefont
  {{Scott}}}\ and\ \bibinfo {author} {\bibfnamefont {M.~J.}\ \bibnamefont
  {{Rees}}},\ }\href@noop {} {\bibfield  {journal} {\bibinfo  {journal}
  {Mon.~Not.~Royal~Astron.~Soc.}\ }\textbf {\bibinfo {volume} {247}},\ \bibinfo
  {pages} {510} (\bibinfo {year} {1990})}\BibitemShut {NoStop}%
\bibitem [{\citenamefont {{Loeb}}\ and\ \citenamefont
  {{Zaldarriaga}}(2004)}]{Loeb:2004lr}%
  \BibitemOpen
  \bibfield  {author} {\bibinfo {author} {\bibfnamefont {A.}~\bibnamefont
  {{Loeb}}}\ and\ \bibinfo {author} {\bibfnamefont {M.}~\bibnamefont
  {{Zaldarriaga}}},\ }\href {\doibase 10.1103/PhysRevLett.92.211301} {\bibfield
   {journal} {\bibinfo  {journal} {Physical Review Letters}\ }\textbf {\bibinfo
  {volume} {92}},\ \bibinfo {eid} {211301} (\bibinfo {year} {2004})},\ \Eprint
  {http://arxiv.org/abs/astro-ph/0312134} {astro-ph/0312134} \BibitemShut
  {NoStop}%
\bibitem [{\citenamefont {{Furlanetto}}\ \emph {et~al.}(2006)\citenamefont
  {{Furlanetto}}, \citenamefont {{Oh}},\ and\ \citenamefont
  {{Briggs}}}]{Furlanetto:2006kx}%
  \BibitemOpen
  \bibfield  {author} {\bibinfo {author} {\bibfnamefont {S.~R.}\ \bibnamefont
  {{Furlanetto}}}, \bibinfo {author} {\bibfnamefont {S.~P.}\ \bibnamefont
  {{Oh}}}, \ and\ \bibinfo {author} {\bibfnamefont {F.~H.}\ \bibnamefont
  {{Briggs}}},\ }\href {\doibase 10.1016/j.physrep.2006.08.002} {\bibfield
  {journal} {\bibinfo  {journal} {Physics Reports}\ }\textbf {\bibinfo {volume}
  {433}},\ \bibinfo {pages} {181} (\bibinfo {year} {2006})},\ \Eprint
  {http://arxiv.org/abs/astro-ph/0608032} {astro-ph/0608032} \BibitemShut
  {NoStop}%
\bibitem [{\citenamefont {{Morales}}\ and\ \citenamefont
  {{Wyithe}}(2010)}]{Morales:2010qy}%
  \BibitemOpen
  \bibfield  {author} {\bibinfo {author} {\bibfnamefont {M.~F.}\ \bibnamefont
  {{Morales}}}\ and\ \bibinfo {author} {\bibfnamefont {J.~S.~B.}\ \bibnamefont
  {{Wyithe}}},\ }\href {\doibase 10.1146/annurev-astro-081309-130936}
  {\bibfield  {journal} {\bibinfo  {journal} {Ann.~Rev.~Astron.~Astrophys.}\
  }\textbf {\bibinfo {volume} {48}},\ \bibinfo {pages} {127} (\bibinfo {year}
  {2010})},\ \Eprint {http://arxiv.org/abs/0910.3010} {arXiv:0910.3010
  [astro-ph.CO]} \BibitemShut {NoStop}%
\bibitem [{\citenamefont {{Pober}}\ \emph {et~al.}(2014)\citenamefont
  {{Pober}}, \citenamefont {{Liu}}, \citenamefont {{Dillon}}, \citenamefont
  {{Aguirre}}, \citenamefont {{Bowman}}, \citenamefont {{Bradley}},
  \citenamefont {{Carilli}}, \citenamefont {{DeBoer}}, \citenamefont
  {{Hewitt}}, \citenamefont {{Jacobs}}, \citenamefont {{McQuinn}},
  \citenamefont {{Morales}}, \citenamefont {{Parsons}}, \citenamefont
  {{Tegmark}},\ and\ \citenamefont {{Werthimer}}}]{Pober:2013lr}%
  \BibitemOpen
  \bibfield  {author} {\bibinfo {author} {\bibfnamefont {J.~C.}\ \bibnamefont
  {{Pober}}}, \bibinfo {author} {\bibfnamefont {A.}~\bibnamefont {{Liu}}},
  \bibinfo {author} {\bibfnamefont {J.~S.}\ \bibnamefont {{Dillon}}}, \bibinfo
  {author} {\bibfnamefont {J.~E.}\ \bibnamefont {{Aguirre}}}, \bibinfo {author}
  {\bibfnamefont {J.~D.}\ \bibnamefont {{Bowman}}}, \bibinfo {author}
  {\bibfnamefont {R.~F.}\ \bibnamefont {{Bradley}}}, \bibinfo {author}
  {\bibfnamefont {C.~L.}\ \bibnamefont {{Carilli}}}, \bibinfo {author}
  {\bibfnamefont {D.~R.}\ \bibnamefont {{DeBoer}}}, \bibinfo {author}
  {\bibfnamefont {J.~N.}\ \bibnamefont {{Hewitt}}}, \bibinfo {author}
  {\bibfnamefont {D.~C.}\ \bibnamefont {{Jacobs}}}, \bibinfo {author}
  {\bibfnamefont {M.}~\bibnamefont {{McQuinn}}}, \bibinfo {author}
  {\bibfnamefont {M.~F.}\ \bibnamefont {{Morales}}}, \bibinfo {author}
  {\bibfnamefont {A.~R.}\ \bibnamefont {{Parsons}}}, \bibinfo {author}
  {\bibfnamefont {M.}~\bibnamefont {{Tegmark}}}, \ and\ \bibinfo {author}
  {\bibfnamefont {D.~J.}\ \bibnamefont {{Werthimer}}},\ }\href {\doibase
  10.1088/0004-637X/782/2/66} {\bibfield  {journal} {\bibinfo  {journal}
  {\apj}\ }\textbf {\bibinfo {volume} {782}},\ \bibinfo {eid} {66} (\bibinfo
  {year} {2014})},\ \Eprint {http://arxiv.org/abs/1310.7031} {arXiv:1310.7031}
  \BibitemShut {NoStop}%
\bibitem [{\citenamefont {{Zaldarriaga}}\ \emph {et~al.}(2004)\citenamefont
  {{Zaldarriaga}}, \citenamefont {{Furlanetto}},\ and\ \citenamefont
  {{Hernquist}}}]{Zaldarriaga:2004rt}%
  \BibitemOpen
  \bibfield  {author} {\bibinfo {author} {\bibfnamefont {M.}~\bibnamefont
  {{Zaldarriaga}}}, \bibinfo {author} {\bibfnamefont {S.~R.}\ \bibnamefont
  {{Furlanetto}}}, \ and\ \bibinfo {author} {\bibfnamefont {L.}~\bibnamefont
  {{Hernquist}}},\ }\href {\doibase 10.1086/386327} {\bibfield  {journal}
  {\bibinfo  {journal} {\apj}\ }\textbf {\bibinfo {volume} {608}},\ \bibinfo
  {pages} {622} (\bibinfo {year} {2004})},\ \Eprint
  {http://arxiv.org/abs/astro-ph/0311514} {astro-ph/0311514} \BibitemShut
  {NoStop}%
\bibitem [{\citenamefont {{Morales}}\ \emph {et~al.}(2012)\citenamefont
  {{Morales}}, \citenamefont {{Hazelton}}, \citenamefont {{Sullivan}},\ and\
  \citenamefont {{Beardsley}}}]{Morales:2012fj}%
  \BibitemOpen
  \bibfield  {author} {\bibinfo {author} {\bibfnamefont {M.~F.}\ \bibnamefont
  {{Morales}}}, \bibinfo {author} {\bibfnamefont {B.}~\bibnamefont
  {{Hazelton}}}, \bibinfo {author} {\bibfnamefont {I.}~\bibnamefont
  {{Sullivan}}}, \ and\ \bibinfo {author} {\bibfnamefont {A.}~\bibnamefont
  {{Beardsley}}},\ }\href {\doibase 10.1088/0004-637X/752/2/137} {\bibfield
  {journal} {\bibinfo  {journal} {\apj}\ }\textbf {\bibinfo {volume} {752}},\
  \bibinfo {eid} {137} (\bibinfo {year} {2012})},\ \Eprint
  {http://arxiv.org/abs/1202.3830} {arXiv:1202.3830 [astro-ph.IM]} \BibitemShut
  {NoStop}%
\bibitem [{\citenamefont {{Pritchard}}\ and\ \citenamefont
  {{Loeb}}(2012)}]{Pritchard:2012uq}%
  \BibitemOpen
  \bibfield  {author} {\bibinfo {author} {\bibfnamefont {J.~R.}\ \bibnamefont
  {{Pritchard}}}\ and\ \bibinfo {author} {\bibfnamefont {A.}~\bibnamefont
  {{Loeb}}},\ }\href {\doibase 10.1088/0034-4885/75/8/086901} {\bibfield
  {journal} {\bibinfo  {journal} {Reports on Progress in Physics}\ }\textbf
  {\bibinfo {volume} {75}},\ \bibinfo {eid} {086901} (\bibinfo {year}
  {2012})},\ \Eprint {http://arxiv.org/abs/1109.6012} {arXiv:1109.6012
  [astro-ph.CO]} \BibitemShut {NoStop}%
\bibitem [{\citenamefont {{Boylan-Kolchin}}\ \emph {et~al.}(2012)\citenamefont
  {{Boylan-Kolchin}}, \citenamefont {{Bullock}},\ and\ \citenamefont
  {{Kaplinghat}}}]{Boylan-Kolchin2012}%
  \BibitemOpen
  \bibfield  {author} {\bibinfo {author} {\bibfnamefont {M.}~\bibnamefont
  {{Boylan-Kolchin}}}, \bibinfo {author} {\bibfnamefont {J.~S.}\ \bibnamefont
  {{Bullock}}}, \ and\ \bibinfo {author} {\bibfnamefont {M.}~\bibnamefont
  {{Kaplinghat}}},\ }\href@noop {} {\bibfield  {journal} {\bibinfo  {journal}
  {Mon.~Not.~Royal~Astron.~Soc.}\ }\textbf {\bibinfo {volume} {422}},\ \bibinfo
  {pages} {1203} (\bibinfo {year} {2012})}\BibitemShut {NoStop}%
\bibitem [{\citenamefont {{Weinberg}}\ \emph {et~al.}(2013)\citenamefont
  {{Weinberg}}, \citenamefont {{Bullock}}, \citenamefont {{Governato}},
  \citenamefont {{Kuzio de Naray}},\ and\ \citenamefont
  {{Peter}}}]{Weinberg:2013fj}%
  \BibitemOpen
  \bibfield  {author} {\bibinfo {author} {\bibfnamefont {D.~H.}\ \bibnamefont
  {{Weinberg}}}, \bibinfo {author} {\bibfnamefont {J.~S.}\ \bibnamefont
  {{Bullock}}}, \bibinfo {author} {\bibfnamefont {F.}~\bibnamefont
  {{Governato}}}, \bibinfo {author} {\bibfnamefont {R.}~\bibnamefont {{Kuzio de
  Naray}}}, \ and\ \bibinfo {author} {\bibfnamefont {A.~H.~G.}\ \bibnamefont
  {{Peter}}},\ }\href@noop {} {\bibfield  {journal} {\bibinfo  {journal} {ArXiv
  e-prints}\ } (\bibinfo {year} {2013})},\ \Eprint
  {http://arxiv.org/abs/1306.0913} {arXiv:1306.0913 [astro-ph.CO]} \BibitemShut
  {NoStop}%
\bibitem [{\citenamefont {{Conrad}}\ \emph {et~al.}(2013)\citenamefont
  {{Conrad}}, \citenamefont {{Louis}},\ and\ \citenamefont
  {{Shaevitz}}}]{Conrad:2013ys}%
  \BibitemOpen
  \bibfield  {author} {\bibinfo {author} {\bibfnamefont {J.~M.}\ \bibnamefont
  {{Conrad}}}, \bibinfo {author} {\bibfnamefont {W.~C.}\ \bibnamefont
  {{Louis}}}, \ and\ \bibinfo {author} {\bibfnamefont {M.~H.}\ \bibnamefont
  {{Shaevitz}}},\ }\href {\doibase 10.1146/annurev-nucl-102711-094957}
  {\bibfield  {journal} {\bibinfo  {journal} {Annual Review of Nuclear and
  Particle Science}\ }\textbf {\bibinfo {volume} {63}},\ \bibinfo {pages} {45}
  (\bibinfo {year} {2013})},\ \Eprint {http://arxiv.org/abs/1306.6494}
  {arXiv:1306.6494 [hep-ex]} \BibitemShut {NoStop}%
\bibitem [{\citenamefont {{de Gouvea}}\ \emph {et~al.}(2013)\citenamefont {{de
  Gouvea}}, \citenamefont {{Pitts}}, \citenamefont {{Scholberg}}, \citenamefont
  {{Zeller}}, \citenamefont {{Alonso}}, \citenamefont {{Bernstein}},
  \citenamefont {{Bishai}}, \citenamefont {{Elliott}}, \citenamefont
  {{Heeger}}, \citenamefont {{Hoffman}}, \citenamefont {{Huber}}, \citenamefont
  {{Kaufman}}, \citenamefont {{Kayser}}, \citenamefont {{Link}}, \citenamefont
  {{Lunardini}}, \citenamefont {{Monreal}}, \citenamefont {{Morfin}},
  \citenamefont {{Robertson}}, \citenamefont {{Tayloe}}, \citenamefont
  {{Tolich}}, \citenamefont {{Abazajian}}, \citenamefont {{Akiri}},
  \citenamefont {{Albright}}, \citenamefont {{Asaadi}}, \citenamefont {{Babu}},
  \citenamefont {{Balantekin}}, \citenamefont {{Barbeau}}, \citenamefont
  {{Bass}}, \citenamefont {{Blake}}, \citenamefont {{Blondel}}, \citenamefont
  {{Blucher}}, \citenamefont {{Bowden}}, \citenamefont {{Brice}}, \citenamefont
  {{Bross}}, \citenamefont {{Carls}}, \citenamefont {{Cavanna}}, \citenamefont
  {{Choudhary}}, \citenamefont {{Coloma}}, \citenamefont {{Connolly}},
  \citenamefont {{Conrad}}, \citenamefont {{Convery}}, \citenamefont
  {{Cooper}}, \citenamefont {{Cowen}}, \citenamefont {{da Motta}},
  \citenamefont {{de Young}}, \citenamefont {{Di Lodovico}}, \citenamefont
  {{Diwan}}, \citenamefont {{Djurcic}}, \citenamefont {{Dracos}}, \citenamefont
  {{Dodelson}}, \citenamefont {{Efremenko}}, \citenamefont {{Ekelof}},
  \citenamefont {{Feng}}, \citenamefont {{Fleming}}, \citenamefont
  {{Formaggio}}, \citenamefont {{Friedland}}, \citenamefont {{Fuller}},
  \citenamefont {{Gallagher}}, \citenamefont {{Geer}}, \citenamefont
  {{Gilchriese}}, \citenamefont {{Goodman}}, \citenamefont {{Grant}},
  \citenamefont {{Gratta}}, \citenamefont {{Hall}}, \citenamefont {{Halzen}},
  \citenamefont {{Harris}}, \citenamefont {{Heffner}}, \citenamefont
  {{Henning}}, \citenamefont {{Hewett}}, \citenamefont {{Hill}}, \citenamefont
  {{Himmel}}, \citenamefont {{Horton-Smith}}, \citenamefont {{Karle}},
  \citenamefont {{Katori}}, \citenamefont {{Kearns}}, \citenamefont
  {{Kettell}}, \citenamefont {{Klein}}, \citenamefont {{Kim}}, \citenamefont
  {{Kim}}, \citenamefont {{Kolomensky}}, \citenamefont {{Kordosky}},
  \citenamefont {{Kudenko}}, \citenamefont {{Kudryavtsev}}, \citenamefont
  {{Lande}}, \citenamefont {{Lang}}, \citenamefont {{Lanza}}, \citenamefont
  {{Lau}}, \citenamefont {{Lee}}, \citenamefont {{Li}}, \citenamefont
  {{Littlejohn}}, \citenamefont {{Lin}}, \citenamefont {{Liu}}, \citenamefont
  {{Liu}}, \citenamefont {{Long}}, \citenamefont {{Louis}}, \citenamefont
  {{Luk}}, \citenamefont {{Marciano}}, \citenamefont {{Mariani}}, \citenamefont
  {{Marshak}}, \citenamefont {{Mauger}}, \citenamefont {{McDonald}},
  \citenamefont {{McFarland}}, \citenamefont {{McKeown}}, \citenamefont
  {{Messier}}, \citenamefont {{Mishra}}, \citenamefont {{Mosel}}, \citenamefont
  {{Mumm}}, \citenamefont {{Nakaya}}, \citenamefont {{Nelson}}, \citenamefont
  {{Nygren}}, \citenamefont {{Orebi Gann}}, \citenamefont {{Osta}},
  \citenamefont {{Palamara}}, \citenamefont {{Paley}}, \citenamefont
  {{Papadimitriou}}, \citenamefont {{Parke}}, \citenamefont {{Parsa}},
  \citenamefont {{Patterson}}, \citenamefont {{Piepke}}, \citenamefont
  {{Plunkett}}, \citenamefont {{Poon}}, \citenamefont {{Qian}}, \citenamefont
  {{Raaf}}, \citenamefont {{Rameika}}, \citenamefont {{Ramsey-Musolf}},
  \citenamefont {{Rebel}}, \citenamefont {{Roser}}, \citenamefont {{Rosner}},
  \citenamefont {{Rott}}, \citenamefont {{Rybka}}, \citenamefont {{Sahoo}},
  \citenamefont {{Sangiorgio}}, \citenamefont {{Schmitz}}, \citenamefont
  {{Shrock}}, \citenamefont {{Shaevitz}}, \citenamefont {{Smith}},
  \citenamefont {{Smy}}, \citenamefont {{Sobel}}, \citenamefont {{Sorensen}},
  \citenamefont {{Sousa}}, \citenamefont {{Spitz}}, \citenamefont {{Strauss}},
  \citenamefont {{Svoboda}}, \citenamefont {{Tanaka}}, \citenamefont
  {{Thomas}}, \citenamefont {{Tian}}, \citenamefont {{Tschirhart}},
  \citenamefont {{Tully}}, \citenamefont {{Van Bibber}}, \citenamefont {{Van de
  Water}}, \citenamefont {{Vahle}}, \citenamefont {{Vogel}}, \citenamefont
  {{Walter}}, \citenamefont {{Wark}}, \citenamefont {{Wascko}}, \citenamefont
  {{Webber}}, \citenamefont {{Weerts}}, \citenamefont {{White}}, \citenamefont
  {{White}}, \citenamefont {{Whitehead}}, \citenamefont {{Wilson}},
  \citenamefont {{Winslow}}, \citenamefont {{Wongjirad}}, \citenamefont
  {{Worcester}}, \citenamefont {{Yokoyama}}, \citenamefont {{Yoo}},\ and\
  \citenamefont {{Zimmerman}}}]{de-Gouvea:2013fr}%
  \BibitemOpen
  \bibfield  {author} {\bibinfo {author} {\bibfnamefont {A.}~\bibnamefont {{de
  Gouvea}}}, \bibinfo {author} {\bibfnamefont {K.}~\bibnamefont {{Pitts}}},
  \bibinfo {author} {\bibfnamefont {K.}~\bibnamefont {{Scholberg}}}, \bibinfo
  {author} {\bibfnamefont {G.~P.}\ \bibnamefont {{Zeller}}}, \bibinfo {author}
  {\bibfnamefont {J.}~\bibnamefont {{Alonso}}}, \bibinfo {author}
  {\bibfnamefont {A.}~\bibnamefont {{Bernstein}}}, \bibinfo {author}
  {\bibfnamefont {M.}~\bibnamefont {{Bishai}}}, \bibinfo {author}
  {\bibfnamefont {S.}~\bibnamefont {{Elliott}}}, \bibinfo {author}
  {\bibfnamefont {K.}~\bibnamefont {{Heeger}}}, \bibinfo {author}
  {\bibfnamefont {K.}~\bibnamefont {{Hoffman}}}, \bibinfo {author}
  {\bibfnamefont {P.}~\bibnamefont {{Huber}}}, \bibinfo {author} {\bibfnamefont
  {L.~J.}\ \bibnamefont {{Kaufman}}}, \bibinfo {author} {\bibfnamefont
  {B.}~\bibnamefont {{Kayser}}}, \bibinfo {author} {\bibfnamefont
  {J.}~\bibnamefont {{Link}}}, \bibinfo {author} {\bibfnamefont
  {C.}~\bibnamefont {{Lunardini}}}, \bibinfo {author} {\bibfnamefont
  {B.}~\bibnamefont {{Monreal}}}, \bibinfo {author} {\bibfnamefont {J.~G.}\
  \bibnamefont {{Morfin}}}, \bibinfo {author} {\bibfnamefont {H.}~\bibnamefont
  {{Robertson}}}, \bibinfo {author} {\bibfnamefont {R.}~\bibnamefont
  {{Tayloe}}}, \bibinfo {author} {\bibfnamefont {N.}~\bibnamefont {{Tolich}}},
  \bibinfo {author} {\bibfnamefont {K.}~\bibnamefont {{Abazajian}}}, \bibinfo
  {author} {\bibfnamefont {T.}~\bibnamefont {{Akiri}}}, \bibinfo {author}
  {\bibfnamefont {C.}~\bibnamefont {{Albright}}}, \bibinfo {author}
  {\bibfnamefont {J.}~\bibnamefont {{Asaadi}}}, \bibinfo {author}
  {\bibfnamefont {K.~S.}\ \bibnamefont {{Babu}}}, \bibinfo {author}
  {\bibfnamefont {A.~B.}\ \bibnamefont {{Balantekin}}}, \bibinfo {author}
  {\bibfnamefont {P.}~\bibnamefont {{Barbeau}}}, \bibinfo {author}
  {\bibfnamefont {M.}~\bibnamefont {{Bass}}}, \bibinfo {author} {\bibfnamefont
  {A.}~\bibnamefont {{Blake}}}, \bibinfo {author} {\bibfnamefont
  {A.}~\bibnamefont {{Blondel}}}, \bibinfo {author} {\bibfnamefont
  {E.}~\bibnamefont {{Blucher}}}, \bibinfo {author} {\bibfnamefont
  {N.}~\bibnamefont {{Bowden}}}, \bibinfo {author} {\bibfnamefont {S.~J.}\
  \bibnamefont {{Brice}}}, \bibinfo {author} {\bibfnamefont {A.}~\bibnamefont
  {{Bross}}}, \bibinfo {author} {\bibfnamefont {B.}~\bibnamefont {{Carls}}},
  \bibinfo {author} {\bibfnamefont {F.}~\bibnamefont {{Cavanna}}}, \bibinfo
  {author} {\bibfnamefont {B.}~\bibnamefont {{Choudhary}}}, \bibinfo {author}
  {\bibfnamefont {P.}~\bibnamefont {{Coloma}}}, \bibinfo {author}
  {\bibfnamefont {A.}~\bibnamefont {{Connolly}}}, \bibinfo {author}
  {\bibfnamefont {J.}~\bibnamefont {{Conrad}}}, \bibinfo {author}
  {\bibfnamefont {M.}~\bibnamefont {{Convery}}}, \bibinfo {author}
  {\bibfnamefont {R.~L.}\ \bibnamefont {{Cooper}}}, \bibinfo {author}
  {\bibfnamefont {D.}~\bibnamefont {{Cowen}}}, \bibinfo {author} {\bibfnamefont
  {H.}~\bibnamefont {{da Motta}}}, \bibinfo {author} {\bibfnamefont
  {T.}~\bibnamefont {{de Young}}}, \bibinfo {author} {\bibfnamefont
  {F.}~\bibnamefont {{Di Lodovico}}}, \bibinfo {author} {\bibfnamefont
  {M.}~\bibnamefont {{Diwan}}}, \bibinfo {author} {\bibfnamefont
  {Z.}~\bibnamefont {{Djurcic}}}, \bibinfo {author} {\bibfnamefont
  {M.}~\bibnamefont {{Dracos}}}, \bibinfo {author} {\bibfnamefont
  {S.}~\bibnamefont {{Dodelson}}}, \bibinfo {author} {\bibfnamefont
  {Y.}~\bibnamefont {{Efremenko}}}, \bibinfo {author} {\bibfnamefont
  {T.}~\bibnamefont {{Ekelof}}}, \bibinfo {author} {\bibfnamefont {J.~L.}\
  \bibnamefont {{Feng}}}, \bibinfo {author} {\bibfnamefont {B.}~\bibnamefont
  {{Fleming}}}, \bibinfo {author} {\bibfnamefont {J.}~\bibnamefont
  {{Formaggio}}}, \bibinfo {author} {\bibfnamefont {A.}~\bibnamefont
  {{Friedland}}}, \bibinfo {author} {\bibfnamefont {G.}~\bibnamefont
  {{Fuller}}}, \bibinfo {author} {\bibfnamefont {H.}~\bibnamefont
  {{Gallagher}}}, \bibinfo {author} {\bibfnamefont {S.}~\bibnamefont {{Geer}}},
  \bibinfo {author} {\bibfnamefont {M.}~\bibnamefont {{Gilchriese}}}, \bibinfo
  {author} {\bibfnamefont {M.}~\bibnamefont {{Goodman}}}, \bibinfo {author}
  {\bibfnamefont {D.}~\bibnamefont {{Grant}}}, \bibinfo {author} {\bibfnamefont
  {G.}~\bibnamefont {{Gratta}}}, \bibinfo {author} {\bibfnamefont
  {C.}~\bibnamefont {{Hall}}}, \bibinfo {author} {\bibfnamefont
  {F.}~\bibnamefont {{Halzen}}}, \bibinfo {author} {\bibfnamefont
  {D.}~\bibnamefont {{Harris}}}, \bibinfo {author} {\bibfnamefont
  {M.}~\bibnamefont {{Heffner}}}, \bibinfo {author} {\bibfnamefont
  {R.}~\bibnamefont {{Henning}}}, \bibinfo {author} {\bibfnamefont {J.~L.}\
  \bibnamefont {{Hewett}}}, \bibinfo {author} {\bibfnamefont {R.}~\bibnamefont
  {{Hill}}}, \bibinfo {author} {\bibfnamefont {A.}~\bibnamefont {{Himmel}}},
  \bibinfo {author} {\bibfnamefont {G.}~\bibnamefont {{Horton-Smith}}},
  \bibinfo {author} {\bibfnamefont {A.}~\bibnamefont {{Karle}}}, \bibinfo
  {author} {\bibfnamefont {T.}~\bibnamefont {{Katori}}}, \bibinfo {author}
  {\bibfnamefont {E.}~\bibnamefont {{Kearns}}}, \bibinfo {author}
  {\bibfnamefont {S.}~\bibnamefont {{Kettell}}}, \bibinfo {author}
  {\bibfnamefont {J.}~\bibnamefont {{Klein}}}, \bibinfo {author} {\bibfnamefont
  {Y.}~\bibnamefont {{Kim}}}, \bibinfo {author} {\bibfnamefont {Y.~K.}\
  \bibnamefont {{Kim}}}, \bibinfo {author} {\bibfnamefont {Y.}~\bibnamefont
  {{Kolomensky}}}, \bibinfo {author} {\bibfnamefont {M.}~\bibnamefont
  {{Kordosky}}}, \bibinfo {author} {\bibfnamefont {Y.}~\bibnamefont
  {{Kudenko}}}, \bibinfo {author} {\bibfnamefont {V.~A.}\ \bibnamefont
  {{Kudryavtsev}}}, \bibinfo {author} {\bibfnamefont {K.}~\bibnamefont
  {{Lande}}}, \bibinfo {author} {\bibfnamefont {K.}~\bibnamefont {{Lang}}},
  \bibinfo {author} {\bibfnamefont {R.}~\bibnamefont {{Lanza}}}, \bibinfo
  {author} {\bibfnamefont {K.}~\bibnamefont {{Lau}}}, \bibinfo {author}
  {\bibfnamefont {H.}~\bibnamefont {{Lee}}}, \bibinfo {author} {\bibfnamefont
  {Z.}~\bibnamefont {{Li}}}, \bibinfo {author} {\bibfnamefont {B.~R.}\
  \bibnamefont {{Littlejohn}}}, \bibinfo {author} {\bibfnamefont {C.~J.}\
  \bibnamefont {{Lin}}}, \bibinfo {author} {\bibfnamefont {D.}~\bibnamefont
  {{Liu}}}, \bibinfo {author} {\bibfnamefont {H.}~\bibnamefont {{Liu}}},
  \bibinfo {author} {\bibfnamefont {K.}~\bibnamefont {{Long}}}, \bibinfo
  {author} {\bibfnamefont {W.}~\bibnamefont {{Louis}}}, \bibinfo {author}
  {\bibfnamefont {K.~B.}\ \bibnamefont {{Luk}}}, \bibinfo {author}
  {\bibfnamefont {W.}~\bibnamefont {{Marciano}}}, \bibinfo {author}
  {\bibfnamefont {C.}~\bibnamefont {{Mariani}}}, \bibinfo {author}
  {\bibfnamefont {M.}~\bibnamefont {{Marshak}}}, \bibinfo {author}
  {\bibfnamefont {C.}~\bibnamefont {{Mauger}}}, \bibinfo {author}
  {\bibfnamefont {K.~T.}\ \bibnamefont {{McDonald}}}, \bibinfo {author}
  {\bibfnamefont {K.}~\bibnamefont {{McFarland}}}, \bibinfo {author}
  {\bibfnamefont {R.}~\bibnamefont {{McKeown}}}, \bibinfo {author}
  {\bibfnamefont {M.}~\bibnamefont {{Messier}}}, \bibinfo {author}
  {\bibfnamefont {S.~R.}\ \bibnamefont {{Mishra}}}, \bibinfo {author}
  {\bibfnamefont {U.}~\bibnamefont {{Mosel}}}, \bibinfo {author} {\bibfnamefont
  {P.}~\bibnamefont {{Mumm}}}, \bibinfo {author} {\bibfnamefont
  {T.}~\bibnamefont {{Nakaya}}}, \bibinfo {author} {\bibfnamefont {J.~K.}\
  \bibnamefont {{Nelson}}}, \bibinfo {author} {\bibfnamefont {D.}~\bibnamefont
  {{Nygren}}}, \bibinfo {author} {\bibfnamefont {G.~D.}\ \bibnamefont {{Orebi
  Gann}}}, \bibinfo {author} {\bibfnamefont {J.}~\bibnamefont {{Osta}}},
  \bibinfo {author} {\bibfnamefont {O.}~\bibnamefont {{Palamara}}}, \bibinfo
  {author} {\bibfnamefont {J.}~\bibnamefont {{Paley}}}, \bibinfo {author}
  {\bibfnamefont {V.}~\bibnamefont {{Papadimitriou}}}, \bibinfo {author}
  {\bibfnamefont {S.}~\bibnamefont {{Parke}}}, \bibinfo {author} {\bibfnamefont
  {Z.}~\bibnamefont {{Parsa}}}, \bibinfo {author} {\bibfnamefont
  {R.}~\bibnamefont {{Patterson}}}, \bibinfo {author} {\bibfnamefont
  {A.}~\bibnamefont {{Piepke}}}, \bibinfo {author} {\bibfnamefont
  {R.}~\bibnamefont {{Plunkett}}}, \bibinfo {author} {\bibfnamefont
  {A.}~\bibnamefont {{Poon}}}, \bibinfo {author} {\bibfnamefont
  {X.}~\bibnamefont {{Qian}}}, \bibinfo {author} {\bibfnamefont
  {J.}~\bibnamefont {{Raaf}}}, \bibinfo {author} {\bibfnamefont
  {R.}~\bibnamefont {{Rameika}}}, \bibinfo {author} {\bibfnamefont
  {M.}~\bibnamefont {{Ramsey-Musolf}}}, \bibinfo {author} {\bibfnamefont
  {B.}~\bibnamefont {{Rebel}}}, \bibinfo {author} {\bibfnamefont
  {R.}~\bibnamefont {{Roser}}}, \bibinfo {author} {\bibfnamefont
  {J.}~\bibnamefont {{Rosner}}}, \bibinfo {author} {\bibfnamefont
  {C.}~\bibnamefont {{Rott}}}, \bibinfo {author} {\bibfnamefont
  {G.}~\bibnamefont {{Rybka}}}, \bibinfo {author} {\bibfnamefont
  {H.}~\bibnamefont {{Sahoo}}}, \bibinfo {author} {\bibfnamefont
  {S.}~\bibnamefont {{Sangiorgio}}}, \bibinfo {author} {\bibfnamefont
  {D.}~\bibnamefont {{Schmitz}}}, \bibinfo {author} {\bibfnamefont
  {R.}~\bibnamefont {{Shrock}}}, \bibinfo {author} {\bibfnamefont
  {M.}~\bibnamefont {{Shaevitz}}}, \bibinfo {author} {\bibfnamefont
  {N.}~\bibnamefont {{Smith}}}, \bibinfo {author} {\bibfnamefont
  {M.}~\bibnamefont {{Smy}}}, \bibinfo {author} {\bibfnamefont
  {H.}~\bibnamefont {{Sobel}}}, \bibinfo {author} {\bibfnamefont
  {P.}~\bibnamefont {{Sorensen}}}, \bibinfo {author} {\bibfnamefont
  {A.}~\bibnamefont {{Sousa}}}, \bibinfo {author} {\bibfnamefont
  {J.}~\bibnamefont {{Spitz}}}, \bibinfo {author} {\bibfnamefont
  {T.}~\bibnamefont {{Strauss}}}, \bibinfo {author} {\bibfnamefont
  {R.}~\bibnamefont {{Svoboda}}}, \bibinfo {author} {\bibfnamefont {H.~A.}\
  \bibnamefont {{Tanaka}}}, \bibinfo {author} {\bibfnamefont {J.}~\bibnamefont
  {{Thomas}}}, \bibinfo {author} {\bibfnamefont {X.}~\bibnamefont {{Tian}}},
  \bibinfo {author} {\bibfnamefont {R.}~\bibnamefont {{Tschirhart}}}, \bibinfo
  {author} {\bibfnamefont {C.}~\bibnamefont {{Tully}}}, \bibinfo {author}
  {\bibfnamefont {K.}~\bibnamefont {{Van Bibber}}}, \bibinfo {author}
  {\bibfnamefont {R.~G.}\ \bibnamefont {{Van de Water}}}, \bibinfo {author}
  {\bibfnamefont {P.}~\bibnamefont {{Vahle}}}, \bibinfo {author} {\bibfnamefont
  {P.}~\bibnamefont {{Vogel}}}, \bibinfo {author} {\bibfnamefont {C.~W.}\
  \bibnamefont {{Walter}}}, \bibinfo {author} {\bibfnamefont {D.}~\bibnamefont
  {{Wark}}}, \bibinfo {author} {\bibfnamefont {M.}~\bibnamefont {{Wascko}}},
  \bibinfo {author} {\bibfnamefont {D.}~\bibnamefont {{Webber}}}, \bibinfo
  {author} {\bibfnamefont {H.}~\bibnamefont {{Weerts}}}, \bibinfo {author}
  {\bibfnamefont {C.}~\bibnamefont {{White}}}, \bibinfo {author} {\bibfnamefont
  {H.}~\bibnamefont {{White}}}, \bibinfo {author} {\bibfnamefont
  {L.}~\bibnamefont {{Whitehead}}}, \bibinfo {author} {\bibfnamefont {R.~J.}\
  \bibnamefont {{Wilson}}}, \bibinfo {author} {\bibfnamefont {L.}~\bibnamefont
  {{Winslow}}}, \bibinfo {author} {\bibfnamefont {T.}~\bibnamefont
  {{Wongjirad}}}, \bibinfo {author} {\bibfnamefont {E.}~\bibnamefont
  {{Worcester}}}, \bibinfo {author} {\bibfnamefont {M.}~\bibnamefont
  {{Yokoyama}}}, \bibinfo {author} {\bibfnamefont {J.}~\bibnamefont {{Yoo}}}, \
  and\ \bibinfo {author} {\bibfnamefont {E.~D.}\ \bibnamefont {{Zimmerman}}},\
  }\href@noop {} {\bibfield  {journal} {\bibinfo  {journal} {ArXiv e-prints}\ }
  (\bibinfo {year} {2013})},\ \Eprint {http://arxiv.org/abs/1310.4340}
  {arXiv:1310.4340 [hep-ex]} \BibitemShut {NoStop}%
\bibitem [{\citenamefont {{Balantekin}}\ and\ \citenamefont
  {{Fuller}}(2013)}]{Balantekin:2013zr}%
  \BibitemOpen
  \bibfield  {author} {\bibinfo {author} {\bibfnamefont {A.~B.}\ \bibnamefont
  {{Balantekin}}}\ and\ \bibinfo {author} {\bibfnamefont {G.~M.}\ \bibnamefont
  {{Fuller}}},\ }\href {\doibase 10.1016/j.ppnp.2013.03.008} {\bibfield
  {journal} {\bibinfo  {journal} {Progress in Particle and Nuclear Physics}\
  }\textbf {\bibinfo {volume} {71}},\ \bibinfo {pages} {162} (\bibinfo {year}
  {2013})},\ \Eprint {http://arxiv.org/abs/1303.3874} {arXiv:1303.3874
  [nucl-th]} \BibitemShut {NoStop}%
\bibitem [{\citenamefont {{Gardner}}\ and\ \citenamefont
  {{Fuller}}(2013)}]{Gardner:2013lr}%
  \BibitemOpen
  \bibfield  {author} {\bibinfo {author} {\bibfnamefont {S.}~\bibnamefont
  {{Gardner}}}\ and\ \bibinfo {author} {\bibfnamefont {G.~M.}\ \bibnamefont
  {{Fuller}}},\ }\href {\doibase 10.1016/j.ppnp.2013.03.001} {\bibfield
  {journal} {\bibinfo  {journal} {Progress in Particle and Nuclear Physics}\
  }\textbf {\bibinfo {volume} {71}},\ \bibinfo {pages} {167} (\bibinfo {year}
  {2013})},\ \Eprint {http://arxiv.org/abs/1303.4758} {arXiv:1303.4758
  [hep-ph]} \BibitemShut {NoStop}%
\bibitem [{\citenamefont {{Abazajian}}\ \emph {et~al.}(2005)\citenamefont
  {{Abazajian}}, \citenamefont {{Bell}}, \citenamefont {{Fuller}},\ and\
  \citenamefont {{Wong}}}]{Abazajian:2005fk}%
  \BibitemOpen
  \bibfield  {author} {\bibinfo {author} {\bibfnamefont {K.}~\bibnamefont
  {{Abazajian}}}, \bibinfo {author} {\bibfnamefont {N.~F.}\ \bibnamefont
  {{Bell}}}, \bibinfo {author} {\bibfnamefont {G.~M.}\ \bibnamefont
  {{Fuller}}}, \ and\ \bibinfo {author} {\bibfnamefont {Y.~Y.~Y.}\ \bibnamefont
  {{Wong}}},\ }\href {\doibase 10.1103/PhysRevD.72.063004} {\bibfield
  {journal} {\bibinfo  {journal} {\prd}\ }\textbf {\bibinfo {volume} {72}},\
  \bibinfo {eid} {063004} (\bibinfo {year} {2005})},\ \Eprint
  {http://arxiv.org/abs/astro-ph/0410175} {astro-ph/0410175} \BibitemShut
  {NoStop}%
\bibitem [{\citenamefont {{Smith}}\ \emph {et~al.}(2006)\citenamefont
  {{Smith}}, \citenamefont {{Fuller}}, \citenamefont {{Kishimoto}},\ and\
  \citenamefont {{Abazajian}}}]{Smith:2006qy}%
  \BibitemOpen
  \bibfield  {author} {\bibinfo {author} {\bibfnamefont {C.~J.}\ \bibnamefont
  {{Smith}}}, \bibinfo {author} {\bibfnamefont {G.~M.}\ \bibnamefont
  {{Fuller}}}, \bibinfo {author} {\bibfnamefont {C.~T.}\ \bibnamefont
  {{Kishimoto}}}, \ and\ \bibinfo {author} {\bibfnamefont {K.~N.}\ \bibnamefont
  {{Abazajian}}},\ }\href {\doibase 10.1103/PhysRevD.74.085008} {\bibfield
  {journal} {\bibinfo  {journal} {\prd}\ }\textbf {\bibinfo {volume} {74}},\
  \bibinfo {eid} {085008} (\bibinfo {year} {2006})},\ \Eprint
  {http://arxiv.org/abs/astro-ph/0608377} {astro-ph/0608377} \BibitemShut
  {NoStop}%
\bibitem [{\citenamefont {{Kneller}}\ \emph {et~al.}(2001)\citenamefont
  {{Kneller}}, \citenamefont {{Scherrer}}, \citenamefont {{Steigman}},\ and\
  \citenamefont {{Walker}}}]{Kneller:2001uq}%
  \BibitemOpen
  \bibfield  {author} {\bibinfo {author} {\bibfnamefont {J.~P.}\ \bibnamefont
  {{Kneller}}}, \bibinfo {author} {\bibfnamefont {R.~J.}\ \bibnamefont
  {{Scherrer}}}, \bibinfo {author} {\bibfnamefont {G.}~\bibnamefont
  {{Steigman}}}, \ and\ \bibinfo {author} {\bibfnamefont {T.~P.}\ \bibnamefont
  {{Walker}}},\ }\href {\doibase 10.1103/PhysRevD.64.123506} {\bibfield
  {journal} {\bibinfo  {journal} {\prd}\ }\textbf {\bibinfo {volume} {64}},\
  \bibinfo {pages} {123506} (\bibinfo {year} {2001})},\ \Eprint
  {http://arxiv.org/abs/astro-ph/0101386} {astro-ph/0101386} \BibitemShut
  {NoStop}%
\bibitem [{\citenamefont {{Shimon}}\ \emph {et~al.}(2010)\citenamefont
  {{Shimon}}, \citenamefont {{Miller}}, \citenamefont {{Kishimoto}},
  \citenamefont {{Smith}}, \citenamefont {{Fuller}},\ and\ \citenamefont
  {{Keating}}}]{Shimon:2010lr}%
  \BibitemOpen
  \bibfield  {author} {\bibinfo {author} {\bibfnamefont {M.}~\bibnamefont
  {{Shimon}}}, \bibinfo {author} {\bibfnamefont {N.~J.}\ \bibnamefont
  {{Miller}}}, \bibinfo {author} {\bibfnamefont {C.~T.}\ \bibnamefont
  {{Kishimoto}}}, \bibinfo {author} {\bibfnamefont {C.~J.}\ \bibnamefont
  {{Smith}}}, \bibinfo {author} {\bibfnamefont {G.~M.}\ \bibnamefont
  {{Fuller}}}, \ and\ \bibinfo {author} {\bibfnamefont {B.~G.}\ \bibnamefont
  {{Keating}}},\ }\href {\doibase 10.1088/1475-7516/2010/05/037} {\bibfield
  {journal} {\bibinfo  {journal} {Journal of Cosmology and Astroparticle
  Physics}\ }\textbf {\bibinfo {volume} {05}},\ \bibinfo {eid} {037} (\bibinfo
  {year} {2010})},\ \Eprint {http://arxiv.org/abs/1001.5088} {arXiv:1001.5088
  [astro-ph.CO]} \BibitemShut {NoStop}%
\end{thebibliography}%

\end{document}